\pdfoutput=1
\documentclass[12pt]{iopart}
\usepackage{iopams}
 
\expandafter\let\csname equation*\endcsname\relax
\expandafter\let\csname endequation*\endcsname\relax
\usepackage{amsmath,amssymb,mathtools}

\usepackage{graphicx,color}
\usepackage{enumerate}

\usepackage[colorlinks,linkcolor=blue,urlcolor=blue,citecolor=blue]{hyperref}

 \newcommand{\erefss}[1]{\eqref{#1}}%
 \newcommand{\erefs}[1]{\eqref{#1}}%
 \newcommand{\frefss}[1]{figures~\ref{#1}} %
 \newcommand{\frefs}[1]{~\ref{#1}} %
 \newcommand{\ffref}[1]{Figure~\ref{#1}} %
 \newcommand{\srefss}[1]{Sections~\eqref{#1}}%
 \newcommand{\srefs}[1]{~\eqref{#1}}%
  \newcommand{\ssref}[1]{Section~\ref{#1}}%
 \newcommand{\aref}[1]{\ref{#1}}%

\newcommand\be{\begin{equation}} 
\newcommand\ee{\end{equation}} 
\begin{document}

\title{Entropy production for partially observed harmonic systems}

\author{Deepak Gupta\textsuperscript{1,}\textsuperscript{2} and Sanjib Sabhapandit\textsuperscript{2}}
\address{Dipartimento di Fisica `G. Galilei', INFN, Universit\'a di Padova, Via Marzolo 8, 35131 Padova, Italy\textsuperscript{1}}
\address{Raman Research Institute,  Bangalore - 560080, India\textsuperscript{2}}
\date{\today}

\begin{abstract}
The probability distribution of the total entropy production in the non-equilibrium steady state follows a symmetry relation called the fluctuation theorem. When a certain part of the system is masked or hidden, it is difficult to infer the exact estimate of the total entropy production. Entropy produced from the observed part of the system shows significant deviation from the steady state fluctuation theorem. This deviation occurs due to the interaction between the observed and the masked part of the system. A naive guess would be that the deviation from the steady state fluctuation theorem may disappear in the limit of small interaction between both parts of the system. In contrast, we investigate the entropy production of a particle in a harmonically coupled Brownian particle system (say, particle A and B) in a heat reservoir at a constant temperature. The system is maintained in the non-equilibrium steady state using stochastic driving. When the coupling between particle A and B is infinitesimally weak, the deviation from the steady state fluctuation theorem for the entropy production of a partial system of a coupled system is studied. Furthermore, we consider a harmonically confined system (i.e., a harmonically coupled system of particle A and B in harmonic confinement). In the weak coupling limit, the entropy produced by the partial system (e.g., particle A) of the coupled system in a harmonic trap satisfies the steady state fluctuation theorem. Numerical simulations are performed to support the analytical results. Part of these results were announced in a recent letter, Europhys. Lett. 115, 60003 (2016).    
\end{abstract}

\maketitle

\noindent\rule{\textwidth}{2pt}
\tableofcontents
\noindent\rule{\textwidth}{2pt}
\markboth{Entropy production for partially observed harmonic systems}{}

\section{Introduction}
\emph{Stochastic thermodynamics} \cite{Sekimoto,Seifert-2,Seifert-ST} aims to extend the classical thermodynamics \cite{Callen} to small-scale systems. These small-scale systems may be a polystyrene bead (Brownian particle), enzymes, DNA, RNA molecules, etc., in a fluctuating environment called heat bath that is in equilibrium with a well-defined temperature $T$. Within the framework of stochastic thermodynamics, the notion of thermodynamical observables such as {\color{black}work, entropy production, heat exchange}, etc., can be defined at the level of a trajectory of a non-equilibrium ensemble. The heat exchange by a small system with the surrounding  equilibrium environment, work done by external forces on the small system, and the change in its internal energy satisfy the \emph{first law of thermodynamics} (energy conservation) even for a single stochastic trajectory of any time duration $\tau$. 
Since the work done on such small systems is comparable to the thermal energy $k_B T$ (where $k_B$ is 
the Boltzmann's constant), once in a while, it is expected 
to observe the reverse action of these systems by consuming heat from the surrounding bath. For 
large scale system, we call it the \emph{violations of the second law of thermodynamics}. 
\emph{Fluctuation theorems} 
\cite{Evans-Cohen,Evans-Searles,Searles-Evans,Searles2001,Gallavotti-Cohen,Gallavotti1995,Kurchan,Lebowitz-Spohn,Crooks1998,Crooks-1,Crooks-2,Seifert-1,Seifert-2,Harris} are the relations that measure this violation in terms of the ratio of the probability of positive entropy production to that of negative entropy production. There has been  a great amount of research on understanding the validity of fluctuation theorems for various stochastic quantities such as {\color{black}heat exchange, work}, power injection, entropy production, etc.
 \cite{SS-1,SS-2,Apal-1,Apal-2,Verley-1,VanZon-1,VanZon-2,VanZon-3,Wang,Visco,Kundu,Farago2002,Mazonka,Fogedby2011,Fogedby2012,Ciliberto-1,Ciliberto-2,Ciliberto-3,Chun-shang}.  

Consider a system in contact with a heat bath of
constant temperature, and is maintained out of equilibrium using a steady supply of energy provided by an external driving such as time-dependent external fields, stochastic forces, periodic driving, shear flow, etc. Similarly, a system can also be driven away from equilibrium by connecting its ends to heat and/or particle reservoirs of different temperatures and/or chemical potentials. In general, a stochastic quantity $A$ such as {\color{black}work, heat exchange}, power injection, entropy production, etc., is a functional of the fluctuating trajectories. Hence, these quantities are described by probability distributions. When the probability distribution of $A$ satisfies the relation 
\begin{equation}
\dfrac{P(A=+a\tau)}{P(A=-a\tau)}\sim e^{a \tau},
\label{SSFT}
\end{equation}   
for large time $\tau$, we say that a fluctuation theorem holds for $A$. Here sign "$\sim$" implies the logarithmic equality:
\begin{equation}
\lim_{\tau\to\infty}\dfrac{1}{\tau}\ln \dfrac{P(A=+a\tau)}{P(A=-a\tau)}=a.
\end{equation} 
The quantity $A$ is an extensive quantity that scales with the observation time $\tau$. Therefore, it is clear from \eref{SSFT} that, as the observation time gets longer, the system will appear as time irreversible,  which is consistent with the second law of thermodynamics. As an example, consider a thermal conductor connected to a temperature gradient, and let $A$ be the amount of heat that flows for a duration $\tau$. The quantity $A$ is taken to be positive when heat flows down the temperature gradient, and it is considered negative when the heat
flows in the opposite direction. {\color{black}When the observable $A$ is measured along the stochastic trajectories emanating from the steady state ensemble, the resulting fluctuation theorem is called the steady state fluctuation theorem (SSFT). It is shown in Refs. \cite{Seifert-1,Seifert-2} that, if $A$ denotes the total entropy production, incorporating the change in the system entropy and the entropy change of the surrounding bath in a given observation time, then the SSFT is valid for $A$ for all time, i.e., in this case, the sign "$\sim$" is replaced by the equality sign "$=$" in \eref{SSFT}. }

Consider a system of $n$ interacting degrees of freedom (DOFs). The time scale of relaxation of these DOFs is much larger than that of the bath DOFs. The state of the system at a time $t\in [0,\tau]$ is represented by $x(t):=(x_1(t),x_2(t),\dots,x_n(t))$ in the phase space, which evolves according to a stochastic dynamics. In practice, there can be some technical difficulties due to which one cannot access the whole system. Suppose $m$-DOFs of the system are experimentally observed which we call a subsystem or partial system, i.e., $x_s(t):=(x_1(t),x_2(t),\dots,x_m(t))$ where $m<n$. In such case, the observable statistics of a partial system might not be same as that of a complete system. In fact, a lot of work has been done in the area of partial measurement. {\color{black}For example, Shiraishi \emph{et al.} \cite{Shiraishi} showed that the partial entropy production for a subset of all transitions satisfies the integral fluctuation theorem. Similarly, Kawaguchi \emph{et al.} \cite{Kawaguchi} proved the integral fluctuation theorem for the hidden entropy production. Rahav \emph{et al.} \cite{Rahav-Jarzynski} defined an entropy production for coarse-grained dynamics (i.e., \emph{coarse-grained entropy}) and the deviation from the fluctuation theorem was studied. The effect of coarse-graining on the entropy production can also be seen in Refs. \cite{Puglisi,Yli}. Mehl \emph{et al.} \cite{JMehl} experimentally studied the deviation from the fluctuation theorem for the total entropy production by observing one of the beads (partial entropy production as defined later) in the non-equilibrium steady state of coupled paramagnetic bead system with the interaction strength. Similarly, Borrelli \emph{et al.} \cite{Borrelli} investigated the fluctuation theorem for the entropy production of one of the single-electron boxes in a system of coupled single-electron boxes. In Ref. \cite{Lacoste}, Lacoste \emph{et al.} showed that the Gallavotti-Cohen symmetry \cite{Lebowitz-Spohn} is present in a purely ratchet model and this symmetry is preserved for flashing ratchet only when one includes both chemical and mechanical DOFs in the description. In an experimental work, Ribezzi-Crivellari \emph{et al.} \cite{Ribezzi-Crivellari} inferred the full work distribution from the partial work measurement using the Crooks fluctuation theorem.} Similar studies on partial observation can also be found in Refs. \cite{Christian,Chun,Patrick,Kuan,Polettini,Todd,Uhl,Kahlen,eff-espo,Simone,Esposito}. In all above references, the cause of the partial observation was either due to coarse-graining or inaccessibility of certain DOFs. Within the context of fluctuation of entropy production, the total entropy production of a subsystem may not obey the SSFT when the interactions among DOFs of the complete system are significantly large. In contrast to a naive understanding, we
\cite{Deepak} have given a mechanism under which such partial measurement leads to a new fluctuation theorem for total entropy production of a partial system \emph{in the limit of vanishing interaction} among the observed and hidden DOFs. Moreover, we have briefly reported a technique to nullify the effect of the weak coupling of hidden variables on the observed ones using harmonic confinement. Similarly, in Ref.  \cite{Deepak-HT}, we have shown the deviation from the steady state fluctuation theorem for entropy production of a partial system in a heat transport model in the weak coupling limit. 

In this paper, we consider two different model systems where two Brownian particles are interacting harmonically with each other  in the absence of a confining potential (model 1) and in the presence of harmonic confinement (model 2), in a heat bath at temperature $T$. In contrast to Ref. \cite{Deepak-HT}, each particle is driven using an external stochastic Gaussian force. The given system generates entropy. Here, we consider two definitions of total entropy production of one of the particles in the coupled system: partial and apparent entropy production (see \sref{def}). In the non-equilibrium steady state, fluctuation theorem for both definitions of entropy productions and also for both model systems is tested. 
Part of the results were announced earlier in~\cite{Deepak} without giving details.

The remaining of the paper is organized as follows. {\color{black}We first give a brief outline of the paper in \sref{outline}}. In \sref{model}, we present model 1.  The definitions of partial and apparent entropy production $\Delta S^A_{tot}$ are introduced in \sref{def}. \ssref{fp-eqn} contains the Fokker-Planck equation for the moment generating function of the partial and apparent entropy production and its formal solution in the large time limit. Further, we invert the moment generating function $\langle e^{ -\lambda \Delta S^A_{tot}}\rangle\approx g(\lambda)e^{(\tau/\tau_\gamma) \mu(\lambda)}$, to obtain the probability density function $P(\Delta S^A_{tot})$. In \sref{mu-delta-lambda}, we show the computation of $\mu(\lambda)$ in the limit of coupling strength tending to zero ($\delta\to0$). The large deviation function $I(s)$, asymmetry function $f(s)=I(s)-I(-s)$, and the fluctuation theorem are discussed in \sref{ldf-ft}. In \sref{num-sim}, we give the comparison of numerical simulations with the analytical predictions. We discuss the results for model 2 in \sref{model2}. The method of numerical simulation is explained in \sref{method-sim}. Finally, we summarize our paper in \sref{summ}. In \aref{appendix-1}, we present the complete calculation to obtain the moment generating function of the partial and apparent entropy production for model 1. {\color{black} A discussion on branch point singularities of $\mu(\lambda)$ and the contour that distinguishes regions of different possibilities of singularities is given in \aref{BP-EQ-C}. } We discuss the large deviation function and its continuity properties in \aref{section-LDF} and \aref{cont-LDF}, respectively, in the limit $\delta\to0$. The continuity properties of the asymmetry function $f(s)$ in the limit $\delta\to0$ are shown in \aref{cont-asymm-function}. The computation of the moment generating function of the partial and apparent entropy production for model 2 is given in \aref{appendix}.

\begin{figure}
\includegraphics[width=.5\textwidth]{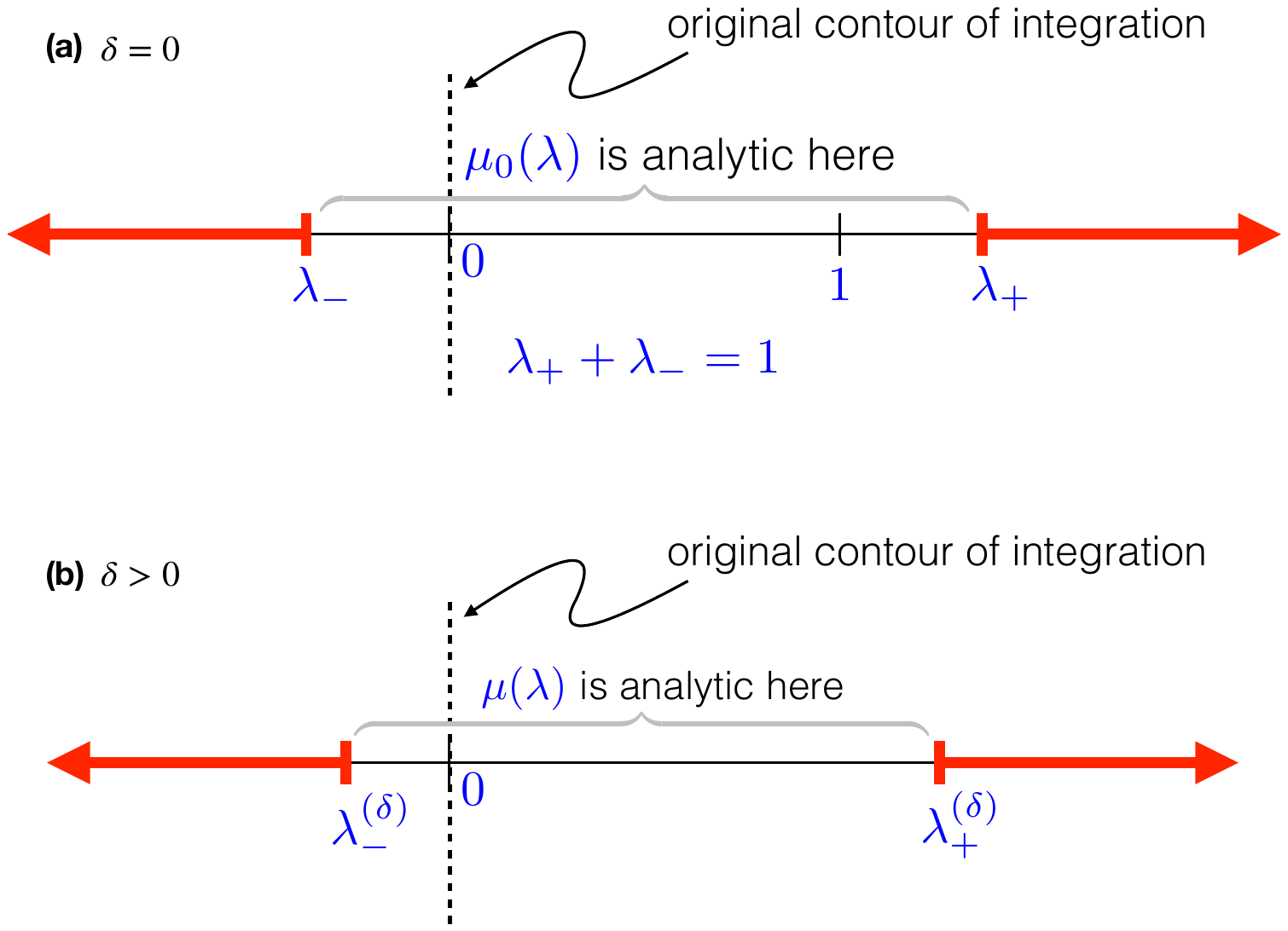}
\includegraphics[width=.55\textwidth]{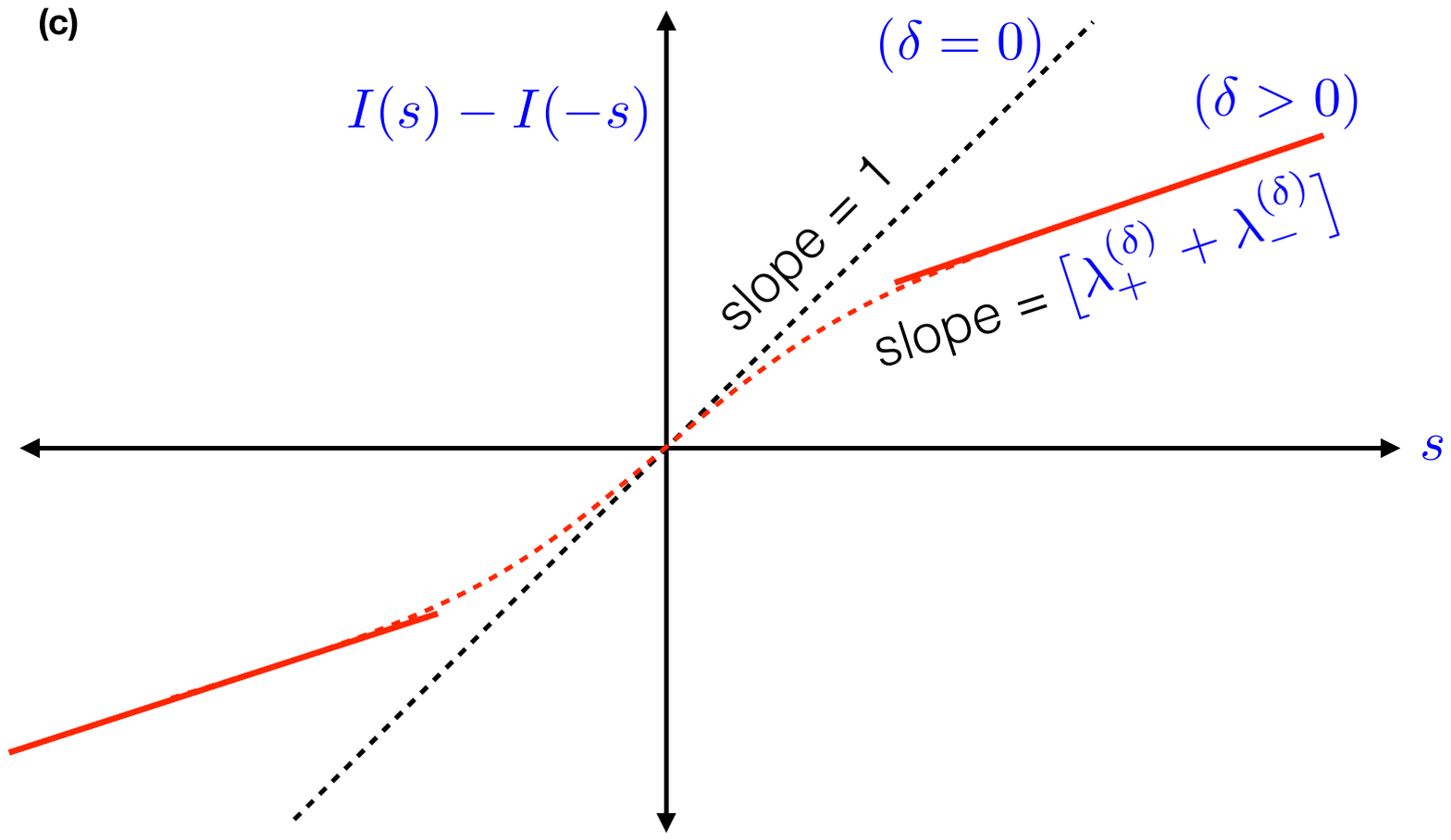}
\caption{\label{branch-points-fig} (a) For $\delta=0$, $\mu_0(\lambda)$ has two branch-points at $\lambda_\pm$ respectively and the two branch-cuts in $(-\infty, \lambda_-)$ and $(\lambda_+,\infty)$ are shown by the red thick line. (b) For $\delta >0$, the branch-points shift to new locations $\lambda_\pm^{(\delta)}$ respectively.  (c) For $\delta=0$, the fluctuation theorem dictates that $I(s)-I(-s)=s$, i.e., the slope is unity. On the other hand, for $\delta >0$, the slope as $s\to\infty$,  tends to $\lambda_+^{(\delta)}+\lambda_-^{(\delta)}$, which need not be unity. 
}
\end{figure}

\section{Outline}
\label{outline}
Let us first outline this paper, before jumping into the details.  We consider a system (model 1) of two Brownian particles  (say A and B) with a harmonic coupling (with a dimensionless coupling parameter $\delta$) and driven by external random forces [see \fref{system}], which evolves by the Langevin equations \eqref{Fd-1}-\eqref{Fd-3}. We consider the entropy production due to only one of the particles (particle A) defined by \eref{pep}, which we term  as the \emph{partial entropy production}. We also introduce another entropy production, that we term as \emph{apparent entropy production}, which is defined by considering only particle A, uncoupled particle (as if the second particle is not present), although in the computation the full dynamics is taken into account.  In fact, for the ease of computation, both the definitions of the entropy production can be combined  into a single expression given by  \eref{Stot-model1}.  At large time, the probability density function of the time averaged entropy production  $s=(\tau/\tau_\gamma)^{-1} \Delta S^A_{tot}$, of the partial system,  follows the large deviation form $p(s) \sim \exp\bigl[(\tau/\tau_\gamma) I(s) \bigr]$,  given by \eref{LDform.0}.  To compute the large deviation function $I(s)$, we first obtain the moment generating function as  $\langle e^{ -\lambda \Delta S^A_{tot}}\rangle\approx g(\lambda)e^{(\tau/\tau_\gamma) \mu(\lambda)}$ for large $\tau$, where $\mu(\lambda)$ is given by \eref{mu-int} and the prefactor $g(\lambda)$ is given by \eref{g-lambda}. For $\delta=0$, i.e., when particle A is uncoupled from  B, $\mu(\lambda)$, which we denote it by $\mu_0(\lambda)$ and is given by \eref{mu0-ext},  has two branch-point $\lambda_\pm$ respectively [see \fref{branch-points-fig}(a)] with $\lambda_+ +\lambda_-=1$. Both $\mu_0(\lambda)$ and $g_0(\lambda)$ [i.e., $g(\lambda)$ for $\delta=0$, given by \eref{g0-1}] follow the Gallavotti-Cohen symmetry $\mu_0(\lambda) = \mu_0(1-\lambda)$ and $g_0(\lambda)=g_0(1-\lambda)$, which  ensures that the fluctuation theorem is satisfied (at least for large $\tau$). The large deviation function $I(s)$ obtained by the Legendre transform of $\mu_0(\lambda)$ follows the symmetry $I(s)-I(-s)=s$ [see \fref{branch-points-fig}(c)]. Moreover, $I(s)\to \lambda_\mp s$ as $s\to \pm\infty$, giving $I(s)-I(-s) \to [\lambda_+ + \lambda_-] s$ as $s\to \infty$, which is consistent with the above large deviation symmetry as $\lambda_+ +\lambda_-=1$. Now, for $\delta>0$, the branch points shift to new locations $\lambda_\pm^{(\delta)}$ respectively [see \fref{branch-points-fig}(b)] and  $I(s)\to \lambda_\mp^{(\delta)} s$ as $s\to \pm\infty$, giving $I(s)-I(-s) \to [\lambda_+^{(\delta)} + \lambda_-^{(\delta)}] s$ as $s\to \infty$. However, now $\lambda_+^{(\delta)} + \lambda_-^{(\delta)}\not=1$ [see \fref{branch-points-fig}(c)], and hence, the fluctuation theorem is not satisfied as $s\to\infty$.  Naively, one would expect that $\lambda_\pm^{(\delta)} \to \lambda_\pm$ as $\delta\to 0$.  However, it turns out to be the case of singular perturbation [see \sref{mu-delta-lambda}], where $\lambda_\pm^{(\delta)}$ as $\delta\to 0$, do not always tend to $\lambda_\pm $, rather, in some parameter regime, either one or both of $\lambda_\pm^{(\delta)}$ tend to new values $\tilde{\lambda}_\pm$. Whenever it happens, it makes the eventual slope of $I(s)-I(-s)$ as $s\to \infty$ different from unity, which is a deviation from the fluctuation theorem.  We next confine [see \fref{system-pic}] the above two particle coupled system in a harmonic potential (model 2, see \sref{model2}). In this case, we always find that $\lambda_\pm^{(\delta)} \to \lambda_\pm$ as $\delta\to 0$, and consequently, the fluctuation theorem is satisfied [see \fref{results}].
\section{Model 1}
\label{model}
\begin{figure}
  \begin{center}
    \includegraphics[width=0.5\textwidth]{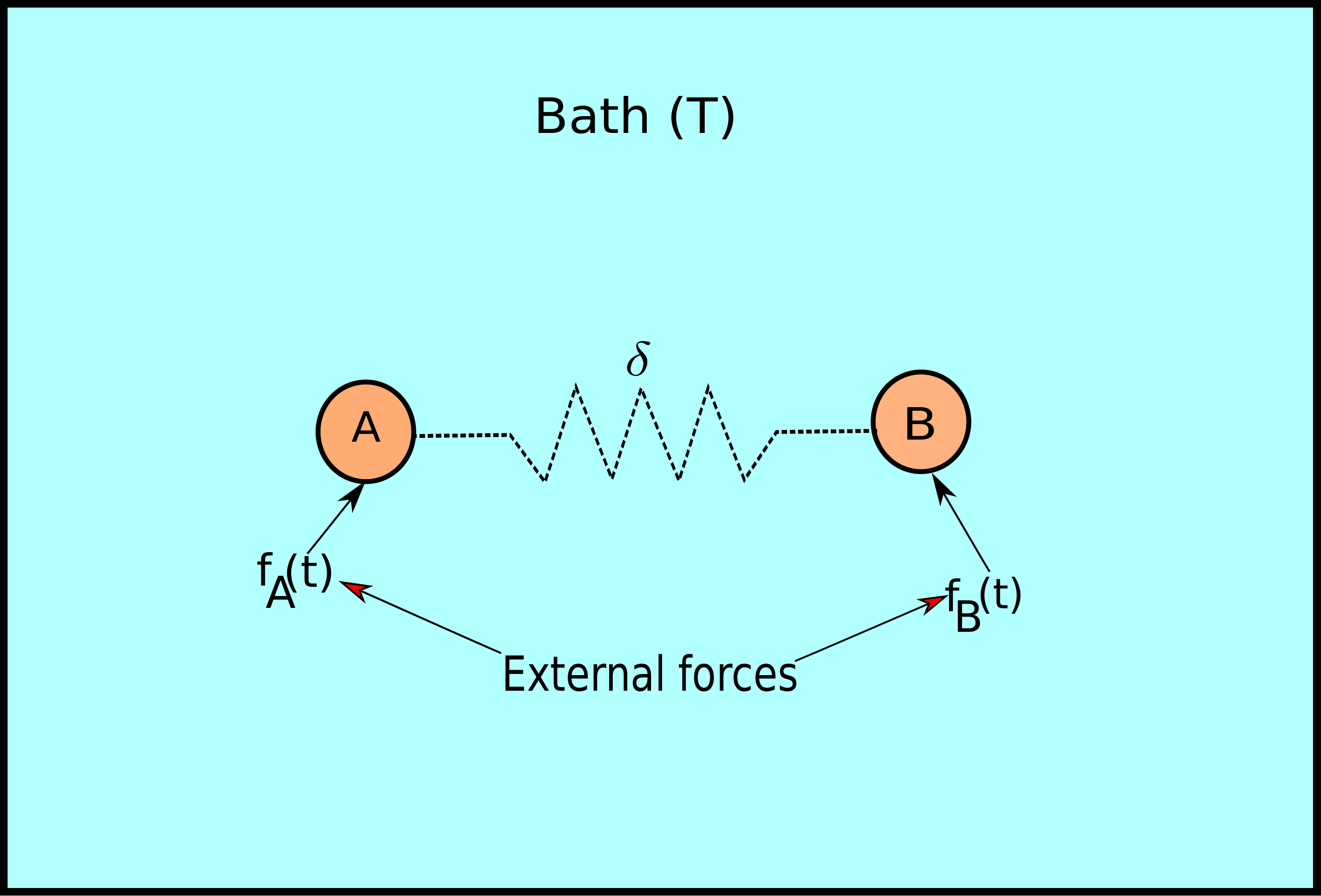}
    \caption{\label{system} Two Brownian particles (A and B) are coupled with a spring of coupling parameter $\delta=2km/\gamma^2$ (dimensionless). The whole system is in contact with a heat bath of constant temperature $T$. The external stochastic Gaussian forces $f_A(t)$ and $f_B(t)$ are acting on the particles A and B, respectively.}
  \end{center}
\end{figure}

Consider two Brownian particles (say, A and B) in an aqueous medium at a constant temperature $T$. For simplicity, we consider the motion along one dimension. Both of these particles are interacting with each other with a harmonic spring of stiffness $k$. The schematic diagram of the system is shown in \fref{system}. The Hamiltonian $\mathcal{H}(y,v_A,v_B)$ of the coupled Brownian particle system is  
\begin{equation}
\mathcal{H}(y,v_A,v_B)=\dfrac{1}{2}mv_A^{2}+\dfrac{1}{2}mv_B^{2}+\dfrac{1}{2}ky^{2},
\label{Ham1}
\end{equation}
where $m$ is the mass of each particle, $y=x_A-x_B$ is the relative position of particle A with respect to particle B, and $v_A$ and $v_B$ are velocities of particle A and B, respectively.

The given system is maintained in the non-equilibrium steady state using external stochastic Gaussian forces. Let $f_A(t)$ and $f_B(t)$ be the external forces acting on particle A and B, respectively. The dynamics of the coupled Brownian particle system is described by the following underdamped Langevin equations \cite{Zwanzig}
\begin{align}
&\dot{y}=v_A(t)-v_B(t),\label{Fd-1}\\
&m\dot{v}_A=-\gamma v_A(t)+\eta_A(t)-k y(t)+f_A(t),\label{Fd-2}\\
&m\dot{v}_B=-\gamma v_B(t)+\eta_B(t)+k y(t)+f_B(t),\label{Fd-3} 
\end{align}
where dot represents a time derivative and $\gamma$ the dissipation constant. The thermal Gaussian noises $\eta_A(t)$ and $\eta_B(t)$ are acting on the particles A and B, respectively, from the heat bath. These thermal Gaussian noises have mean zero and correlation $\langle\eta_{i}(t)\eta_{j}(t')\rangle=2D\delta_{ij}\delta(t-t')$, where $D=\gamma T$.
Similarly, the external forces $f_A(t)$ and $f_B(t)$ have mean zero and correlation $\langle f_A(t)f_A(t^{\prime})\rangle=2 D \theta\delta(t-t^{\prime})$ and $\langle f_B(t)f_B(t^{\prime})\rangle$=$2 D\theta\alpha^{2}\delta(t-t^{\prime})$. In this paper, we consider two different choices of external forces: (1) both external forces are independent of each other, and (2) the force on particle B is correlated with that on particle A: $f_B(t)=\alpha f_A(t)$. Moreover, the external force $f_i(t)$ is uncorrelated with the thermal Gaussian noise $\eta_i(t)$ for all time. Notice that a physical system where the external forces due an electric field would be correlated with each other, could be a coupled colloidal particle system with different electric charges on them. Here, we consider three dimensionless parameters: (1) strength of force $\theta$ acting on particle A relative to that of bath, (2) strength of force $\alpha^2$ acting on particle B relative to that on particle A, and (3) coupling parameter $\delta=2km/\gamma^{2}$. For convenience, we set Boltzmann's constant $k_B=1$ throughout the calculation.
\section{Partial and apparent entropy production}
\label{def}
Incomplete information can be of two types: (1) the observer knows the full system, but intentionally observes the part of the system (i.e. partial information), and (2) the observer is not aware of a hidden part of the system (i.e. apparent information). In both scenarios, actual information of the system is lost. In this paper, we have two models of stochastically driven coupled Brownian particle system (particle A and B) in a heat bath (\srefss{model} and \srefs{model2}). The observable in this paper is the total entropy production of particle A in the coupled Brownian particle system in steady state.  

Let us first consider model 1.
The total entropy production due to particle A in the coupled Brownian particle system (\emph{partial entropy production}) is
\begin{equation}
\Delta \bar{S}^A_{tot}=-\dfrac{Q^A}{T}-\ln \dfrac{P^A_{ss}[v_A(\tau)]}{P^A_{ss}[v_A(0)]}.
\label{pep}
\end{equation}  
In the above equation, $Q^A=\int_0^\tau dt\ [\eta_A(t)-\gamma v_A(t)]v_A(t)$ is the heat absorbed by the Brownian particle A of the coupled system from the heat bath, and $P^A_{ss}(v_A)$ is the steady state probability distribution of the velocity of particle A obtained after integrating the joint distribution $P_{ss}(y,v_A,v_B)$ obtained from \erefss{Fd-1}--\erefs{Fd-3} over the relative distance $y$ and the velocity of particle B. Thus,
\begin{equation}
P^A_{ss}[v_A(\tau)]=\frac{e^{-v^2_A(\tau)/(2 H_P)}}{\sqrt{2\pi H_P}},
\label{ss-P}
\end{equation}
where $H_P$ is given by
\begin{align}
H_P&=\lim_{\tau\to\infty}\langle [v_A(\tau)-\langle v_A(\tau)\rangle]^2 \rangle\nonumber\\
&=\dfrac{D[(2+\theta+\alpha^{2}\theta)mk+2(1+\theta)\gamma^{2}]}{2m\gamma(\gamma^{2}+mk)},
\end{align}
for both choices of external forces. In \eref{pep}, the first and second terms on the right-hand side are, respectively, the entropy change in the bath and the system entropy production due to particle A in the coupled system shown in \fref{system} \cite{Seifert-1,Seifert-2}.

Using \erefss{Fd-2}, \erefs{pep} and \erefs{ss-P}, we rewrite the partial entropy production $\Delta \bar{S}^A_{tot}$ as
\begin{align}
\Delta \bar{S}^A_{tot}=\dfrac{1}{T}\int_0^\tau dt\ [f_A(t)-k y(t)]v_A(t)-\dfrac{1}{2}\bigg[\dfrac{m}{T}-\dfrac{1}{H_P}\bigg][v_A^2(\tau)-v_A^2(0)].
\label{FullPEP}
\end{align}
In the above equation, the integral shown on the right-hand side follows the Stratonovich rule of integration  \cite{Sekimoto}.

In the case of \emph{apparent entropy production}, the observer is not aware of the hidden particle present in the system. Let us call particle B the hidden particle. Therefore, he/she constructs the total entropy production for particle A as follows. Since, for the observer, particle A is the only one
present in the heat bath (she/he being unaware of the presence of particle
B), its velocity evolves according to the following underdamped
Langevin equation:
\begin{equation}
m\dot{v}_A=-\gamma v_A(t)+\eta_A(t)+f_A(t).
\label{Pdynamics}
\end{equation}
Therefore, apparent entropy production for particle A is obtained by substituting $k=0$ in \eref{FullPEP}:
\begin{align}
\Delta \tilde{S}^A_{tot}=\dfrac{1}{T}\int_{0}^{\tau}dt\ f_A(t) v_A(t)-\dfrac{1}{2}\bigg[\dfrac{m}{T}-\dfrac{1}{H_A}\bigg][v_A^2(\tau)-v_A^2(0)].
\label{FullAEP}
\end{align}
In the above equation, 
\begin{equation}
H_A=H_P\big|_{k=0}=\dfrac{D(1+\theta)}{m \gamma}.
\end{equation}
Both definitions of entropy production [see \erefss{FullPEP} and \erefs{FullAEP}] can be jointly written as
\begin{align}
\Delta S^A_{tot}=\dfrac{1}{T}\int_0^\tau dt\ f_A(t)v_A(t)-\dfrac{\Pi k}{T}\int_0^\tau dt\ y(t) v_A(t)-\dfrac{1}{2}\bigg[\dfrac{m}{T}-\dfrac{1}{H}\bigg][v_A^2(\tau)-v_A^2(0)],
\label{Stot-model1}
\end{align}
where $\Pi=1$ and $\Pi=0$ correspond to partial and apparent entropy production, respectively, and $H=\Pi H_P+(1-\Pi) H_A$.

It is important to note that in both types of incomplete information, the underlying dynamics of the coupled system is given by \erefss{Fd-1}--\erefs{Fd-3} \cite{Deepak}. Therefore, we compute the distribution of $\Delta S^A_{tot}$ \erefs{Stot-model1} subject to the actual dynamics given by \erefss{Fd-1}--\erefs{Fd-3}. 

The column vector $U=(y,v_A,v_B)^T$ is linearly dependent on thermal Gaussian noises and external stochastic Gaussian forces [see \erefss{Fd-1}--\erefs{Fd-3}]. Therefore, the steady state distribution of $U$ is a Gaussian distribution [see \eref{SS-U}]. Notice that the quantity $\Delta S^A_{tot}$ depends on thermal and external noises quadratically [see \erefss{Fd-1}--\erefs{Fd-3}, and \erefs{Stot-model1}]. Therefore, the distribution $P(\Delta S^A_{tot})$ is expected to be non-Gaussian.

We define
\begin{equation}
W=\dfrac{1}{T}\left[\int_0^\tau dt\ f_A(t)v_A(t)-\Pi k\int_0^\tau dt\ y(t) v_A(t)\right].
\label{Wtot}
\end{equation}
While the first term inside the bracket is the work done by the stochastic external force $f_A(t)$ on the Brownian particle A \cite{Gomez-2010,SS-2}, the second term  is the interaction energy. Both of these terms are scaled by the  temperature $T$ of the heat bath.

In this paper, we give two model systems (\srefss{model} and \srefs{model2}) driven away from the equilibrium. The quantity of interest is the entropy production from a part  (e.g., particle A) of the coupled system. The steady state fluctuation theorem \erefs{SSFT} would not be satisfied for the partial and apparent entropy productions as we are observing a part of the coupled system. An interesting question arises what would happen if one takes the limit of coupling tending to zero? Therefore, the aim of this paper is to understand the steady state fluctuation theorem for such entropy productions in the \emph{weak coupling limit} ($\delta\to0$).
\section{Probability density function and fluctuation theorem}
\label{fp-eqn}
The entropy production $\Delta S^A_{tot}$ \erefs{Stot-model1} can be written as
\begin{equation}
\Delta S^A_{tot}=W-\dfrac{1}{2}\bigg[\dfrac{m}{T}-\dfrac{1}{H}\bigg][v_A^2(\tau)-v_A^2(0)],
\label{Stot-2}
\end{equation}
where the quantity $W$ is a functional of the stochastic trajectories and the second term on the right-hand side depends on the initial $v_A(0)$ and the final velocity $v_A(\tau)$ of the Brownian particle A. Therefore, the conditional moment generating function for $\Delta S^A_{tot}$ is 
\begin{align}
Z(\lambda,U,\tau|U_{0})&=\big\langle e^{-\lambda \Delta S^A_{tot}} \delta[U-U(\tau)]\big\rangle_{U_0}\nonumber\\&=Z_{W}(\lambda,U,\tau|U_0)~e^{(\lambda/2)(m T^{-1}-H^{-1})(U^T\Sigma U-U_0^T\Sigma U_0)},
\label{RCF}
\end{align}
where $Z_W(\lambda,U,\tau|U_0)=\big\langle e^{-\lambda W}\delta[U-U(\tau)] \big\rangle_{U_0}$, and $\Sigma$ is a square matrix whose elements are given by $\Sigma_{ij}=\delta_{i2}\delta_{ij}$ with $\{i,j\}=\{1,2,3\}$. In the above equation, angular brackets represent the average over the set of all trajectories from the fixed initial variable $U_0$ to the final variable $U$, and $\lambda$ is the conjugate variable with respect to $\Delta S_{tot}^A$ in the Fourier transform.

The evolution of the restricted moment generating function $Z_W(\lambda,U,\tau|U_0)$ is governed by the following Fokker-Planck equation \cite{Risken} 
\begin{equation}
\dfrac{\partial Z_W(\lambda,U,\tau|U_{0})}{\partial \tau}=\mathcal{L}_{\lambda}Z_W(\lambda,U,\tau|U_{0})
\label{WFPE}
\end{equation}
with the initial condition $Z_W(\lambda,U,\tau=0|U_0)=\delta(U-U_0)$. 
In the above equation, the Fokker-Planck operator $\mathcal{L}_{\lambda}$ is 
\begin{align}
\mathcal{L}_{\lambda}&=\dfrac{1}{m}\sum_{i=A,B}\left[\dfrac{\partial \mathcal{H}}{\partial x_{i}}\dfrac{\partial}{\partial v_{i}}-\dfrac{\partial \mathcal{H}}{\partial v_{i}}\dfrac{\partial}{\partial x_{i}}\right]+\dfrac{D(1+\theta)}{m^{2}}\dfrac{\partial^{2}}{\partial v_A^{2}}+\dfrac{\gamma v_A}{m}(1+2\lambda\theta)\dfrac{\partial }{\partial v_A}\nonumber\\&+\dfrac{\gamma }{m}(v_B+2C\lambda \alpha\theta v_A)\dfrac{\partial }{\partial v_B}
+\gamma\bigg[\dfrac{2}{m}+\dfrac{\lambda}{D}\bigg(\Pi k y v_A+\dfrac{D\theta}{m}\bigg)\bigg]+\dfrac{\lambda^2\gamma^2v^2_{_{A}}\theta}{D}\nonumber\\&+\dfrac{D(1+\theta \alpha^{2})}{m^{2}}\dfrac{\partial^{2}}{\partial v_B^{2}}+\dfrac{2CD\theta\alpha}{m^2}\dfrac{\partial^2}{\partial v_A\partial v_B},
\end{align}
where we have used the correlation parameter $C$ defined as
\begin{equation}
C=
\begin{cases}
0\hspace{0.8cm} \text{for}\ \langle f_A(t)f_B(t^\prime)\rangle=0\quad \forall\quad t,t^\prime,\\
1\hspace{0.8cm} \text{for}\ f_B(t)=\alpha f_A(t).
\end{cases} 
\label{C-parameter}
\end{equation}
It is difficult to obtain the solution of Fokker-Planck equation \erefs{WFPE}. Nevertheless, a formal solution of it in the large time limit ($\tau\to\infty$) can be written as
\begin{align}
Z_W(\lambda,U,\tau|U_{0})= \chi(U_{0},\lambda)\Psi(U,\lambda)e^{(\tau/\tau_{\gamma}) \mu(\lambda)}+\dots,
\label{factor}
\end{align}
where $\tau_\gamma=m/\gamma$ is the characteristic time, $\mu(\lambda)$ is the largest eigenvalue of the Fokker-Planck operator $\mathcal{L}_{\lambda}$, and the corresponding right eigenfunction is $\Psi(U,\lambda)$:
$\mathcal{L}_{\lambda}\Psi(U,\lambda)=\mu(\lambda)\Psi(U,\lambda).$
In \eref{factor}, $\chi(U_{0},\lambda)$ is the projection of initial state onto the left eigenvector of Fokker-Planck operator $\mathcal{L}_\lambda$ corresponding to the largest eigenvalue $\mu(\lambda)$. These left and right eigenfunctions satisfy the normalization condition:
$\int dU\ \chi(U,\lambda)\Psi(U,\lambda)=1.$

The moment generating function $Z(\lambda)$ for $\Delta S^A_{tot}$ is obtained by integrating the restricted moment generating function $Z(\lambda,U,\tau|U_0)$ given in \eref{RCF}, over the initial state $U_0$ sampled from the steady state distribution $P_{ss}(U_0)$ and the final state $U$:
\begin{align}
Z(\lambda)&=\int dU \int dU_{0}\ P_{ss}(U_{0})\ Z(\lambda,U,\tau|U_{0})\nonumber\\&=g(\lambda)\ e^{(\tau/\tau_{\gamma}) \mu(\lambda)}+\dots,
\label{chfun}
\end{align}
where
\begin{align}
g(\lambda)=&\int dU \int dU_{0}\ P_{ss}(U_{0})\chi(U_{0},\lambda)\Psi(U,\lambda)~ e^{(\lambda/2)(m T^{-1}-H^{-1})(U^T\Sigma U-U_0^T\Sigma U_0)}.
\end{align}

To obtain the probability density function for $\Delta S^A_{tot}$, we invert the moment generating function $Z(\lambda)$ using  the inverse Fourier transformation in the complex $\lambda$ space
\begin{equation}
P(\Delta S^A_{tot}=s\tau/\tau_{\gamma})=\dfrac{1}{2\pi i}\int_{-i\infty}^{+i\infty} d\lambda\ Z(\lambda)\ e^{\lambda s\tau/\tau_{\gamma}},
\end{equation}
where  $s=\Delta S^A_{tot} \tau_\gamma/\tau$ is the scaled variable.
The contour of integration in the above equation is along the imaginary axis through the origin of the complex $\lambda$-plane. In the large-$\tau$ limit ($\tau\gg\tau_{\gamma}$), using \eref{chfun}, we get
\begin{equation}
P(\Delta S^A_{tot}=s\tau/\tau_{\gamma})\approx\dfrac{1}{2\pi i}\int_{-i\infty}^{+i\infty} d\lambda\ g(\lambda)e^{(\tau/\tau_{\gamma}) h_s(\lambda)},
\label{intPDF}
\end{equation}
in which $h_s(\lambda)=\mu(\lambda)+\lambda s$.

The above integral can be approximated using the saddle-point method provided both $\mu(\lambda)$ and $g(\lambda)$ are analytic function of $\lambda$ \cite{Touchette}. Thus,
\begin{equation}
P(\Delta S^A_{tot}=s\tau/\tau_{\gamma})\approx \dfrac{g(\lambda^*)e^{(\tau/\tau_\gamma) h_s(\lambda^*)}}{\sqrt{2 \pi \tau/\tau_\gamma |h_s^{\prime \prime}(\lambda^{*})}|},
\label{saddle}
\end{equation}
where $\lambda^*(s)$ is the saddle point solution of
\begin{equation}
 \dfrac{\partial \mu(\lambda)}{\partial \lambda}\bigg|_{\lambda=\lambda^*}=-s,
 \label{saddle-point}
 \end{equation}
 and 
\begin{equation}
 h_s^{\prime\prime }(\lambda^*):=\dfrac{\partial^2 h_s(\lambda)}{\partial \lambda^2}\bigg|_{\lambda=\lambda^*}.
 \end{equation}
 The fluctuation theorem estimates the ratio of the probability of positive total entropy production and that of negative total entropy production where the latter one is
\begin{align}
P(\Delta S^A_{tot}=-s\tau/\tau_\gamma)\approx\dfrac{1}{2\pi i}\int_{-i\infty}^{+i\infty} d\lambda\ g(\lambda)e^{(\tau/\tau_\gamma) [\mu(\lambda)-\lambda s]}.
\end{align}
If both functions $\mu(\lambda)$ and $g(\lambda)$ satisfy Gallavotti-Cohen (GC) symmetry \cite{Lebowitz-Spohn}, i.e., $\mu(\lambda)=\mu(1-\lambda)$ and $g(\lambda)=g(1-\lambda)$, after some simplifications, we obtain
\begin{align}
P(\Delta S^A_{tot}=-s\tau/\tau_\gamma)\approx \dfrac{e^{-s\tau/\tau_\gamma}}{2\pi i}\int_{1-i\infty}^{1+i\infty} d\lambda\ g(\lambda)e^{(\tau/\tau_\gamma) [\mu(\lambda)+\lambda s]},\nonumber
\end{align}
where the contour of integration is along the imaginary axis at $\mathrm{Re} (\lambda)=1$  of the complex $\lambda$-plane. 
In the absence of singularity in $\mu(\lambda)$ and $g(\lambda)$ between $(1-i\infty,1+i\infty)$ and $(-i\infty,+i\infty)$, the contour of integration can be shifted from $\mathrm{Re} (\lambda)=1$  to the origin of the complex $\lambda$-plane. Therefore, we get
\begin{equation}
P(\Delta S^A_{tot}=-s\tau/\tau_\gamma)\approx \dfrac{e^{-s\tau/\tau_\gamma }}{2\pi i}\int_{-i\infty}^{+i\infty} d\lambda\ g(\lambda)e^{(\tau/\tau_\gamma) [\mu(\lambda)+\lambda s]}.
\label{negintPDF}
\end{equation}
From \erefss{intPDF} and \erefs{negintPDF}, we obtain the relation
\begin{equation}
\dfrac{P(\Delta S^A_{tot}=s\tau/\tau_\gamma)}{P(\Delta S^A_{tot}=-s\tau/\tau_\gamma)}\approx e^{s\tau/\tau_\gamma}.
\label{FT-def}
\end{equation}
The above relation is the fluctuation theorem for $\Delta S_{tot}^A$ in the steady state for a time segment $\tau$. Notice that the sign "$\approx$" indicates that the above identity is true for a large time $\tau$. Therefore, the criteria for any observable to satisfy the fluctuation theorem in the steady state are: (1) the corresponding $\mu(\lambda)$ and prefactor $g(\lambda)$ should be analytic functions for $\lambda\in[0,1]$, and (2) both of these functions must obey GC symmetry.

\section{Moment generating function}
\label{mu-delta-lambda}
The computation of $Z(\lambda)$ can be done using a method developed in \cite{Kundu} and detailed in \aref{appendix-1}. 
We obtain [see \eref{mu-int-1}]
\begin{equation}
\mu(\lambda)=-\dfrac{1}{4\pi}\int_{-\infty}^{\infty}du\ \ln \bigg[1+\dfrac{h(u,\lambda)}{q(u)} \bigg].
\label{mu-int}
\end{equation}
The prefactor term $g(\lambda)$ is given in \eref{g-lambda}.

In the above equation, the function $h(u,\lambda)$ for the first choice of forces $(\text{i.e.}, C=0)$ is 
\begin{align}
h(u,\lambda)=4\theta \lambda(1-\lambda)\big[u^{4}+(1-\delta)u^{2}+\delta^2(2-\Pi)/4]-\lambda\delta^2\big\{\lambda \alpha^2\theta^2(\Pi-1)^2+\lambda \Pi^2\nonumber\\-\Pi\theta[\alpha^2+\lambda\{1-\Pi(1+\alpha^2)\}]\big\}
\label{hu-1}
\end{align}
whereas for the second choice of forces (i.e., $C=1)$
\begin{align}
h(u,\lambda)=4\theta \lambda(1-\lambda)\big[u^{4}+(1-\delta)u^{2}+\delta^2/2]-4 \theta \lambda(1-\lambda\Pi)\delta \alpha u^2
-\lambda \delta^2\big[\lambda \Pi^2\nonumber\\+\theta\{\Pi-\Pi \lambda(2-\Pi)+\alpha(\lambda \Pi-1)(2+\alpha\Pi)\}\big].
\label{hu-2}
\end{align}
The function 
\begin{equation}
q(u)=(1+u^2)\bigl[u^{2}+(u^2-\delta)^2\bigr]
\label{qu-1}
\end{equation}
is same for both choices of external forces and also for both definitions of entropy production.

In the case of a single Brownian particle (see \fref{system} with $\delta$=0) in the heat bath at a temperature $T$ and driven by an external Gaussian white noise, $\mu(\lambda)$ given in \eref{mu-int} reduces to
\begin{equation}
\mu_0(\lambda)=-\dfrac{1}{4\pi}\int_{-\infty}^{\infty}du\ \ln\left[1+\dfrac{4\theta \lambda(1-\lambda)}{u^2+1}\right],
\end{equation} 
which can be solved exactly, and it is given by
\begin{equation}
\mu_0(\lambda)=[1-\nu(\lambda)]/2, \quad \quad \text{where}
\label{mu0-ext}
\end{equation} 
\begin{equation}
\nu(\lambda)=\sqrt{4 \theta (\lambda_{+}-\lambda)(\lambda-\lambda_{-})} 
\label{lambda0-pm}
\end{equation}
in which
\begin{equation*}
\lambda_{\pm}=[1\pm\sqrt{1+{\theta}^{-1}}]/2.
\end{equation*}
In the above equation \erefs{lambda0-pm}, the branch points $\lambda_+>1$ and $\lambda_-<0$. Moreover, the branch points obey  $\lambda_-+\lambda_+=1$.
In this case, one can also compute the prefactor term exactly, and it is given by 
\begin{equation}
g_0(\lambda)=\dfrac{2 \sqrt{\nu(\lambda)}}{1+\nu(\lambda)},
\label{g0-1}
\end{equation}
Here, both $\mu_0(\lambda)$ and $g_0(\lambda)$ are analytic functions of $\lambda\in(\lambda_{-},\lambda_{+})$ and both of them obey GC symmetry, i.e., $\mu_0(\lambda)=\mu_0(1-\lambda)$ and $g_0(\lambda)=g_0(1-\lambda)$. Hence, the SSFT is satisfied (see \sref{fp-eqn}). However, the function $\mu(\lambda)$ does not satisfy the GC symmetry in the presence of coupling ($\delta\neq 0$). Therefore, the steady state fluctuation theorem would not be satisfied for both partial and apparent entropy productions.

In the following, we show how to compute the integral given in \eref{mu-int} (i.e. for $\delta\neq0$). We rewrite this integral using integration by parts
\begin{equation}
\mu(\lambda)=\dfrac{1}{4\pi} 
\int_{-\infty}^{+\infty}du\,
\frac{u \bigl[h'(u,\lambda) q(u) - h(u,\lambda)
q'(u) \bigr]}{q(u) \bigl[q(u)+h(u,\lambda)\bigr]}, 
\label{mu2}
\end{equation}
where $'$ represents a derivative with respect to $u$. In the
above integrand, the factors in the denominator are polynomials in the
variable $u^2$. Therefore, the denominator can be factored in terms of
the roots of the polynomials:
\begin{align*}
q(u)+h(u,\lambda)=\prod_{j} (u^2 -u_j^2)\quad \text{and}\quad q(u)=\prod_{k} (u^2-w_k^2).
\end{align*}
This gives a set of simple poles in the complex $u$-plane. Consequently, evaluating the integral by using the residue
theorem, and using $q(u_j) + h(u_j,\lambda)=0$ and $q(w_k)=0$, gives
\begin{equation}
\mu(\lambda)=\frac{i}{2} \left[ 
\sum_j u_j -\sum_k w_k
\right] , 
\label{mu-in terms of roots}
\end{equation}
where $\{u_j\}$ and $\{w_k\}$ are the zeros of the polynomials $[q(u)
+ h(u,\lambda)]$ and $q(u)$ respectively, that lies on the upper
half of the complex $u$-plane.  Clearly, $\{u_j\}$ are functions of
$\lambda$ while $\{w_k\}$ are independent of $\lambda$. The computation of
$\mu(\lambda)$ using these residues is quite cumbersome. Nevertheless,
one can compute $\mu(\lambda)$ numerically provided the range of $\lambda$ is known within
which $\mu(\lambda)$ remains a real function.
The range of $\lambda$ depends on the choice of $\theta$ and $\alpha$ for given $\delta$.
In the absence of coupling, the saddle point $\lambda_0^*(s)$ [see \eref{saddle-point condition0}] stays within the branch point singularities of $\mu_0(\lambda)$, i.e., $\lambda_0^*(s)\to\lambda_\pm$ as the scaled parameter $s\to\mp\infty$. Therefore, we expect that the saddle point $\lambda^*(s)$ [see \eref{saddle-point}] hits either of the branch point singularities of $\mu(\lambda)$ as $s\to\pm\infty$.   
To obtain these singularities for $\delta>0$, we analyze the argument of
the logarithm in the integrand of \eref{mu-int}, i.e., the terms $q(u)$ and
$q(u)+h(u,\lambda)$.  In all the four cases, we find that
the function $q(u)$ given in \eref{qu-1}, is always positive for any real
$u\in(-\infty, \infty)$. We also find that, 
\begin{align}
h(u,\lambda)= b(u) \lambda - p(u)\lambda^2, 
\label{h}
\end{align}
where the functions $b(u)$ and $p(u)$ are different for all four
cases [see \erefss{hu-1} and \erefs{hu-2}]. In all cases, $p(u)$ is always positive for any
real $u\in(-\infty, \infty)$. The two roots 
\begin{equation}
\lambda_\pm(u)=\dfrac{b(u) \pm \sqrt{b^2(u) + 4 p(u) q(u)}}
{2 p(u)}, 
\label{lambda-pm-u}
\end{equation}
of the quadratic polynomial $q(u) + h(u, \lambda)$, are always real
for any real $u\in(-\infty, \infty)$. Moreover, $\lambda_+(u)$ is
positive and $\lambda_-(u)$ is negative. For a given $u$, the
integrand in \eref{mu-int} is real only within the range $\lambda \in
(\lambda_-(u), \lambda_+(u) )$. Therefore, $\mu(\lambda)$ is real only
within the range $\lambda \in
\bigl(\lambda_-^{(\delta)}, \lambda_+^{(\delta)}\bigr)$, where 
\begin{equation*}
\lambda_-^{(\delta)}=\max_{u} \lambda_-(u) \quad\text{and} \quad
\lambda_+^{(\delta)}=\min_u \lambda_+(u).
\end{equation*}
The equation for the extremum is given by
\begin{equation}
\left.\dfrac{\partial \lambda_{\pm}(u)}{\partial u}\right|_{u=u_{\pm}^{*}}=0.
\label{extremum}
\end{equation}
Consequently, {\color{black}the solution of the equation}
\begin{align}
&\bigg[\sqrt{b^{2}(u^{*}_{\pm})+4p(u^*)q(u^*_{\pm})}\pm b(u^*_{\pm})\bigg]
\bigg[p(u^*_{\pm})b^{\prime}(u^*_{\pm})-p^{\prime}(u^*_{\pm})b(u^*_{\pm})\bigg]\nonumber\\&~~~~~~~~~~~~~~~~~~~~~~~~~~~~~~~~
\pm 2p(u^*_{\pm})\bigg[p(u^*_{\pm})q^{\prime}(u^*_{\pm})-p^{\prime}(u^*_{\pm})q(u^*_{\pm})\bigg]=0.
\label{ustar}
\end{align}
{\color{black}gives $u^*_\pm$. }
Notice that the above equation is true for all $\delta$. In the weak coupling limit ($\delta\to 0$), we can find $u^*_\pm$ using \eref{ustar}. This gives $\lambda_\pm(u)\to \lambda_\pm^{(\delta)}$ as $u\to u^*_\pm$.
\begin{figure*}
\begin{tabular}{cc}
 \includegraphics[width=.42\textwidth]{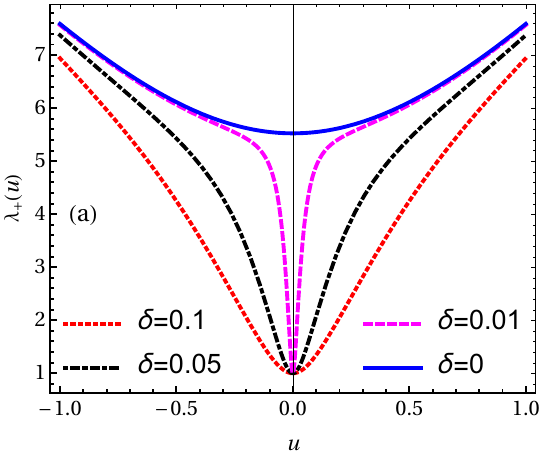}~~~~~~~~&
  \includegraphics[width=.435\textwidth]{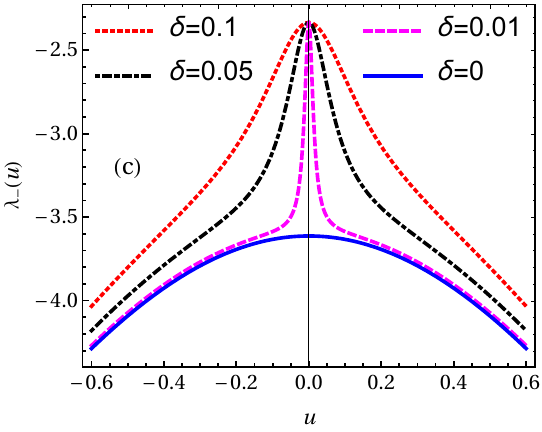}\\
   \includegraphics[width=.43\textwidth]{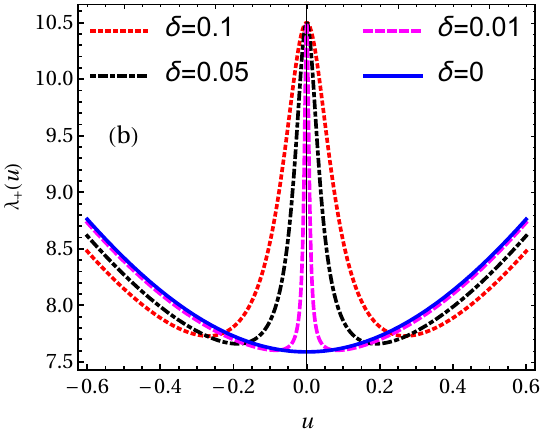}~~~~~~~~&
    \includegraphics[width=.43\textwidth]{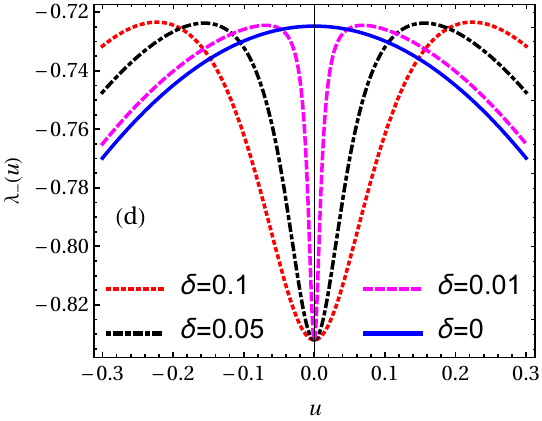}
    \end{tabular}
 \caption{\label{lamb-var}{\color{black}Different possibilities of the variation of $\lambda_\pm(u)$ [see \eref{lambda-pm-u}] with respect to $u$ are shown for different values of $\theta$ and $\alpha$ in which dotted, dashed, and dot-dashed lines correspond to $0<\delta<1$ (coupling is present) whereas solid lines indicate the $\delta=0$ case (coupling is absent). All dotted, dashed, and dot-dashed lines correspond to (a) $\lambda_+''(0)>0$, (b) $\lambda_+''(0)<0$, (c) $\lambda_-''(0)<0$, and (d) $\lambda_-''(0)>0$ [see \erefss{condition-1} and \erefs{condition-2}].}}
 \end{figure*}

\begin{table}[h!]
  \begin{center}
  \scriptsize
    \begin{tabular}{c|c|c} 
       & Partial entropy production  & Apparent entropy production \\
       &&\\
       & ($\Pi=1$) & ($\Pi=0$) \\
       &&\\
      \hline
      &&\\
      Uncorrelated forces& Branch point singularities: $(\lambda_-,\tilde{\lambda}_+)$ in region I  & Branch point singularities: $(\lambda_-,{\lambda}_+)$ in region I\\
      &  \hspace{3.7cm}$(\tilde{\lambda}_-,\tilde{\lambda}_+)$ in region II&\hspace{3.7cm}$({\lambda}_-,\tilde{\lambda}_+)$ in region II\\
      &&\hspace{3.8cm}$(\tilde{\lambda}_-,\tilde{\lambda}_+)$ in region III\\
      &&\\
      ($C=0$)&Equation of contour: $r_1(\theta,\alpha,\delta)=0$& Equations of contour: $r_2(\theta,\alpha,\delta)=0$\\
      &&\hspace{3.cm}$r_3(\theta,\alpha,\delta)=0$\\
      &&\\
      \hline
      &&\\
       Correlated forces& Branch point singularities: $(\lambda_-,\tilde{\lambda}_+)$ in region I  & Branch point singularities: $(\lambda_-,{\lambda}_+)$ in region I \\
       &\hspace{3.7cm}$(\tilde{\lambda}_-,\tilde{\lambda}_+)$ in region II&\hspace{3.7cm}$(\tilde{\lambda}_-,{\lambda}_+)$ in region II\\
      &&\\
      ($C=1$)& Equation of contour: $r_4(\theta,\alpha,\delta)=0$& Equation of contour: $r_5(\theta,\alpha,\delta)=0$\\
      &&\\
    \end{tabular}
  \end{center}
   \caption{{\color{black}Branch point singularities and equations of contours for phase diagrams in \fref{phase-dig}. Their explicit values are given in \aref{BP-EQ-C}. }  }
    \label{tb:table}
\end{table}
\begin{figure*}
\begin{tabular}{cc}
 \includegraphics[width=0.45\textwidth]{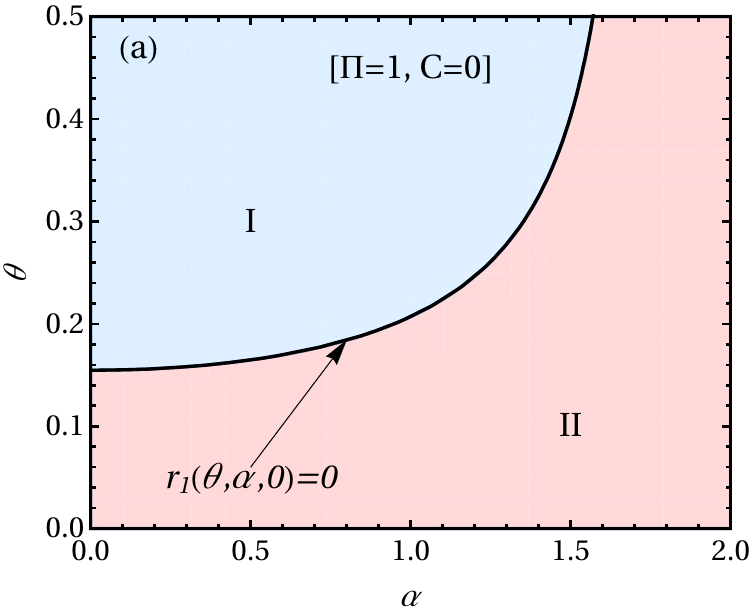}~~~~~~~~~&
  \includegraphics[width=0.45\textwidth]{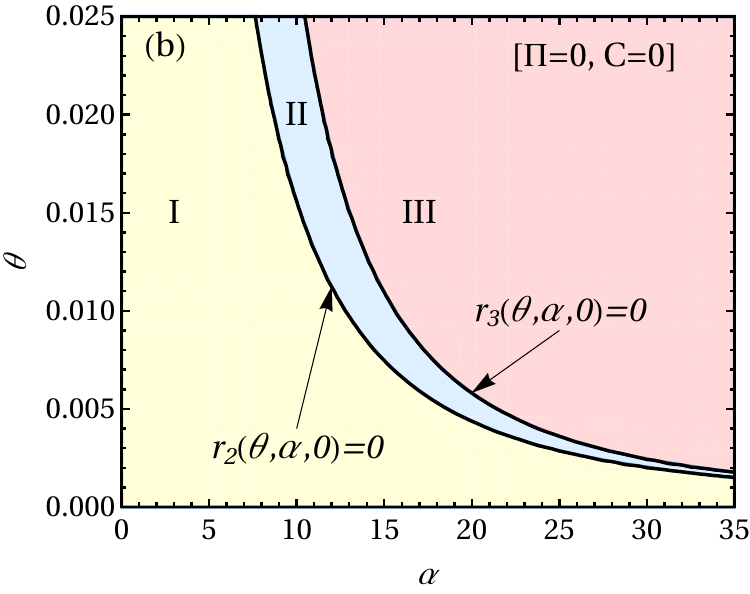}\\
   \includegraphics[width=0.45\textwidth]{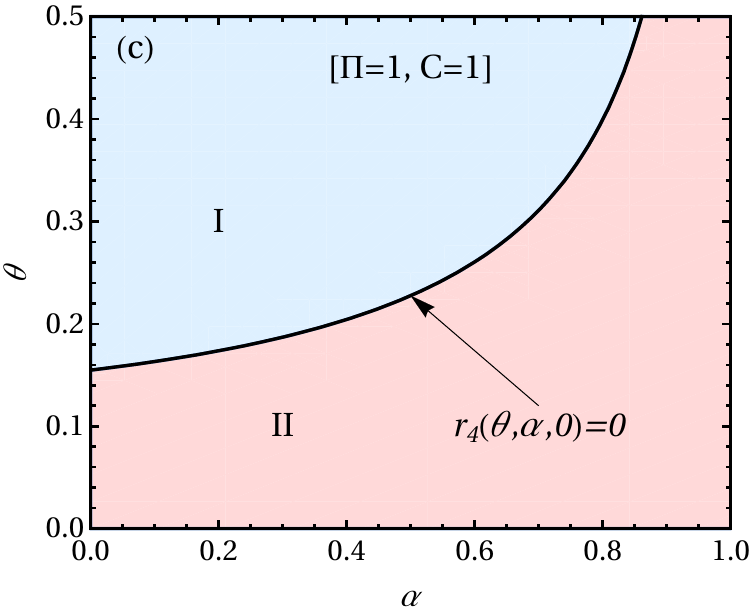}~~~~~~~~&
    \includegraphics[width=0.45\textwidth]{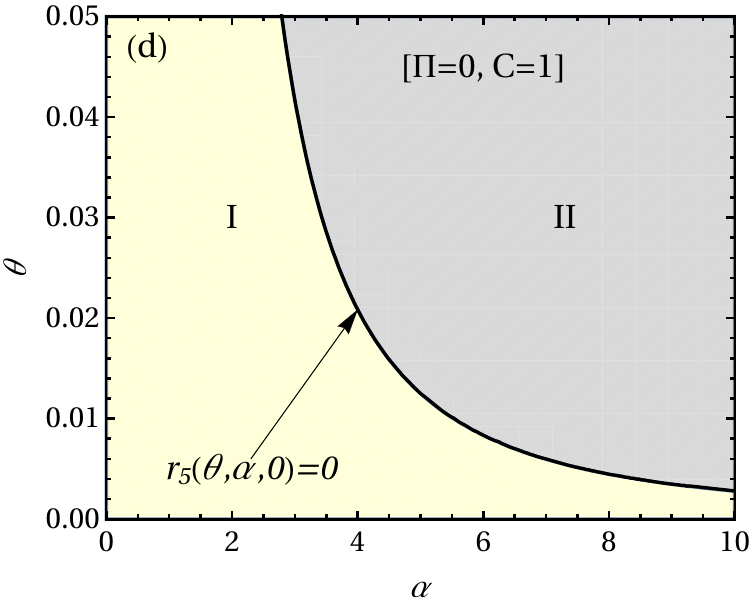}
    \end{tabular}
 \caption{\label{phase-dig} The phase diagrams corresponding to different choices of $\Pi$ (where $\Pi=1$ and $\Pi=0$ represent the case for partial and apparent entropy production, respectively) and $C$ [see \eref{C-parameter}] are shown in the limit of weak coupling (i.e., $\delta\to0$). The axis  (not shown) that corresponds to $\delta$ is perpendicular to the plane of the paper. {\color{black}Black solid contour for given $\Pi$ and $C$ (see \tref{tb:table}) 
separates regions of different branch point singularities in $\mu(\lambda)$. }}
 \end{figure*}

{\color{black}In \fref{lamb-var}, we plot $\lambda_\pm(u)$ given in \eref{lambda-pm-u}, for different values of $\theta$ and $\alpha$ against $u$. 
For $0<\delta <1$ (dotted, dashed, and dot-dashed lines), we find that $\lambda_+(u)$ has either one minimum located at $u=0$ [see \fref{lamb-var}(a)]}
 or two minima
located at $u= \pm u^*_+$ [see \fref{lamb-var}(b)] (where $u^*_+\to 0$ as $\delta\to 0$) depending
on the parameters $\theta$ and $\alpha$. This is determined by whether
\begin{equation}
\frac{\partial^2 \lambda_+(u)}{\partial u^2}\Big|_{u=0}
>0 \quad\text{or}\quad
\frac{\partial^2 \lambda_+(u)}{\partial u^2}\Big|_{u=0}
<0.
\label{condition-1}
\end{equation}
In the first case, we get $\lambda_+^{(\delta)} \to \tilde{\lambda}_+
< \lambda_+$ as $\delta\to 0$, whereas in the latter case
$\lambda_+^{(\delta)}$ converges to $\lambda_+$ as $\delta\to
0$. Similarly, we get $\lambda_-^{(\delta)}\to \tilde{\lambda}_-
> \lambda_-$ [see \fref{lamb-var}(c)] or $\lambda_-$ [see \fref{lamb-var}(d)] depending on whether
\begin{equation}
\frac{\partial^2 \lambda_-(u)}{\partial u^2}\Big|_{u=0}
<0 \quad\text{or}\quad
\frac{\partial^2 \lambda_-(u)}{\partial u^2}\Big|_{u=0}
>0.
\label{condition-2}
\end{equation}
{\color{black}Setting $\lambda_\pm^{\prime\prime}(0)=0$ gives the contour in $(\alpha,\theta)$ space separating different possibilities of
pair of branch point singularities $(\lambda^{(\delta)}_-,\lambda_+^{(\delta)})$ of $\mu(\lambda)$, and these singularities and the equations of contours are given in \tref{tb:table}. Using these equations (see \tref{tb:table}) in the limit $\delta\to0$, the phase diagrams can be obtained for different possibilities of $\Pi$ and $C$ and are shown in \fref{phase-dig}.}

{\color{black} In the following, we show how $\mu(\lambda)$ behaves near the branch point singularities. We write $\mu(\lambda)=\mu_\mathrm{a}(\lambda)
+\mu_\mathrm{s}(\lambda)$  near the
branch point, where $\mu_\mathrm{a}(\lambda)$ and
$\mu_\mathrm{s}(\lambda)$ are the analytic and the singular part of
$\mu(\lambda)$, respectively. } 
We now analyze the roots of $q(u)+h(u,\lambda)$ with respect to the
variable $u$, near a branch point.  Let us consider the case
$\lambda_+^{(\delta)} \to \tilde\lambda_+$. Using \eref{h} and writing
$\tilde\lambda_+ -\lambda=\epsilon$ near the branch point (where
$\epsilon>0$), we get
\begin{equation}
A(u) -\epsilon B(u) -O(\epsilon^2)=0, 
\label{SBP-eqn}
\end{equation}
where
\begin{align*}
A(u)&=q(u) + b(u)\tilde\lambda_+ -p(u) \tilde\lambda_+^2, \\
B(u)&= b(u) -2p(u)\tilde\lambda_+.
\end{align*}
The left side of \eref{SBP-eqn} is a polynomial in $u^2$.  Since the
minimum is located at $u=0$, the branch point $\tilde\lambda_+$
satisfies the equation $A(0)=0$.  Therefore, two of the roots
$u_{1\pm}^2$ of \eref{SBP-eqn} are of $O(\epsilon)$ for small
$\epsilon$, which are given by
\begin{equation*}
u_{1\pm}^2=\frac{2B(0)}{A''(0)}\,\epsilon +O(\epsilon^2). 
\end{equation*}
On the other hand, differentiating the equation 
\begin{equation*}
q(u) + b(u) \lambda_+(u) -p(u)\lambda_+^2(u)=0
\end{equation*}
with respect to $u$, and using the condition $\lambda'_+(0)=0$, we get
$B(0)/A''(0)=-1/\lambda''_+(0)$. Since, $\lambda''_+(0) >0$, we finally
get 
\begin{equation}
u_{1\pm} =\pm i\frac{\sqrt{2\epsilon}}{\sqrt{\lambda''_+(0)}}
+O(\epsilon). 
\label{u1}
\end{equation}
The other roots are of $O(1)$ which, at the leading order, satisfy the
reduced equation $u^{-2}A(u)=0$. Similarly, we can find the behavior
near $\tilde{\lambda}_-$. Thus, from \eref{mu-in terms of roots}, we
find the nature of the singularities near a branch point as:
\begin{equation}
\mu_\mathrm{s}(\lambda) =
\begin{cases}
-
\dfrac{1}{\sqrt{2\lambda_+''(0)}}
\sqrt{\tilde\lambda_+
-\lambda}&\text{as}~\lambda\to \tilde\lambda_+, \\[5mm]
 -
\dfrac{1}{\sqrt{-2\lambda_-''(0)}}
\sqrt{\lambda-\tilde\lambda_-}
&\text{as}~\lambda\to \tilde\lambda_-.
\end{cases}
\label{asymp-mu}
\end{equation}
Note that $\pm\lambda_\pm''(0)$ diverges as $\delta\to 0$. Therefore,
$\mu_\mathrm{s}(\lambda)$ goes to zero as $\delta\to 0$.

In the other two cases where $\lambda_\pm^{(\delta)} \to \lambda_\pm$,
the singular behaviors of $\mu(\lambda)$ near the singularities in the
limit $\delta\to 0$ are same as that of $\mu_0(\lambda)$. In all
cases, away from the singularities, $\mu(\lambda)\to \mu_0(\lambda)$ as
$\delta\to 0$.

In order to invert the moment generating function $Z(\lambda)$, we need $g(\lambda)$ which is difficult to obtain [see \eref{g-lambda}]. Since we are interested in the weak coupling limit $(0<\delta\ll1)$, we can approximate the prefactor $g(\lambda)$ with that of the free particle $g_0(\lambda)$ (see \aref{section-LDF}):
\begin{align*}
g(\lambda)\approx g_0(\lambda),
\end{align*}
where $g_0(\lambda)$ is given in \eref{g0-1}.

In the region I of \frefss{phase-dig}(b) and \frefs{phase-dig}(d), the branch point singularities of $\mu(\lambda)$ are $\lambda_\pm$ in the limit of coupling tending to zero. We expect that the function $\mu(\lambda)$ would obey the GC symmetry and both $\mu(\lambda)$ and $g_0(\lambda)$ are analytic functions for $\lambda\in (\lambda_-,\lambda_+)$ in the weak coupling limit. Therefore, the steady state fluctuation theorem might hold for apparent entropy production in those regions (see \sref{fp-eqn}).

\section{Large deviation function, asymmetry function and fluctuation theorem}
\label{ldf-ft}
The probability density function $p(s)$ can be written as [see \eref{saddle}]
\begin{equation}
p(s)=P(\Delta S^{A}_{tot}=s\tau/\tau_\gamma)\ \tau/\tau_\gamma,
\label{pdf-1}
\end{equation}
and its large deviation form is \cite{Touchette} 
\begin{equation}
p(s)\sim e^{(\tau/\tau_\gamma) I(s)}.
\label{LDform.0}
\end{equation}
In the above equation, $I(s):=h_s(\lambda^*)$ is the large deviation function and is given 
for all cases in \aref{section-LDF}. The continuity
properties of $I(s)$ in the limit $\delta\to0$ are shown in \aref{cont-LDF}.
When the probability density function $p(s)$ obeys the fluctuation theorem, it satisfies
\begin{equation}
\lim_{\tau/\tau_\gamma \to \infty} \dfrac{\tau_\gamma}{\tau}\ln\dfrac{p(s)}{p(-s)}=s.
\label{eqn-1}
\end{equation} 
One can define an asymmetry function $f(s)$ as
\begin{equation}
f(s)=\dfrac{\tau_\gamma}{\tau}\ln\dfrac{p(s)}{p(-s)}.
\label{finite-time}
\end{equation}
In the large time limit $(\tau\to\infty)$, 
\begin{equation} 
f(s)=I(s)-I(-s).
\label{large-time}
\end{equation}
The asymmetry function $f(s)$ and its continuity properties for all cases in the limit $\delta\to0$ are discussed in \aref{cont-asymm-function}.
\begin{figure*}\center
\begin{tabular}{ccc}
 \includegraphics[width=.35\textwidth]{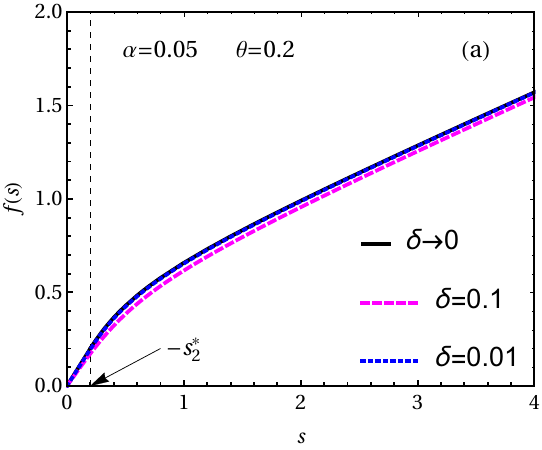}~~&
   \includegraphics[width=.370\textwidth]{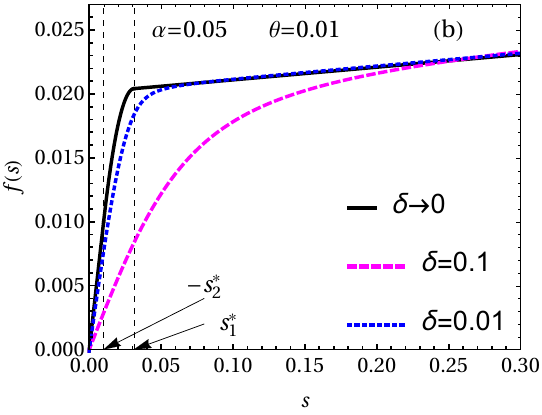}~~&
      \includegraphics[width=.350\textwidth]{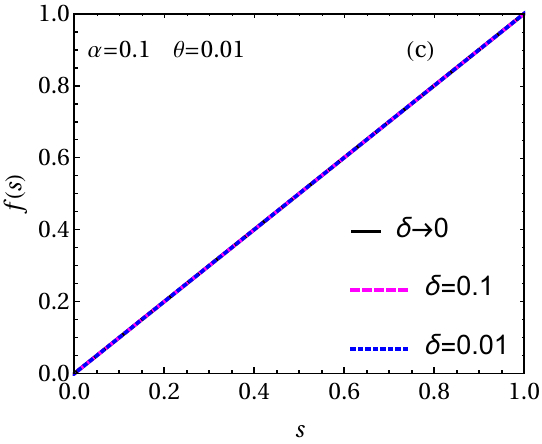}~~\\
        \includegraphics[width=.350\textwidth]{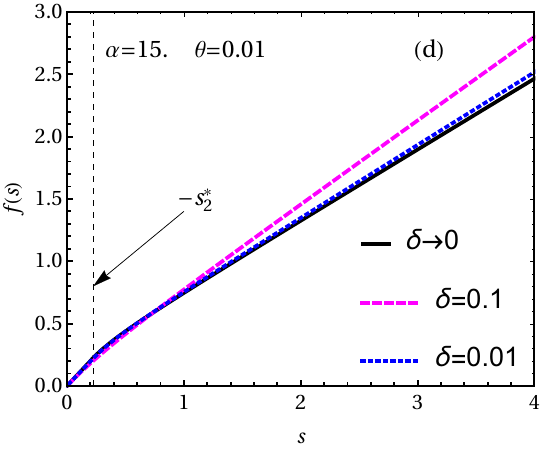}~~&
          \includegraphics[width=.370\textwidth]{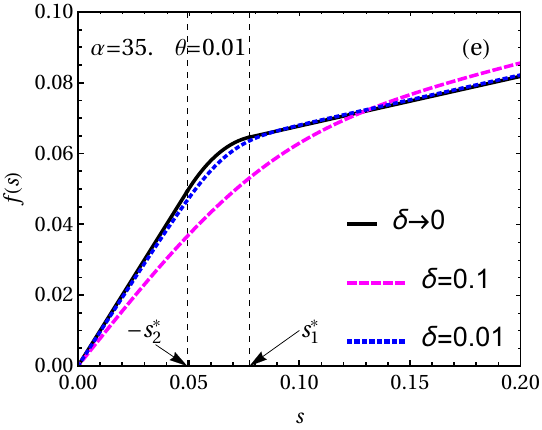}~~&
            \includegraphics[width=.35\textwidth]{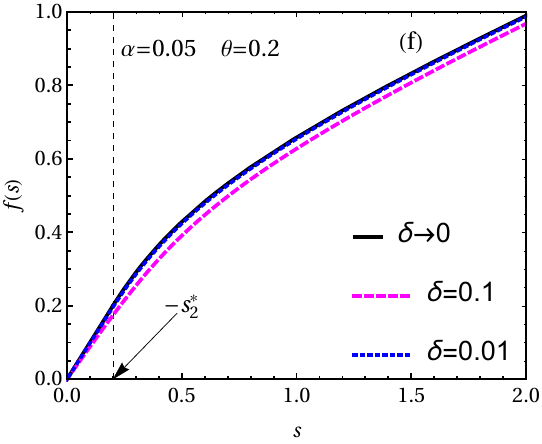}~~\\
              \includegraphics[width=.37\textwidth]{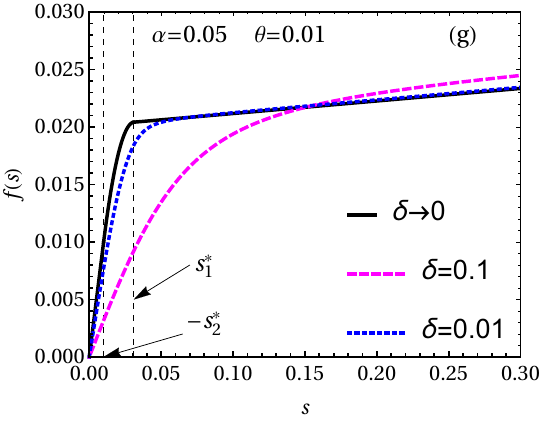}~~&
                \includegraphics[width=.35\textwidth]{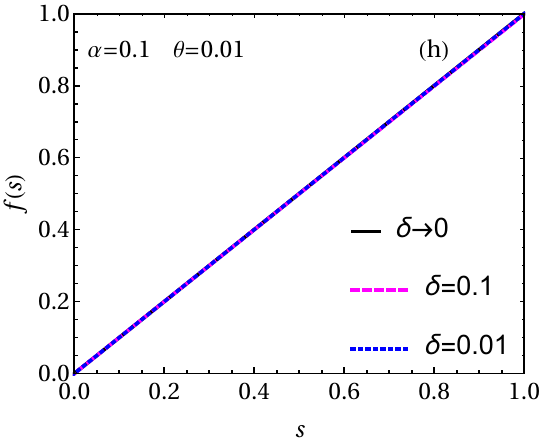}~~&
                  \includegraphics[width=.35\textwidth]{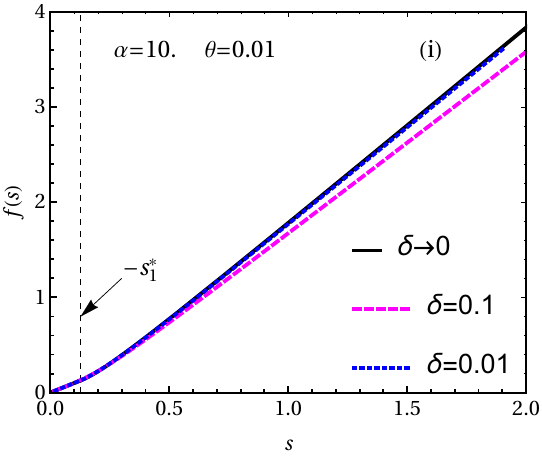}~~
            \end{tabular}
   \caption{\label{fig-set-1}{\color{black} The analytical asymmetry function $f(s)$ given in \eref{large-time} against the scaled variable $s=\Delta S^A_{tot}\tau_\gamma/\tau$ for $s>0$. Partial entropy production in (a)--(b) $(C=0)$ and (f)--(g) $(C=1)$, and apparent entropy production in (c)--(e) ($C=0$) and (h)--(i) $(C=1)$, where $s_1^*$ and $s_2^*$ are given in \aref{section-LDF}. These plots are shown for coupling parameter $\delta=0.1$ (magenta dashed line) and $\delta=0.01$ (blue dotted line). The asymmetry function in the limit $\delta\to 0$ (black solid line) is also shown for respective cases (see \aref{cont-asymm-function}).}}
 \end{figure*}

{\color{black}When $p(s)$ obeys the fluctuation theorem, we find that $f(s)=s$ for all $s$. We plot the asymmetry functions $f(s)$ (for $s>0$) given in \eref{large-time} against the scaled variable $s=\Delta S^A_{tot}\tau_\gamma/\tau$ for partial entropy production [in \frefss{fig-set-1}(a) and \frefs{fig-set-1}(b)] and apparent entropy production [in \frefss{fig-set-1}(c)--\frefs{fig-set-1}(e)] for the first choice of external forces (i.e., $C=0$). Similarly, for the second choice of external forces (i.e., $C=1)$, we plot the asymmetry functions $f(s)$ given in \eref{large-time} against the scaled variable $s=\Delta S^A_{tot}\tau_\gamma/\tau$ for partial entropy production [in \frefss{fig-set-1}(f) and \frefs{fig-set-1}(g)] and apparent entropy production [in \frefss{fig-set-1}(h) and \frefs{fig-set-1}(i)]. These plots are shown for fixed $\delta=0.1$ (magenta dashed line) and $\delta=0.01$ (blue dotted line). Moreover, in the limit $\delta \to0$ (black solid line), the asymmetry functions $f(s)$ given in \aref{cont-asymm-function} is shown for each case. Therefore, we see that as the coupling parameter $\delta$ decreases, the asymmetry function $f(s)$ converges to that of $\delta\to0$ case.} 

Thus, the asymptotic expression for the asymmetry function $f(s)$ in the limit $\delta \to 0$ and for large $s$ ($s\to\infty$) are given as (see \aref{cont-asymm-function})
\begin{align}
&f(s)=\begin{cases}
\mu_0(\lambda_-) -\mu_0(\tilde\lambda_+) +(\lambda_-+\tilde\lambda_+)s  &\text{region I of \fref{phase-dig}(a)},\\
\mu_0(\tilde\lambda_-) -\mu_0(\tilde\lambda_+)+(\tilde\lambda_-+\tilde\lambda_+)s &\text{region II of \fref{phase-dig}(a)},\\
\mu_0(\lambda_-) -\mu_0(\lambda_+)+(\lambda_-+\lambda_+)s &\text{region I of \fref{phase-dig}(b)},\\
\mu_0(\lambda_-) -\mu_0(\tilde\lambda_+)+(\lambda_-+\tilde\lambda_+)s &\text{region II of \fref{phase-dig}(b)},\\
\mu_0(\tilde\lambda_-) -\mu_0(\tilde\lambda_+) +(\tilde\lambda_-+\tilde\lambda_+)s &\text{region III of \fref{phase-dig}(b)},\\
\mu_0(\lambda_-) -\mu_0(\tilde\lambda_+) +(\lambda_-+\tilde\lambda_+)s &\text{region I of \fref{phase-dig}(c)},\\
\mu_0(\tilde\lambda_-) -\mu_0(\tilde\lambda_+) +(\tilde\lambda_-+\tilde\lambda_+)s &\text{region II of \fref{phase-dig}(c)},\\
\mu_0(\lambda_-) -\mu_0(\lambda_+) +(\lambda_-+\lambda_+)s &\text{region I of \fref{phase-dig}(d)},\\
\mu_0(\tilde\lambda_-) -\mu_0(\lambda_+) +(\tilde\lambda_-+\lambda_+)s &\text{region II of \fref{phase-dig}(d)},
\end{cases}
\end{align}
{\color{black}where $\lambda$-s are given in \aref{BP-EQ-C}. Note that the asymmetry function is an odd function of $s$, i.e., $f(-s)=-f(s)$.} 
From the above equation and {\color{black}\frefss{fig-set-1}(c) and \frefs{fig-set-1}(h)}, it is clear that the steady state fluctuation theorem ($f(s)=s$) is satisfied only in the region I of \frefss{phase-dig}(b) and \frefs{phase-dig}(d) in the weak coupling limit. 

\begin{figure*}
\begin{tabular}{cc}
 \includegraphics[width=.45\textwidth]{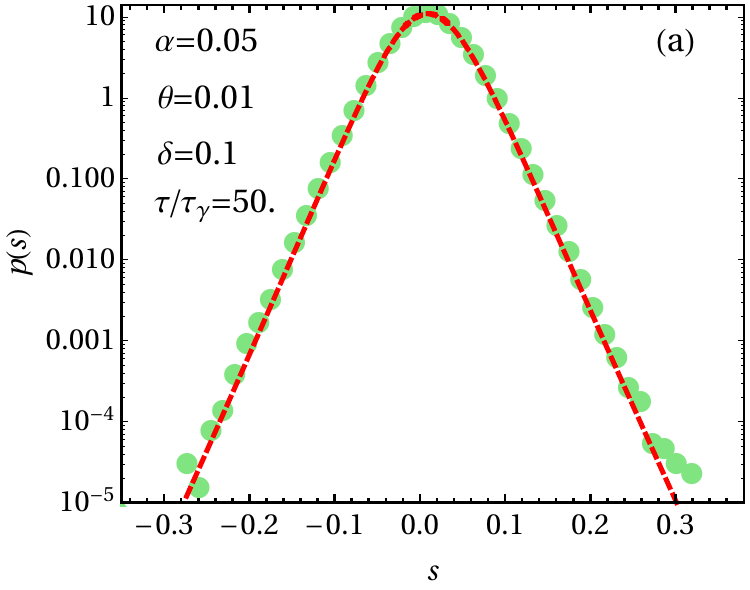}~~~~~~~~&
  \includegraphics[width=.45\textwidth]{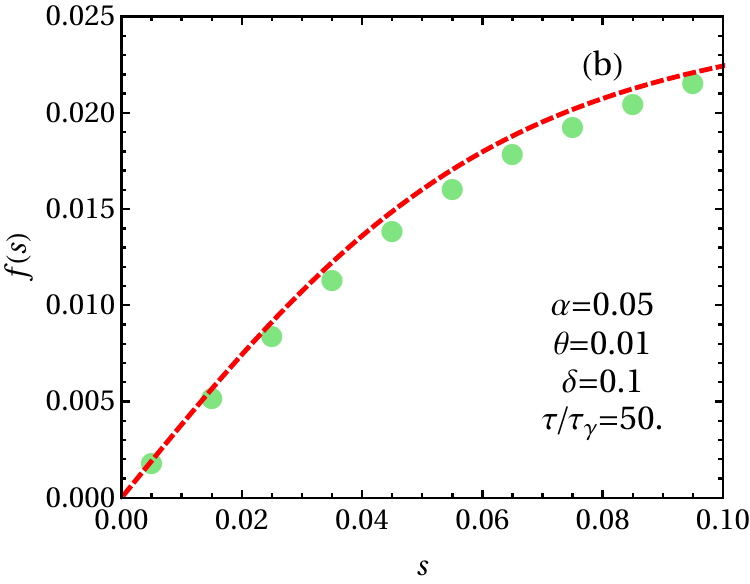}\\
   \includegraphics[width=.45\textwidth]{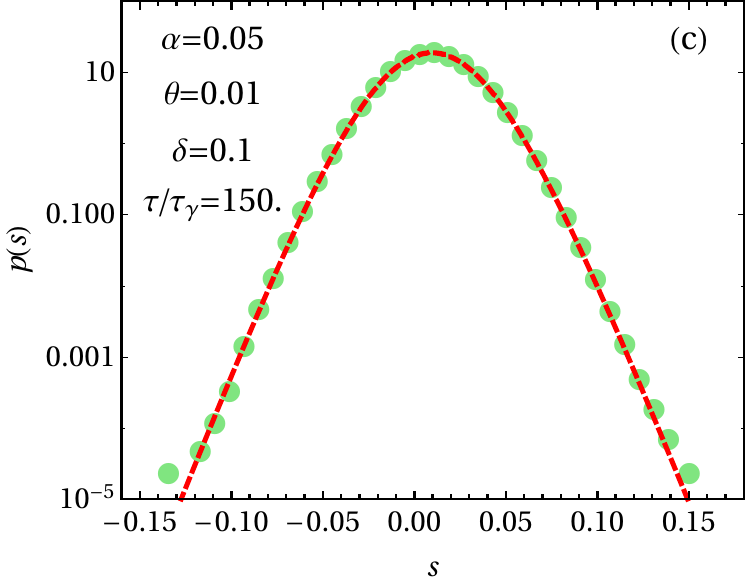}~~~~~~~~&
    \includegraphics[width=.47\textwidth]{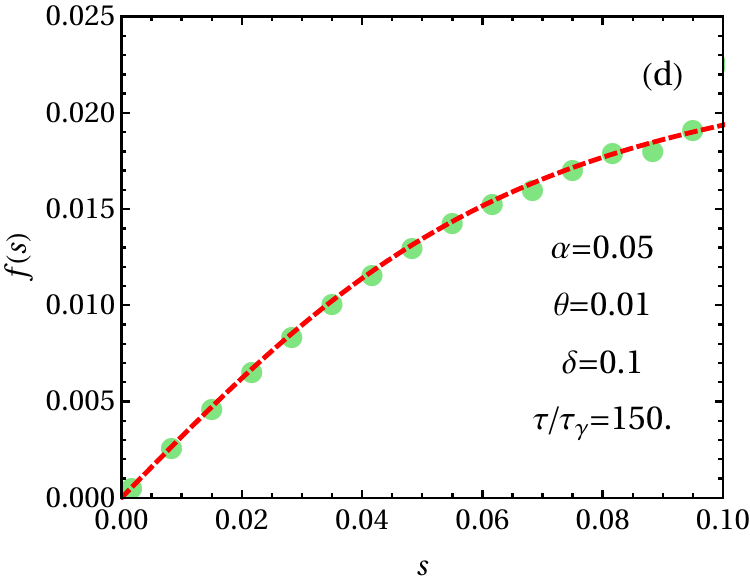}
    \end{tabular}
 \caption{\label{time-comp} A comparison of the analytical probability density function $p(s)$ (red dashed lines) given by \eref{pdf-1} and the asymmetry function $f(s)$ (red dashed line) given by \eref{finite-time} with the numerical simulations {\color{black}(circles)} is shown for time $\tau/\tau_\gamma=50$ [(a) and (b)] and $\tau/\tau_\gamma=150$ [(c) and (d)]. The parameters $\theta$ and $\alpha$ are taken from \fref{phase-dig}(a) for the case of partial entropy production ($\Pi=1$ and $C=0$). The coupling parameter $\delta$ and the temperature of the heat bath are fixed in all of the above figures and taken to be $\delta=0.1$ and $T=1$, respectively. This comparison indicates that as the observation time relative to viscous relaxation time gets longer, the agreement between theoretical prediction and the numerical simulation becomes better.}
 \end{figure*}

\begin{figure*}\center
\begin{tabular}{ccc}
  \includegraphics[width=.35\textwidth]{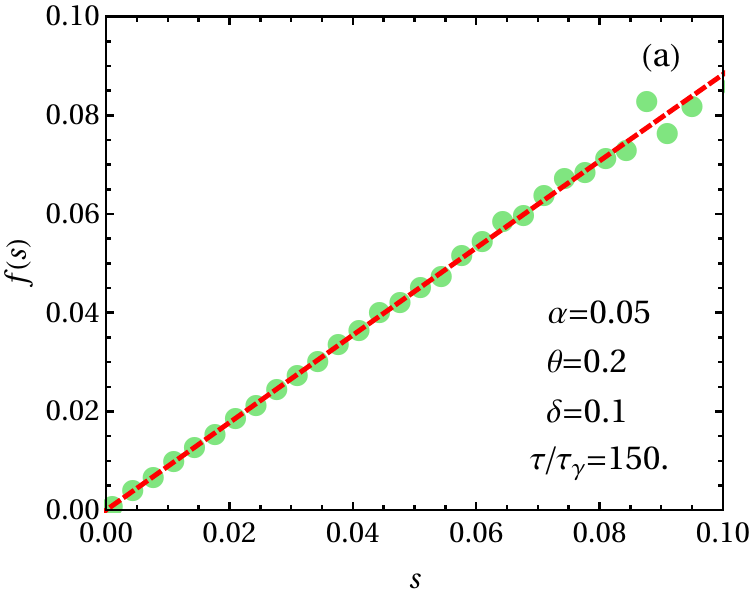}~~&
    \includegraphics[width=.35\textwidth]{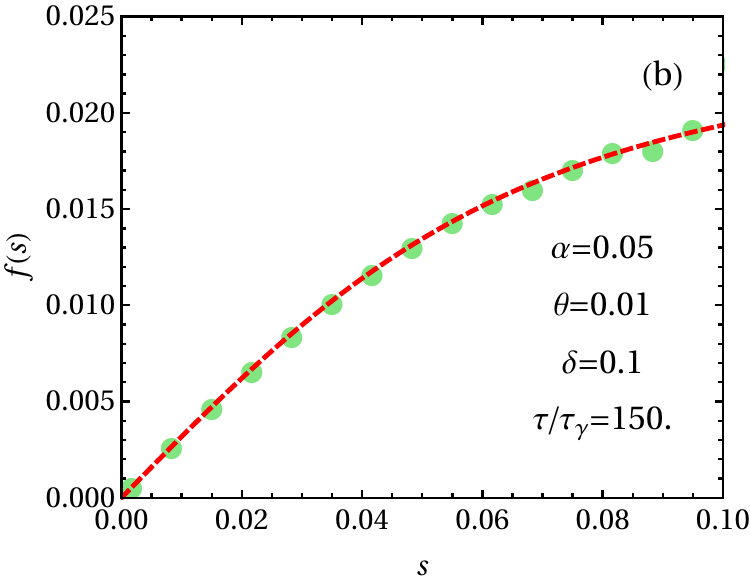}~~&
       \includegraphics[width=.35\textwidth]{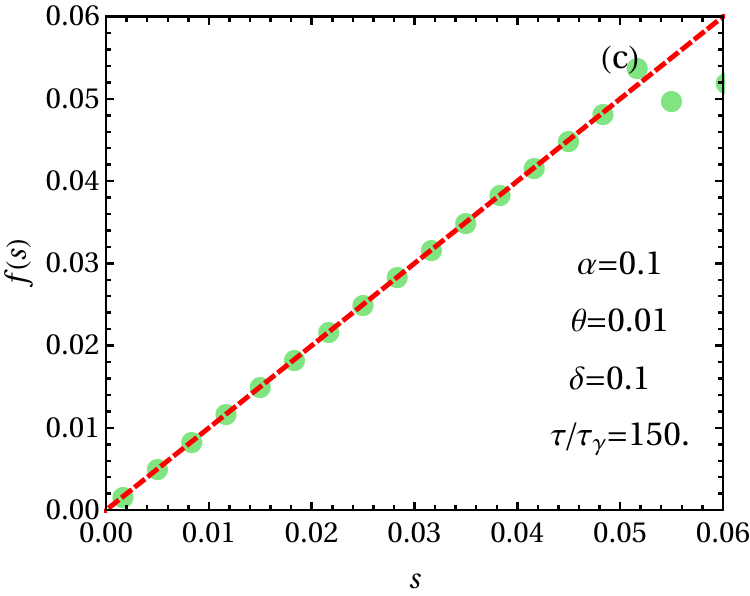}~~\\
         \includegraphics[width=.35\textwidth]{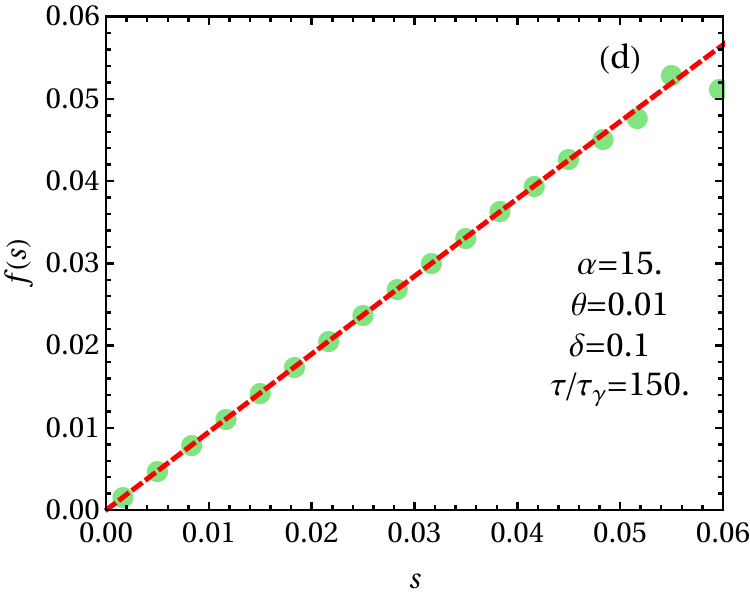}~~&
                  \includegraphics[width=.35\textwidth]{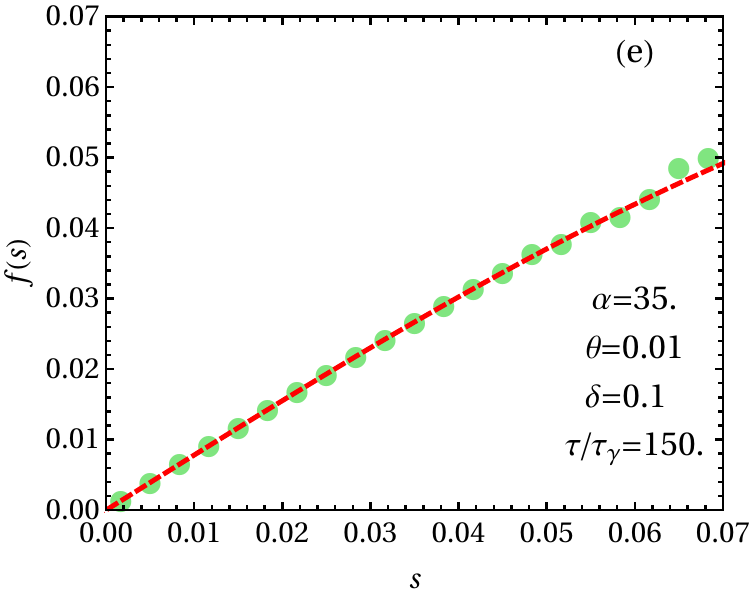}~~&
  \includegraphics[width=.35\textwidth]{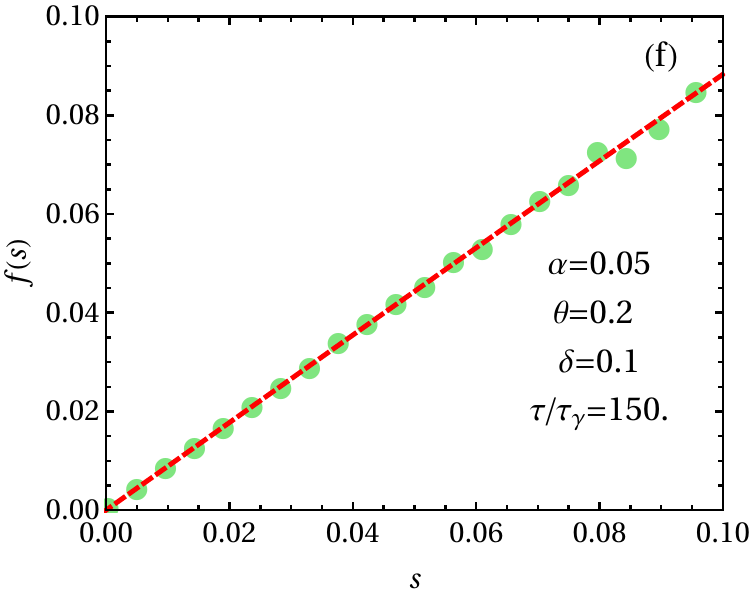}~~\\
    \includegraphics[width=.35\textwidth]{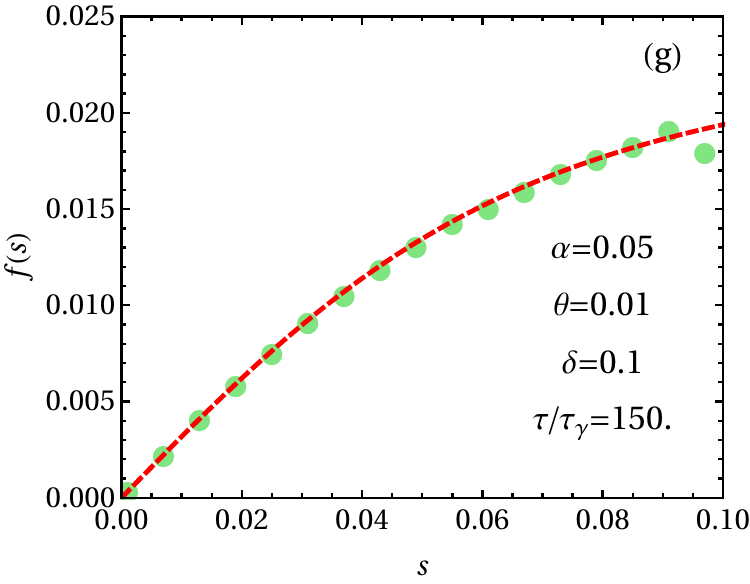}~~&
       \includegraphics[width=.35\textwidth]{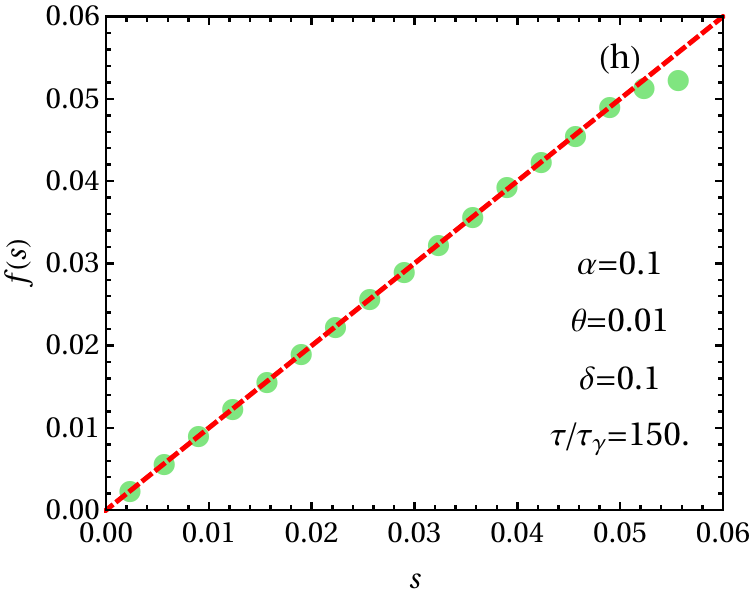}~~&
         \includegraphics[width=.35\textwidth]{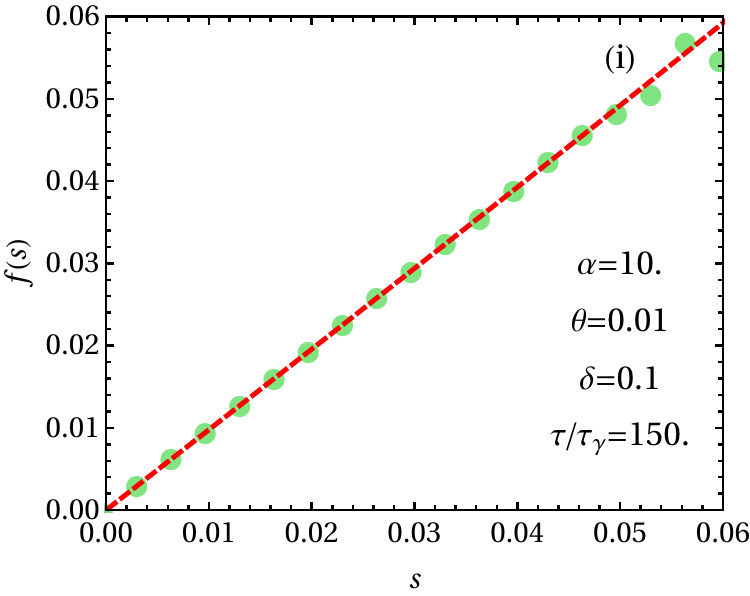}~~
         \end{tabular}
             \caption{\label{fig-set-2} {\color{black} The comparison of analytical asymmetry functions $f(s)$ (red dashed line) given in \eref{finite-time} with the numerical simulation results {\color{black}(circles)}. Partial entropy production in (a)--(b) ($C=0$) and (f)--(g) ($C=1$) and apparent entropy production in (c)--(e) ($C=0$) and (h)--(i) ($C=1$). We take $\delta=0.1$, $\tau/\tau_\gamma=150$, and $T=1$.} }
 \end{figure*}

\section{Numerical simulation}
\label{num-sim}
In \fref{time-comp}, we show a comparison of theoretical predictions of the probability density function $p(s)$ (red dashed line) given by \eref{pdf-1} and the asymmetry function $f(s)$ (red dashed lines) given by \eref{finite-time} with the numerical simulations {\color{black}(circles)} for time $\tau/\tau_\gamma=50$ [(a) and (b)] and $\tau/\tau_\gamma=150$ [(c) and (d)]. The parameters $\alpha$ and $\theta$ are taken from \fref{phase-dig}(a)  ($\Pi=1$ and $C=0$). For all these figures, we set the coupling parameter $\delta=0.1$ and the temperature of the heat bath $T=1$. 

In the following, we give quantitative measures to describe how much the theoretical predictions are closer to the numerical simulation results. In the case of probability density function, we choose two different measures at time $\tau/\tau_\gamma$: (1) the Kullback--Leibler divergence:
\begin{align*}
D_{\text{KL}}(\tau/\tau_\gamma)=\int ds\ p_s(s) \ln \dfrac{p_s(s)}{p_t(s)}, 
\end{align*}
 and (2) the weighted norm:
 \begin{align*}
D_{p}(\tau/\tau_\gamma)=\sqrt{\int ds\ p_s(s) [p_s(s)-p_t(s)]^2}, 
\end{align*}
where $p_s(s)$ is the probability density function obtained from the numerical simulation (after smoothing the data using the appropriate width of the binning) and $p_t(s)$ is the analytical prediction of the probability density function. 
 We find that for time $\tau/\tau_\gamma=50$, $D_{\text{KL}}(50)=0.0874399.... $ and $D_{p}(50)=0.5484667....$ whereas $D_{\text{KL}}(150)=0.0313721...$ and $D_{p}(150)=0.3588365...$ at time $\tau/\tau_\gamma=150$. Similarly, we define a measure in the case of asymmetry function as
 \begin{align*}
D_{f}(\tau/\tau_\gamma)=\sqrt{\int ds\ [f_s(s)-f_t(s)]^2},
 \end{align*} 
 where again the subscripts $t$ and $s$, respectively, refer to the theory and simulation. We find that $D_{f}(50)=0.000304931...$ and $D_{f}(150)=0.0000603556..$ Therefore, we see that as the observation time increases, the distance between theoretical predictions and numerical simulations decreases. Hence, the agreement between these two gets better.

{\color{black}For the first choice of external forces, the comparison of analytical asymmetry function $f(s)$ (red dashed line) given by \eref{finite-time}, with the numerical simulations {\color{black}(circles)} are shown for partial [in \frefss{fig-set-2}(a) and \frefs{fig-set-2}(b)] and apparent entropy production [in \frefss{fig-set-2}(c)--\frefs{fig-set-2}(e)]. Similarly, for the second choice of external forces, we compare the analytical predictions of the asymmetry function $f(s)$ (red dashed lines) given in \eref{finite-time}, with the numerical simulations {\color{black}(circles)} for partial [in \frefss{fig-set-2}(f) and \frefs{fig-set-2}(g)] and apparent entropy production [in \frefss{fig-set-2}(h) and \frefs{fig-set-2}(i)]. Here, we take $\delta=0.1$, $T=1$, and the observation time $\tau/\tau_\gamma=150$. These figures indicate that there is nice agreement between theoretical prediction and numerical simulation. }
\section{Model 2}
\label{model2}
\begin{figure}
\center
\includegraphics[width=0.5\textwidth]{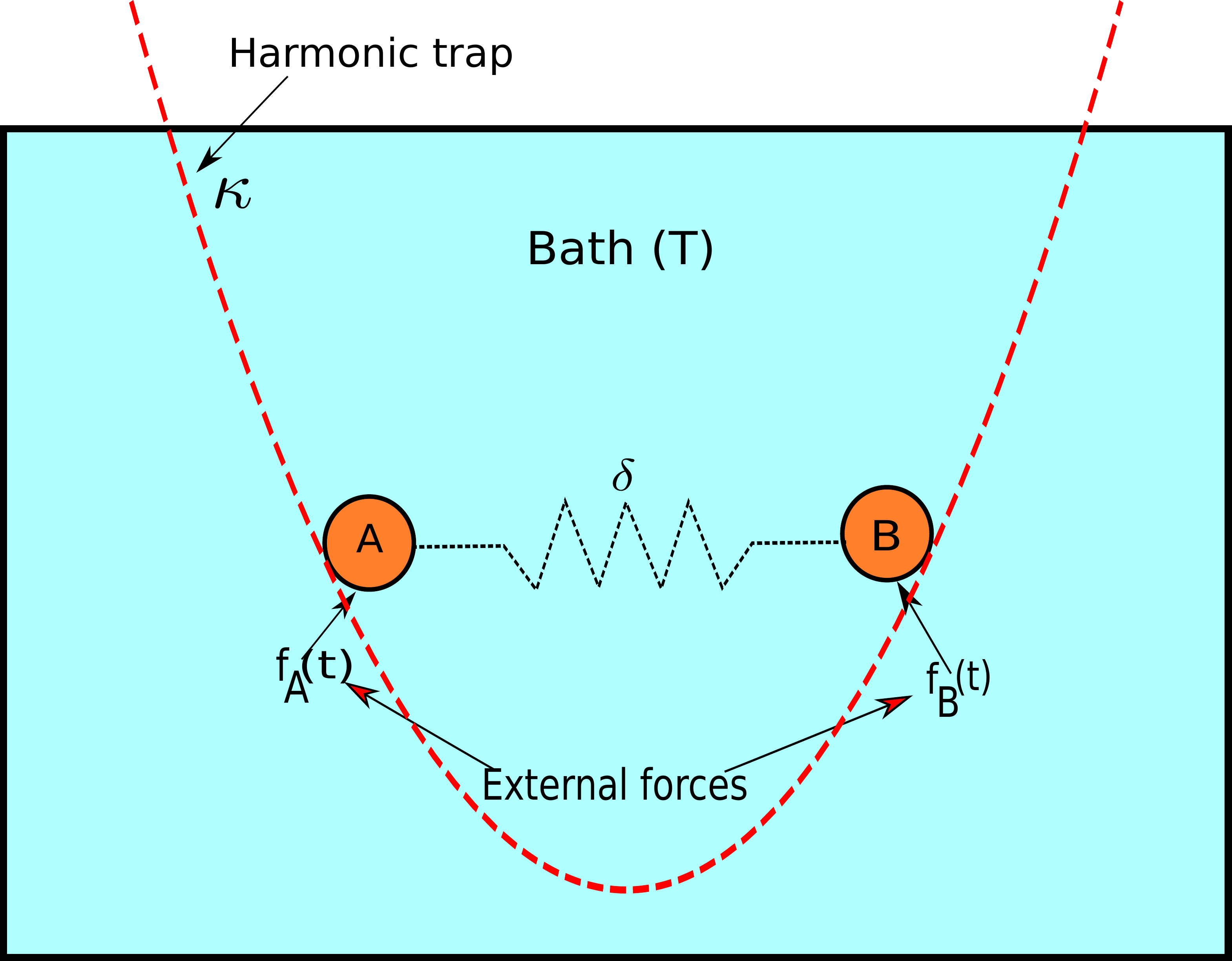}
\caption{\label{system-pic} A harmonically coupled Brownian particle system (particle A and B) in a heat bath of a constant temperature $T$ is shown. Each particle has mass $m$. The coupling strength between particle A and B is $\delta=2km/\gamma^2$ (dimensionless). The whole system in confined in a harmonic trap of strength $\kappa=mk_0/\gamma^2$ (dimensionless). The external stochastic forces $f_A(t)$ and $f_B(t)$ are acting on particle A and B, respectively. }
\end{figure}
In the previous model, we considered a harmonically coupled Brownian particle system (i.e., particle A and B) in a heat bath driven away from equilibrium using external stochastic Gaussian forces. We showed that even in the limit $\delta\to0$, one might see the deviation from the SSFT for partial and apparent entropy production. Here, we consider a system consisting of two Brownian particles (say particle A and B) of mass $m$ coupled by a harmonic spring of stiffness $k$ in a harmonic trap of stiffness $k_0$. The whole system is in contact with a thermal bath at constant temperature $T$. Notice that this model system is different from model 1 (see \fref{system}) because of the presence of a confining potential. The schematic diagram of this model is shown in \fref{system-pic}. The Hamiltonian of this coupled system is given as
\begin{align}
\mathcal{H}_{\kappa}(x_A,x_B,v_A,v_B)=\dfrac{1}{2}mv^2_A+\dfrac{1}{2}mv^2_B+\dfrac{1}{2}k_0x^2_A+\dfrac{1}{2}k_0x^2_B+\dfrac{1}{2}k(x_A-x_B)^2,
\label{hamilt}
\end{align}    
where $m$ is the mass of each particle, and $x_A (x_B)$ and $v_A (v_B$) are the position and velocity of particle A (particle B), respectively. In the above equation, the subscript $\kappa$ indicates the presence of the harmonic trap. Suppose particle A and B are driven using stochastic Gaussian forces $f_A(t)$ and $f_B(t)$, respectively. The Langevin equations of the above coupled system are
 \begin{align}
& \dot{x}_A=v_A,\label{full-dynamics-1}\\
&\dot{x}_B=v_B, \label{full-dynamics-2}\\
&m\dot{v}_A=-\gamma v_A-(k+k_{0})x_A+k x_B+f_A(t)+\eta_A(t),\label{full-dynamics-3}\\  
&m\dot{v}_B=-\gamma v_B+k x_A-(k+k_{0})x_B+f_B(t)+\eta_B(t)\label{full-dynamics-4}.
 \end{align}  
The properties of thermal Gaussian noise $\eta_i(t)$ and external force $f_i(t)$ are introduced in \sref{model}.

For this model system, we can define the partial and apparent entropy production as described in \sref{def}, which can be jointly written as
\begin{align}
\Delta \mathcal{S}^A_{tot}=\mathcal{W}-\dfrac{1}{2}\tilde{\mathcal{U}}^T\tilde{\mathbb{H}}^{-1}\tilde{\mathcal{U}}+\dfrac{1}{2}\tilde{\mathcal{U}}_0^T\tilde{\mathbb{H}}^{-1}\tilde{\mathcal{U}}_0,
\label{general-EP}
\end{align}
where $\tilde{\mathcal{U}}=(x_A,v_A)^T$. In the above equation,
\begin{align}
&\mathcal{W}=\dfrac{1}{T}\int_0^\tau dt\ f_A(t)v_A(t)+\dfrac{\Pi k}{T}\int_0^\tau dt\ x_B(t)v_A(t),\label{W}\\
&\tilde{\mathbb{H}}^{-1}=\Pi(\tilde{\Xi}_P-\tilde{\mathbb{H}}_{P}^{-1})+(1-\Pi)(\tilde{\Xi}_A-\tilde{\mathbb{H}}_{A}^{-1}),
\end{align} where
\begin{equation}
\tilde{\mathbb{H}}_{P,A}=\begin{pmatrix}
\mathbb{H}_{P,A}^{11}&&0\\
0&&\mathbb{H}_{P,A}^{33}
\end{pmatrix},
\label{HP-HA}
\end{equation}
in which subscripts $P$ and $A$ correspond to partial and apparent entropy productions, respectively.

The matrix element $\mathbb{H}_P^{11}$ for the first choice of external forces is given by
 \begin{align*}
 \mathbb{H}^{11}_P&=D m\dfrac{\delta^3(2+\theta+\alpha^2\theta)+16\kappa(1+\theta)(\delta+\kappa)+2\delta^2(2+\theta+\alpha^2\theta)(1+\kappa)}{4 \gamma^3 \kappa(\delta+\kappa) (2\delta+\delta^2+4\kappa)},
 \end{align*}
whereas for second choice of external forces
\begin{align*}
 \mathbb{H}^{11}_P=D m\dfrac{\delta^3\{2+\theta(1+\alpha)^2\}+16\kappa(1+\theta)(\delta+\kappa)+\delta^2(2+\theta+\alpha^2\theta)(1+\kappa)+4\delta\kappa\theta(\delta+2\kappa)}{4 \gamma^3 \kappa(\delta+\kappa)(2\delta+\delta^2+4\kappa)}.
 \end{align*}
On the other hand, $\mathbb{H}^{33}_P$ is the same for both choices of external forces
\begin{align*}
\mathbb{H}^{33}_{P}=\dfrac{D[4(1+\theta)(\delta+2\kappa)+\delta^2(2+\theta+\alpha^2\theta)]}{2 m \gamma(2\delta+\delta^2+4\kappa)}.
\end{align*}
Matrix elements $\mathbb{H}^{11}_A$ and $\mathbb{H}^{33}_A$  are given as
\begin{align*}
\mathbb{H}^{11}_A=\dfrac{Dm(1+\theta)}{\gamma^3 \kappa}, \quad\mathrm{and} \quad \mathbb{H}^{33}_A=\frac{D(1+\theta)}{m\gamma }.
\end{align*}
In the above equations, $\kappa=k_0 m/\gamma^2$ is the strength of the harmonic trap.

The entropy production \erefs{general-EP} depends on the thermal Gaussian noises $\eta_i(t)$ and external stochastic Gaussian forces $f_i(t)$ quadratically. Therefore, its probability density function is expected to be non-Gaussian, and is obtained by inverting the moment generating function which is given as 
\begin{equation}
Z_\kappa(\lambda)=\langle e^{-\lambda \Delta \mathcal{S}^A_{tot} }\rangle=g_\kappa(\lambda)\ e^{(\tau/\tau_\gamma) \mu_\kappa(\lambda)}+\dots,
\label{Z-lambda}
\end{equation}
Complete computation of the moment generating function $Z_\kappa(\lambda)$ in the large time limit is shown in \aref{appendix}. In the above equation, the prefactor $g_\kappa(\lambda)$ is obtained as given in \eref{glambda-eqn}. The function $\mu_\kappa(\lambda)$ in \eref{mu-k} has the following form
\begin{equation}
\mu_\kappa(\lambda)=-\dfrac{1}{4 \pi}\int^\infty_{-\infty} du\ \ln \bigg[1+\dfrac{h_\kappa(u,\lambda)}{q_\kappa(u)}\bigg],
\label{mu-form}
\end{equation}
where $q_\kappa(u)$ is
\begin{equation}
q_\kappa(u)=[(\kappa-u^2)^2+u^2][(\delta+\kappa-u^2)^2+u^2].
\label{qu}
\end{equation} 
The form of $h_\kappa(u,\lambda)$ for the first choice of forces, i.e., for uncorrelated forces, is given as
\begin{align}
h_\kappa(u,\lambda)=4 \theta \lambda(1-\lambda)u^2[u^4+(1-2\kappa-\delta)u^2+\kappa^2
+\kappa\delta
+(2-\Pi)\delta^2/4]\nonumber\\-\lambda\delta^2 u^2[\lambda \alpha^2\theta^2(\Pi-1)^2
+\lambda\Pi^2+\theta \Pi\{\lambda(1+\alpha^2)\Pi-\lambda-\alpha^2\}],
\label{h1}
\end{align}
whereas for the second choice of forces, i.e., for $f_B(t)=\alpha f_A(t)$, is
\begin{align}
h_\kappa(u,\lambda)=4 \theta \lambda(1-\lambda)u^2[u^4+(1-2\kappa-\delta)u^2+\kappa^2+\kappa\delta+(2-\Pi)\delta^2/4]\nonumber\\
-\lambda u^2[\theta \delta\alpha(\lambda\Pi-1)\{4(\kappa-u^2)+\delta(2+\alpha \Pi)\}+\lambda \Pi \delta^2(\Pi-\theta(1-\Pi))].
\label{h2}
\end{align}
In the case of a single Brownian particle confined in a harmonic trap and driven by an external stochastic Gaussian forces in the non-equilibrium steady state, $\mu_\kappa(\lambda)\to\mu^0_\kappa(\lambda)$ is 
\begin{equation}
\mu^0_\kappa(\lambda)=-\dfrac{1}{4\pi}\int_{-\infty}^\infty du\ \ln\ \bigg[1+\dfrac{h^0_\kappa(u,\lambda)}{q^0_\kappa(u)}\bigg] ,
\label{mu0-int}
\end{equation} 
where $h^0_\kappa(u,\lambda)=4\theta\lambda(1-\lambda)u^2$ and $q^0_\kappa(u)=(\kappa-u^2)^2+u^2$. Computation of the above integral yields \eref{mu0-ext}: $\mu^0_\kappa(\lambda)=\mu_0(\lambda)$.
One can also obtain the prefactor $g^0_\kappa(\lambda)$ for this case, which is given by
\begin{equation}
g^0_\kappa(\lambda)=\dfrac{4 \nu(\lambda)}{[1+\nu(\lambda)]^2}.
\label{g0}
\end{equation}
Notice that both functions $\mu_\kappa^{0}(\lambda)$ and $g_\kappa^{0}(\lambda)$ are independent of the strength of the harmonic trap $\kappa$.
In \erefss{mu0-int} and \erefs{g0}, the superscript $0$ corresponds to $\delta=0$ (coupling is absent). Both $\mu^0_\kappa(\lambda)$ and $g^0_\kappa(\lambda)$ are analytic functions for $\lambda\in(\lambda_-,\lambda_+)$. Moreover, they obey GC Symmetry. Therefore, in this case, the total entropy production satisfies the SSFT.

But in our case, $\mu_\kappa(\lambda)$ does not obey GC symmetry for large coupling $\delta$. Thus, one may not expect the validity of the SSFT for the total entropy production of the partial system (partial and apparent entropy production) for large value of the coupling parameter $\delta$.

The analytical computation of the prefactor $g_\kappa(\lambda)$ is difficult to obtain as it requires the computation of matrices $\bar{H}_1(\lambda)$, $\bar{H}_2(\lambda)$, and $\bar{H}_3(\lambda)$ [see \erefss{h1-k}--\erefs{h3-k}], which is not illuminating to us. Since our aim is to understand the fluctuation theorem for the partial system in the weak coupling limit ($\delta\to 0$),  we approximate $g_\kappa(\lambda)$ by the prefactor of the moment generating function of a stochastically driven single Brownian particle in a harmonic trap, i.e. $g_\kappa(\lambda)\approx g^0_\kappa(\lambda)$ [see \eref{g0}]\cite{Deepak}. 

\begin{figure}
\center
\includegraphics[width=.45\textwidth]{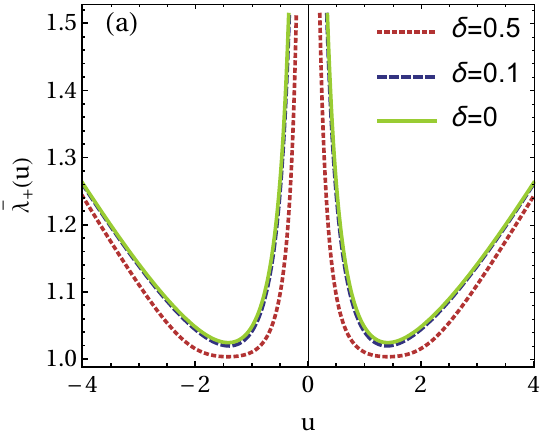}~~~~~
\includegraphics[width=.45\textwidth]{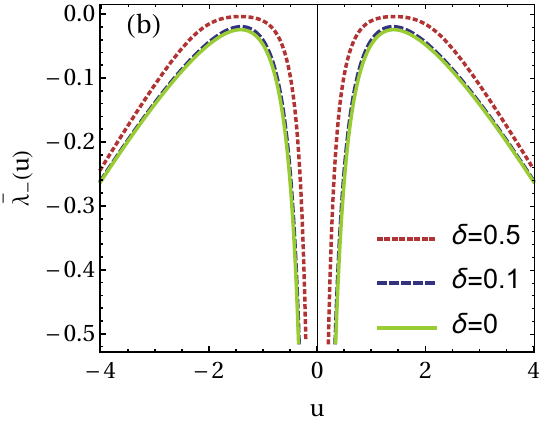}
\caption{\label{lambda-u}{\color{black}Variation of $\bar{\lambda}_\pm(u)$ against $u$ for $0<\delta<1$ (dotted and dashed lines) and $\delta=0$ (solid lines) at a trap strength $\kappa=2$.}}
\end{figure}

In the following, we used the method described in \sref{mu-delta-lambda} to evaluate the integral for $\mu_\kappa(\lambda)$. First consider the integral of $\mu^0_\kappa(\lambda)$ for a single Brownian particle  in a harmonic trap given in \eref{mu0-int}. The arguments in the logarithm in the integrand of $\mu^0_\kappa(\lambda)$ are $q^0_\kappa(u)$ and $[h^0_\kappa(u,\lambda)+q^0_\kappa(u)]$. While the function $q^0_\kappa(u)$ is always positive for all real values of $u$, the function $[h^0_\kappa(u,\lambda)+q^0_\kappa(u)]$ may have any sign. To understand the sign, we solve the quadratic equation
\begin{align}
h^0_\kappa(u,\lambda)+q^0_\kappa(u)&=0\nonumber\\
K_1(u)\lambda^2-K_2(u)\lambda-q^0_\kappa(u)&=0,
\end{align}  
in $\lambda$ where $K_1(u)=K_2(u)=4\theta u^2$. The roots of the above quadratic equation are
 \begin{equation}
 \bar\lambda^0_\pm(u)=\dfrac{K_2(u)\pm\sqrt{K_2^2(u)+4 K_1(u)q^0_\kappa(u)}}{2 K_1(u)}.
 \end{equation}
 In \frefss{lambda-u}(a) and \frefs{lambda-u}(b), we show the variation of $\bar\lambda^0_\pm(u)$  (solid lines) against $u$  at fixed trap strength $\kappa=2$.
 In the complex $\lambda$-plane, we see that $\mu^0_\kappa(\lambda)$ is a real function for $\lambda\in(\lambda_{\max},\lambda_{\min})$ where $\lambda_{\min}=\min\{\bar\lambda_+^0(u)\}$ and $\lambda_{\max}=\max\{\bar\lambda_-^0(u)\}$. In this case, the extrema of functions $\bar\lambda^0_\pm(u)$ occur at $u^*=\pm\sqrt{\kappa}$ (see \fref{lambda-u}). Therefore,
 \begin{equation}
 \bar\lambda^0_{\pm}(u^*=\pm\sqrt{\kappa})=\lambda_\pm=[1\pm\sqrt{1+\theta^{-1}}]/2.
 \label{lambda0+-}
 \end{equation}
 This implies $\mu^0_\kappa(\lambda)$ is a real function within $(\lambda_-,\lambda_+)$.
Similarly, we can find the domain within which $\mu_\kappa(\lambda)$ is a real quantity. 
For $\delta\neq0$ (coupling is present), the argument of the logarithm in integrand of \eref{mu-form} are $q_\kappa(u)$ and $[h_\kappa(u,\lambda)+q(u)]$ where $q_\kappa(u)$ is clearly a positive function for all real values of $u$. To see the domain, we write the quadratic equation
\begin{align}
h_\kappa(u,\lambda)+q_\kappa(u)&=0\nonumber\\
K_3(u)\lambda^2-K_4(u)\lambda-q_\kappa(u)&=0,
\label{quad-eqn}
\end{align}   
 in $\lambda$. The function $q_\kappa(u)$ is given in \eref{qu}, and one can find $K_3(u)$ and $K_4(u)$ from \erefss{h1} and \erefs{h2} for both definitions of entropy production and for both choices of external forces. The roots of the quadratic equation \erefs{quad-eqn} are
\begin{equation}
\bar\lambda_\pm(u)=\dfrac{K_4(u)\pm\sqrt{K_4^2(u)+4 K_3(u)q_\kappa(u)}}{2 K_3(u)}.
\label{lambda-delta-u}
\end{equation} 
{\color{black}Figures \ref{lambda-u}(a) and \frefs{lambda-u}(b) show the variation of $\bar{\lambda}_\pm(u)$ (dotted and dashed lines) against $u$ for $0<\delta<1$ at fixed trap strength $\kappa=2$. We have also compared the given curve with $\delta=0$ case (solid lines).} It is clear from \fref{lambda-u} that $\bar{\lambda}_\pm(u)$ converge to $\bar\lambda^0_\pm(u)$ in the limit of coupling tending to zero. Therefore, one can use perturbation theory to evaluate $u^*$ in the limit $\delta\to0$. For the first choice of forces and both definitions of entropy production,  we see that
\begin{equation}
u^*=\pm \bigg[\sqrt{\kappa}+\dfrac{\delta}{4\sqrt{\kappa}}+O(\delta^2)\bigg],
\end{equation}
whereas for the second choice of forces,
\begin{equation}
u^*=
\begin{cases}
\pm\bigg[\sqrt{\kappa}+\dfrac{\delta(1-\alpha)}{4\sqrt{\kappa}}+O(\delta^2)\bigg]\hspace{1.6cm} &\text{PEP},\\
\pm\bigg[\sqrt{\kappa}+\dfrac{\delta[1+2\alpha(\theta+\sqrt{\theta(1+\theta)})]}{4\sqrt{\kappa}}+O(\delta^2)\bigg] &\text{AEP},
\end{cases}
\end{equation}
where PEP and AEP stand for partial and apparent entropy production.

For each case, we substitute $u^*$ in \eref{lambda-delta-u}. In the limit $\delta\to 0$, we get $\bar{\lambda}_\pm(u^*)\to\lambda_\pm$ where $\lambda_\pm$ are given in \eref{lambda0+-}. When $\lambda\in(\lambda_-,\lambda_+)$, $g^0_\kappa(\lambda)$ is also an analytic function. Therefore, one can directly use the saddle-point approximation to compute the probability density function for $\Delta \mathcal{S}^A_{tot}$ (see \sref{fp-eqn}) and it has the following large deviation form \cite{Touchette}
\begin{equation}
p(s)\sim e^{(\tau/\tau_\gamma)K(s)},
\label{prob-den}
\end{equation} 
where $K(s)=\mu_\kappa(\bar\lambda^*)+\bar\lambda^* s$ in which $\bar\lambda^*$ is the saddle point obtained by solving the equation 
\begin{equation}
\dfrac{\partial \mu_\kappa(\lambda)}{\partial \lambda}\bigg|_{\lambda=\bar\lambda^*}=-s.
\end{equation}
Therefore, the asymmetry function in the large time limit is given by
\begin{equation}
f(s)=K(s)-K(-s).
\label{fs}
\end{equation}
\begin{figure*}\center
 \begin{tabular}{cc}
    \includegraphics[width=0.45  \textwidth]{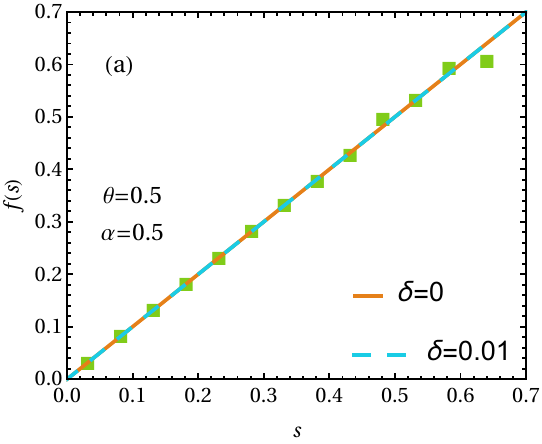}~~~~~~~~~
    \includegraphics[width=0.45 \textwidth]{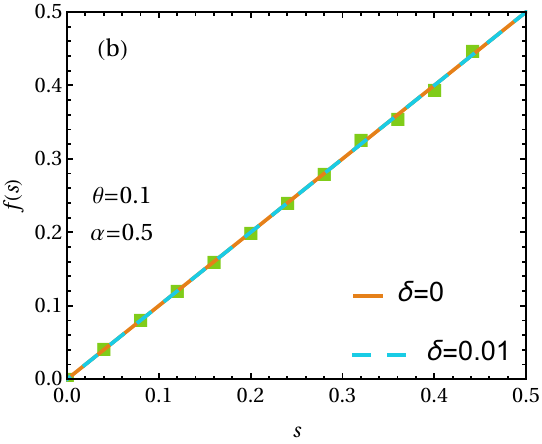}\\
    \includegraphics[width=0.45  \textwidth]{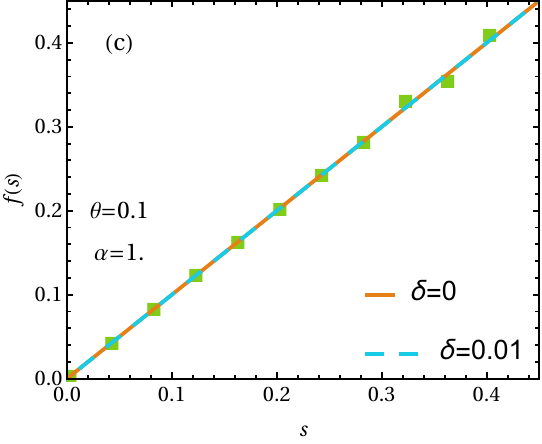}~~~~~~~~~
    \includegraphics[width=0.45 \textwidth]{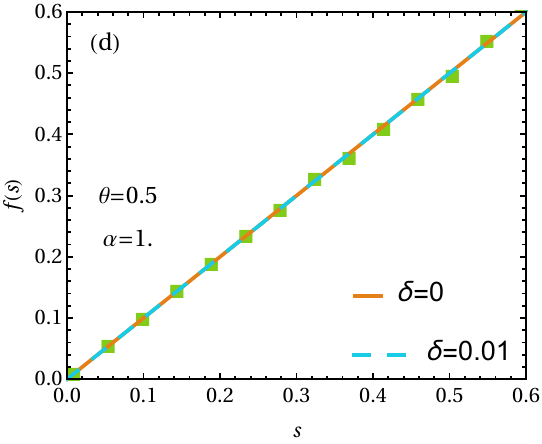}
 \end{tabular}
 \caption{\label{results}{\color{black}The asymmetry functions $f(s)$ (for $s>0$) against the scaled variable $s=\Delta \mathcal{S}^A_{tot}\tau_\gamma/\tau$. (a) Partial entropy production for the first choice of external forces, (b) apparent entropy production for the first choice of external forces, (c) partial entropy production for the second choice of external forces, and (d) apparent entropy production for the second choice of external forces. The solid lines correspond to the case when there is no coupling between particle A and B ($\delta=0$), i.e., $f(s)=s$ whereas the squares and dashed lines, respectively, correspond to numerical simulations and analytical predictions. For all above figures, we choose $\kappa=2.0$, $\delta=0.01$, $T=1$, and $\tau/\tau_\gamma=20$. A discussion on the time scale above which the analytical results are expected to match with the numerical simulations is given in \sref{summ}.}}
\end{figure*} 

\subsection{Numerical simulation}
\label{final-results}
{\color{black}In \fref{results}, we compare the analytical results (dashed lines) of asymmetry function $f(s)$ using \erefss{fs} with the numerical simulation results (squares) for partial entropy production for the first choice of external forces (a), apparent entropy production for the first choice of external forces (b), partial entropy production for the second choice of external forces (c), and apparent entropy production for the second choice of external forces (d). All of these results are obtained for fixed trap strength $\kappa=2.0$, coupling parameter $\delta=0.01$ and the observation time relative to relaxation time $\tau/\tau_\gamma=20$, and the temperature of the bath $T=1$. A discussion on the time scale above which the analytical results are expected to match with the numerical simulations is given in \sref{summ}. \ffref{results} shows a very good agreement between theoretical predictions and numerical simulation. }

From \frefss{results}(a)--(d), it is clear that the slope of asymmetry function $f(s)$ is unity in the limit $\delta \to 0$ which indicates that both definitions of total entropy production of partial system satisfy the SSFT.
\section{Method of simulation}
\label{method-sim}
To obtain the probability density function for $\Delta S_{tot}^A$, i.e.,  $p(s)=P(\Delta S_{tot}^A)\tau/\tau_\gamma $, in the numerical simulation, we first discretize the Langevin equations given in \erefss{Fd-1}--\erefs{Fd-3}, up to an order of time segment $\Delta \tau$. Discretization of the total entropy production for particle A in the coupled system given in \eref{Stot-model1} is as follows. While the first two terms of \eref{Stot-model1}, i.e., $W$ in \eref{Wtot}, are functional of stochastic trajectories, the third term is the boundary term which depends on the initial $v_A(0)$ and the final variable $v_A(\tau)$ of particle A. Notice that for both definitions of entropy production, the initial conditions are drawn from the steady state distribution of the full system [see \eref{SS-U}].  The functional $W$ given in \eref{Wtot}, which involves integrals that follow Stratonovich rule of integration, is discretized as following
\begin{align}
W=\sum_{n=0}^{(\tau/\Delta\tau)-1}[\sqrt{2\gamma T \theta/\Delta \tau}\tilde{f}^{\Delta \tau}_A(t_n)-\Pi k y(t_n )]~[v_A(t_n+\Delta \tau)+v_A(t_n)]\dfrac{\Delta \tau}{2 T},
\end{align}
where $t_n=n\Delta \tau$, and $\tilde{f}_A^{\Delta \tau}(t_n)$ is a Gaussian random variable with mean zero and variance one at each time step $t_n$.

We construct the histogram for $\Delta S_{tot}^A$ for both definitions of entropy production and also for both choices of external forces using $W$ given in the above equation, and the boundary terms as given in \eref{Stot-model1} for $R$ number of realizations. Similarly, we can obtain the probability density function $p(s)=P(\Delta \mathcal{S}_{tot}^A)\tau/\tau_\gamma$ for model 2 in the numerical simulation. 

 \section{Summary}
 \label{summ}
We have considered  two model systems: (model 1) a system of two Brownian particles coupled by a harmonic spring of stiffness $k$, and (model 2) a harmonically coupled Brownian particle system in a harmonic confinement of stiffness $k_0$. In both cases, the system is in contact with a heat bath of constant temperature. Each particle is driven by an external Gaussian white noise. Here we have considered two different choices of forces: (1) both forces are uncorrelated with each other, and (2) both of them are correlated with each other. The strength of the force acting on particle A relative to strength of the noise from the bath is $\theta$, whereas the ratio of strength of the force acting on particle B to that on particle A is $\alpha^2$. A dimensionless coupling parameter $\delta=2km/\gamma^2$ is also introduced. For model 2,  the strength of the harmonic trap is expressed as $\kappa=k_0m/\gamma^2$ (dimensionless). In the presence of external driving, the given systems reach a non-equilibrium steady state and produce entropy. Evidently, the total entropy production from the combined system (A+B) satisfies the fluctuation theorem. The central question we asked in this paper is whether the fluctuation theorem holds for a partial system in the weak coupling limit ($\delta\to0$). Therefore, we have focused on total entropy production due to one of the particle (partial and apparent entropy production) in both coupled systems. For model 1, we have plotted phase diagrams (see \fref{phase-dig}) in $(\alpha,\theta)$ plane in the limit $\delta\to0$, and the SSFT is satisfied for apparent entropy production only in region I of \frefss{phase-dig}(b) and \frefs{phase-dig}(d). In the case of model 2, both definitions of entropy production satisfy the SSFT in the weak coupling limit. Numerical simulations are also done to verify the analytical results and they are in good agreement.

To understand why the trap helps to nullify the effect of weak coupling of the hidden DOFs on the observed ones, 
let us consider the overdamped case from \erefss{full-dynamics-3}--\erefs{full-dynamics-4}, where,
in the presence of the trap, the relative spacing $y=(x_A-x_B)$ evolves according to overdamped Langevin equation given as
\begin{equation}
\dot{y}=-\dfrac{\delta+\kappa}{\tau_\gamma}y+\dfrac{\eta(t)+f(t)}{\gamma}.
\end{equation}
Here, $\eta(t)=\eta_A(t)-\eta_B(t)$ and $f(t)=f_A(t)-f_B(t)$ are the thermal Gaussian noise and external force in the relative frame, respectively. 

First consider the case when there is no harmonic confinement ($\kappa=0$).  In this case the force due to the coupling becomes important 
when 
 $y\sim O(\tau_\gamma/\delta)$.  
 The typical time-scale above which we can see the effect of coupling is given by the diffusive scale $t_y\sim y^2\sim O(\tau_\gamma^2/\delta^2)$ as, for $y \ll O(\tau_\gamma/\delta)$, the effect of coupling is negligible.
Therefore, when the observation time $\tau$ is much larger than $t_y$, we see a finite contribution to the medium entropy production from the term $\delta y$ as it becomes comparable to the external force $f_A$ even in the weak coupling limit ($\delta \to 0$) [see \eref{Stot-model1}].

On the other hand ($\kappa\neq0$ and $\delta \ll \kappa$), the variance of $y$ is proportional to $\tau_\gamma^2/(\delta+\kappa)$ in the limit $\tau\gg \tau_\gamma/\kappa$. Therefore, in the presence of the harmonic confinement, $y$ typically scales as $y\sim O(\tau_\gamma/\sqrt{\kappa})$ when $\tau\gg \tau_\gamma/\kappa$,  
and the force from the coupling term $\delta y\sim O(\delta \tau_\gamma /\sqrt{\kappa})$ is much smaller than the external force $f_A(t)$ for $\tau\gg\tau_\gamma/\kappa$. Thus, in this limit ($\delta \to 0$), the contribution to the medium entropy production from the term $\delta y$ is vanishingly small. Therefore, the SSFT holds for both definitions of entropy production in the presence of a harmonic confinement in the weak coupling limit (i.e., $\delta\ll\kappa$).

The asymmetry function $f(s)$ [see \eref{large-time}] may have negative slope \cite{Deepak-HT} but in the present case, it has only positive slope. The physical origin of the positive and negative slopes is not clear to us. It might be dependent on the choice of the system studied. Therefore, it would be interesting to understand the physical origin of this behavior in future.

\makeatletter%
\def\@sect#1#2#3#4#5#6[#7]#8{\ifnum #2>\c@secnumdepth
  \let\@svsec\@empty\else
  \refstepcounter{#1}\edef\@svsec{\csname the#1\endcsname. }\fi
  \@tempskipa #5\relax
  \ifdim \@tempskipa>\z@
  \begingroup #6\relax
  \noindent{\hskip #3\relax\@svsec}{\interlinepenalty \@M #8\par}%
  \endgroup
  \csname #1mark\endcsname{#7}\addcontentsline
          {toc}{#1}{\ifnum #2>\c@secnumdepth \else
            \protect\numberline{}{\csname the#1\endcsname}\fi
            . #7}\else
          \def\@svsechd{#6\hskip #3\relax  
            \@svsec #8\csname #1mark\endcsname
                    {#7}\addcontentsline
                    {toc}{#1}{\ifnum #2>\c@secnumdepth \else
                      \protect\numberline{}{\csname the#1\endcsname}\fi
                      . #7}}\fi
          \@xsect{#5}}
\makeatother%

\appendix

\section{Calculation of the moment generating function: Model 1}
\label{appendix-1}
In this section, we compute the moment generating function $Z(\lambda)$ of partial and apparent entropy production for model 1 for both choices of external forces using the method described in \cite{Kundu}. 
The model system shown in \fref{system} evolves according to underdamped Langevin \erefss{Fd-1}--\erefs{Fd-3}. We rewrite these equations in the matrix form as
\begin{align}
&\dot{y}=A^{T} V(t), \label{mat-1}\\
&m\dot{V}=-\gamma V(t)-k A y(t)+F(t)+\xi(t), \label{mat-2}
\end{align} 
where $V=(v_A,v_B)^T$, $A=(1,-1)^T$, $F=(f_A,f_B)^T$, and $\xi=(\eta_A,\eta_B)^T$.

Finite time Fourier transform and its inverse for a time-dependent quantity $Q(t)$ are defined as
\begin{align}
\tilde{Q}(\omega_{n})&=\dfrac{1}{\tau}\int_{0}^{\tau}dt\ Q(t)e^{-i \omega_{n}t}, \label{ftft-1}\\ 
Q(t)&=\sum_{n=-\infty}^{\infty} \tilde{Q}(\omega_{n})e^{i \omega_{n}t}, \label{ftft-2}
\end{align}
with $\omega_{n}=2\pi n/\tau$.

Using \eref{ftft-1}, we write  \erefss{mat-1} and \erefs{mat-2} as
 \begin{align}
&\tilde{y}(\omega_n)=A^{T}G(\omega_n) [\tilde{F}(\omega_n)+ \tilde{\xi}(\omega_n)]-\dfrac{A^{T}G(\omega_n)}{2 \tau}[(\gamma+i m \omega_n)A\Delta y+2m \Delta V], \label{y-tl}\\
&\tilde{V}(\omega_n)=i \omega_n G(\omega_n) [\tilde{F}(\omega_n)+ \tilde{\xi}(\omega_n)]+\dfrac{G(\omega_n)}{\tau}[k A \Delta y-i \omega_n m \Delta V]\label{v-tl},
 \end{align} 
where $\Delta y=y(\tau)-y(0)$, $\Delta V=V(\tau)-V(0)$, $\Phi=kAA^{T}$, and $G(\omega_n)=[(i\gamma \omega_n-m\omega_n^{2})\mathrm{I}+\Phi]^{-1}$ in which I is the identity matrix. From the above equations, we can obtain $\tilde{y}(\omega_n)$ and $\tilde{v}_A(\omega_n)$ as  
\begin{align}
&\tilde{y}(\omega_n)=[G_{11}(\omega_n)-G_{12}(\omega_n)][\tilde{\eta}_A(\omega_n)+\tilde{f}_A(\omega_n)-\tilde{\eta}_B(\omega_n)-\tilde{f}_B(\omega_n)]-\dfrac{q_1^T\Delta U}{\tau}, \label{y-o}\\
&\tilde{v}_A(\omega_n)=i\omega_n[G_{11}(\omega_n)\{\tilde{\eta}_A(\omega_n)+\tilde{f}_A(\omega_n)\}+G_{12}(\omega_n)\{\tilde{\eta}_B(\omega_n)+\tilde{f}_B(\omega_n)\}]+\dfrac{q^T_2\Delta U}{\tau}, \label{va-o}
\end{align}
where 
\begin{align}
q_1^T&=\bigg[\dfrac{\gamma+im\omega_n}{2}A^TG(\omega_n)A,m\{G_{11}(\omega_n)-G_{12}(\omega_n)\},~m\{G_{12}(\omega_n)-G_{11}(\omega_n)\}\bigg],\nonumber\\
q_2^T&=[k\{G_{11}(\omega_n)-G_{12}(\omega_n)\},-im\omega_n G_{11}(\omega_n),-im\omega_n G_{12}(\omega_n)],\nonumber
\end{align}
and $\Delta U=(\Delta y,\Delta V^T)^T$. In the above equations, 
the Green's function matrix elements are $G_{11}(\omega_n)=(i \gamma \omega_n-m\omega_n^{2}+k)[i\omega_n(\gamma+im\omega_n)(2k-m \omega_n^{2}+i\gamma \omega_n)]^{-1}$ and $G_{12}(\omega_n)$=$k[i\omega_n(\gamma+im\omega_n)(2k-m \omega_n^{2}+i\gamma \omega_n)]^{-1}$.

Therefore, the row vector $U^T(\tau)=[y(\tau),V^T(\tau)]$ is 
\begin{align}
U^{T}(\tau)&=\lim_{\epsilon \to 0}\sum_{n=-\infty}^{\infty}e^{-i \omega_n \epsilon} \tilde{U}^{T}(\omega_n)\nonumber\\&=\lim_{\epsilon \to 0}\sum_{n=-\infty}^{\infty}e^{-i \omega_n \epsilon} [\tilde{y}(\omega_n),\tilde{V}^{T}(\omega_n)].
\label{U}
\end{align}
Using \erefss{y-tl} and \erefs{v-tl}, we see that
\begin{align}
&\sum_{n=-\infty}^{\infty} e^{-i\omega_n\epsilon}\ \dfrac{A^{T}G}{2 \tau}\left[(\gamma+im \omega_n)A\Delta y+2m \Delta V\right],\nonumber\\
&\sum_{n=-\infty}^{\infty}  e^{-i\omega_n\epsilon}\ \dfrac{1}{\tau}\left[k A^{T}\ \Delta y-i \omega_n m \Delta V^{T}\right]G^{T},\nonumber
\end{align}
go to zero as $\tau\to\infty$. This is because, in the large-$\tau$ limit, the summations can be converted into integrals. As all the poles lie in the upper half of the complex $\omega$-plane, the contribution to integrals is zero. Therefore, \eref{U} reduces to
\begin{align}
U^{T}(\tau)&=\lim_{\epsilon \to 0}\sum_{n=-\infty}^{\infty} e^{-i \omega_n \epsilon} [ \{\tilde{F}^{T}(\omega_n)+\tilde{\xi}^{T}(\omega_n)\}G^{T}A, i\omega_n\{\tilde{F}^{T}(\omega_n)+\tilde{\xi}^{T}(\omega_n)\}G^{T}] \nonumber\\
&=\lim_{\epsilon \to 0}\sum_{n=-\infty}^{\infty} e^{-i \omega_n \epsilon}\ [(1-C)\{q^{T}_{3}(\tilde{\eta}_A+\tilde{f}_A)+q^{T}_{4}(\tilde{\eta}_B+\tilde{f}_B)\}+C(\tilde{\eta}_Al^T_{1}+\tilde{\eta}_Bl^T_{2}+\tilde{f}_Al^T_{3})],
\label{U-final}
\end{align} 
where 
\begin{align}
q_{3}^{T}&=l_{1}^{T}=(G_{11}-G_{12},i\omega_n G_{11},i\omega_n G_{12}),\nonumber\\
q_{4}^{T}&=l_{2}^{T}=(G_{12}-G_{11},i\omega_n G_{12},i\omega_n G_{11}),\nonumber\\
l_{3}^{T}&=[(1-\alpha)(G_{11}-G_{12}),i\omega_n(G_{11}+\alpha G_{12}),i\omega_n(G_{12}+\alpha G_{11})],\nonumber
\end{align}
and the parameter $C$ is given in \eref{C-parameter}. For convenience, we define $G_{ij}=[G(\omega_n)]_{ij}$, $G^*_{ij}=[G(-\omega_n)]_{ij}$, $\tilde{f}_i=\tilde{f}_i(\omega_n)$, $\tilde{f}^*_i=\tilde{f}_i(-\omega_n)$, $\tilde{\eta}_i=\tilde{\eta}_i(\omega_n)$, and $\tilde{\eta}_i^*=\tilde{\eta}_i(-\omega_n)$.

From \eref{U-final}, we see that $U$ depends linearly on the thermal Gaussian noises and external stochastic Gaussian forces. Therefore, the mean and correlation are sufficient to obtain the probability density function of it. Computing the average over thermal Gaussian noises and external stochastic Gaussian forces yields
\begin{align}
\langle U(\tau)\rangle=&0, \label{mean}                \\
\langle U(\tau) U^{T}(\tau)\rangle=&\dfrac{D}{\pi}\int_{-\infty}^{\infty}d\omega\ \big[(1-C)\{(1+\theta)q_{3}q_{3}^{\dagger}+(1+\alpha^{2}\theta)q_{4}q_{4}^{\dagger}\}+C(l_1l_1^\dagger+l_2l_2^\dagger+\theta l_3l_3^\dagger)\big],\label{corr}
\end{align}
where $\dagger$ denotes the transpose of the matrix and complex conjugation operation $*$ (i.e., $\omega_n\to-\omega_n$).

Thus, the steady state distribution of the coupled system is
\begin{equation}
P_{ss}(U)=\frac{e^{-\frac{1}{2}U^TM^{-1}U}}{\sqrt{(2\pi)^3 \det M}}, \quad \text{where}\quad M_{ij}=\langle U(\tau) U^{T}(\tau)\rangle_{ij}.\label{SS-U}
\end{equation}
The functional $W$ given in \eref{Wtot}, can be written as
\begin{align}
W=W_1-W_2,  \quad \quad \text{where}\\\label{Wtot-2}
W_1=\dfrac{1}{T}\int_0^\tau dt\ f_A(t)v_A(t),\\
W_2=\dfrac{\Pi k}{T}\int_0^\tau dt\ y(t) v_A(t). 
\end{align}
Using \eref{ftft-2}, we write $W_1$ as
\begin{align}
W_1=\dfrac{\tau}{2T}\sum_{n=-\infty}^{\infty}[\tilde{f}_A(\omega_n)\tilde{v}_A(-\omega_n)+\tilde{f}_A(-\omega_n)\tilde{v}_A(\omega_n)].
\end{align}
Substituting $\tilde{v}_A(\omega_n)$ from \eref{va-o} in the above equation yields
\begin{align}
W_1=\dfrac{\tau}{2T}\sum_{n=-\infty}^\infty\bigg[i\omega_n\{G_{11}\tilde{f}_A^*(\tilde{\eta}_A+\tilde{f}_A)+G_{12}\tilde{f}_A^*(\tilde{\eta}_B+\tilde{f}_B)-G^*_{11}\tilde{f}_A(\tilde{\eta}^*_A+\tilde{f}^*_A)
\nonumber\\
-G^*_{12}\tilde{f}_A(\tilde{\eta}^*_B+\tilde{f}^*_B\}+\dfrac{f_Aq_2^\dagger\Delta U}{\tau}+\dfrac{f^*_A\Delta U^Tq_2}{\tau}\bigg].
\end{align}
Similarly, using \eref{ftft-2}, we write $W_2$ as
\begin{align}
W_2=\dfrac{\Pi k\tau}{2T}\sum_{n=-\infty}^{\infty}[\tilde{y}(\omega_n)\tilde{v}_A(-\omega_n)+\tilde{y}(-\omega_n)\tilde{v}_A(\omega_n)].
\end{align}
Substituting $\tilde{y}(\omega_n)$ and $\tilde{v}_A(\omega_n)$ from \erefss{y-o} and \erefs{va-o}, respectively, in the above equation, we get
\begin{align}
W_2=&\dfrac{\Pi k\tau}{2T}\sum_{n=-\infty}^{\infty}\bigg[i\omega_n[\{G_{11}(\tilde{\eta}_A+\tilde{f}_A)+G_{12}(\tilde{\eta}_B+\tilde{f}_B)\}(G^*_{11}-G^*_{12})(\tilde{\eta}^*_A+\tilde{f}^*_A-\tilde{\eta}^*_B-\tilde{f}^*_B)\nonumber\\&
-\{G^*_{11}(\tilde{\eta}^*_A+\tilde{f}^*_A)+G^*_{12}(\tilde{\eta}^*_B+\tilde{f}^*_B)\}(G_{11}-G_{12})(\tilde{\eta}_A+\tilde{f}_A-\tilde{\eta}_B-\tilde{f}_B)]\nonumber\\&
+\dfrac{q_2^\dagger \Delta U}{\tau}(G_{11}-G_{12})(\tilde{\eta}_A+\tilde{f}_A-\tilde{\eta}_B-\tilde{f}_B)-\dfrac{i\omega_n}{\tau}q_1^\dagger\Delta U[G_{11}(\tilde{\eta}_A+\tilde{f}_A)+G_{12}(\tilde{\eta}_B+\tilde{f}_B)]\nonumber\\&
+\dfrac{i\omega_n}{\tau}\Delta U^Tq_1[G^*_{11}(\tilde{\eta}^*_A+\tilde{f}^*_A)+G^*_{12}(\tilde{\eta}^*_{B}+\tilde{f}^*_B)]+\dfrac{\Delta U^Tq_2}{\tau}(G^*_{11}-G^*_{12})(\tilde{\eta}^*_A+\tilde{f}^*_A-\tilde{\eta}^*_B-\tilde{f}^*_B)\nonumber\\&
-\dfrac{\Delta U^T(q_1q_2^\dagger+q_2q_1^\dagger)\Delta U}{\tau^2}\bigg].
\end{align}
The restricted moment generating function $Z_W(\lambda,U,\tau|U_0)$ for $W$ is
\begin{align}
Z_{W}(\lambda,U,\tau|U_{0})&=\left\langle e^{-\lambda W}\delta [U-U(\tau)]\right\rangle_{U_0}\nonumber\\&=\int \frac{d^{3}\sigma}{(2\pi)^{3}}e^{i \sigma^{T}U}\left\langle e^{E(\tau)}\right\rangle_{U_0},
\label{ZW}
\end{align}
where the angular brackets represent the average over the set of all trajectories for fixed initial $U_0$ and the final variable $U$. In the above equation, we have used the integral representation of the Dirac delta function and $E(\tau)=-\lambda W-i \sigma^{T} U(\tau)$. Substituting $W$ and $U(\tau)$ from \erefss{Wtot-2} and \erefs{U-final}, respectively, in $E(\tau)$ yields
\begin{align}
E(\tau)&=\sum_{n=1}^{\infty}\left[-\dfrac{\lambda \tau}{T}\zeta^{T}_{n}C_{n}\zeta^{*}_{n}+\zeta^{T}_{n}\alpha_{n}+\alpha_{-n}^{T}\zeta^{*}_{n}-\dfrac{\lambda \Pi k}{T\tau}|q_{n}|^{2} \right]\nonumber\\&-\dfrac{\lambda \tau}{2T}\zeta_{0}^{T}C_{0}\zeta_{0}+\zeta^{T}_{0}\alpha_{0}-\dfrac{\lambda \Pi k}{2T\tau}q_{0}^{2},
\end{align}
where $C_n=C_n^{\mathrm{I}}-\Pi k C_n^{\mathrm{II}}$ and $|q_n|^2=\Delta U^T(q_1q_2^\dagger+q_2q_1^\dagger)\Delta U$.

When the correlation parameter $C=0$, the matrices $C_n^{\mathrm{I}}$ and $C_n^{\mathrm{II}}$ are 
\begin{align}
&C^{\mathrm{I}}_{n}=\begin{bmatrix}
   0&0&i\omega_n G_{11}&0\\
   0&0&i\omega_n G_{12}&0\\
    -i\omega_n G^*_{11}&-i\omega_n G^*_{12}&i\omega_n [G_{11}- G^*_{11}]&-i\omega_n G^*_{12}\\
0&0&i\omega_n G_{12}&0\\
  \end{bmatrix},\nonumber\\
&C^{\mathrm{II}}_{n}=\begin{bmatrix}
   C_{11}&&C_{12}&&C_{13}&&C_{14}\\
   C^*_{12}&&C_{22}&&C_{23}&&C_{24}\\
    C^*_{13}&&C^*_{23}&&C_{33}&&C_{34}\\
C^*_{14}&&C^*_{24}&&C^*_{34}&&C_{44}\\
  \end{bmatrix}.\nonumber
\end{align}
The matrix elements of $C_n^{\mathrm{II}}$ are given as
\begin{align*}
&C_{11}=i\omega_n[G^*_{11}G_{12}-G_{11}G^*_{12}],\\
&C_{12}=i\omega_n[|G_{12}|^2-|G_{11}|^2],\\
&C_{11}=C_{13}=C_{33}=-C_{22},\\
&C_{12}=C_{14}=C_{34},\\
&C^*_{12}=C_{23},\\
&C_{22}=C_{24}=C_{44}.
\end{align*}
The column vector $\alpha_n$ is
\begin{align}
\alpha_{n}=-\dfrac{\lambda}{T}\begin{pmatrix}
a_{11}^T\Delta U\\
a_{21}^T\Delta U\\
a_{31}^T\Delta U\\
a_{41}^T\Delta U
 \end{pmatrix}
 -ie^{-i \omega \epsilon}\begin{pmatrix}
q_{3}^{T}\sigma\\
q_{4}^{T}\sigma\\
q_{3}^{T}\sigma\\
q_{4}^{T}\sigma
 \end{pmatrix},\nonumber
\end{align}
in which
\begin{align*}
a_{11}^T&=\Pi k[i\omega_n G_{11}q_1^\dagger-(G_{11}-G_{12})q_2^\dagger],\\
a_{21}^T&=\Pi k[i\omega_n G_{12}q_1^\dagger-(G_{12}-G_{11})q_2^\dagger],\\
a_{31}^T&=\Pi k[i\omega_n G_{11}q_1^\dagger-(G_{11}-G_{12})q_2^\dagger]+q_2^\dagger,\\
a_{41}^T&=\Pi k[i\omega_n G_{12}q_1^\dagger-(G_{12}-G_{11})q_2^\dagger],\\
\end{align*}
and the row vector containing thermal Gaussian noises and external Gaussian stochastic forces in the frequency domain is $\zeta^T=(\tilde{\eta}_{_A},\tilde{\eta}_{_B},\tilde{f}_{_A},\tilde{f}_{_B})$.

When the correlation parameter $C=1$, the matrices $C_n^{\mathrm{I}}$ and $C_n^{\mathrm{II}}$ are
\begin{align*}
&C^{\mathrm{I}}_{n}=\begin{bmatrix}
   0&0&i\omega_n G_{11}\\
   0&0&i\omega_n G_{12}\\
    -i\omega_n G^*_{11}&-i\omega_n G^*_{12}&i\omega_n[\{G_{11}-G^*_{11}\}+\\
    &&\alpha\{G_{12}-G^*_{12}\}]\\
  \end{bmatrix},\\
&C^{\mathrm{II}}_{n}=\begin{bmatrix}
   C_{11}&&C_{12}&&C_{13}\\
   C_{12}^*&&C_{22}&&C_{23}\\
    C^*_{13}&&C_{23}^*&&C_{33}
  \end{bmatrix},
\end{align*}
where the matrix elements of $C_n^{\mathrm{II}}$ are
\begin{align*}
C_{11}&=i\omega_n[G^*_{11}G_{12}-G_{11}G^*_{12}],\\
C_{12}&=i\omega_n[|G_{12}|^2-|G_{11}|^2],\\
C_{13}&=i\omega_n[G_{12}(G_{11}^*+\alpha G_{12}^*)-G_{11}(G_{12}^*+\alpha G_{11}^*)],\\
C_{22}&=-i\omega_n[G^*_{11}G_{12}-G_{11}G^*_{12}],\\
C_{23}&=i\omega_n[|G_{11}|^2-|G_{12}|^2+\alpha(G_{11}G_{12}^*-G_{12}G_{11}^*)],\\
C_{33}&=i\omega_n(1-\alpha^2)[G^*_{11}G_{12}-G_{11}G^*_{12}].\\
\end{align*}
The column vector $\alpha_n$ is
\begin{align}
\alpha_{n}=-\dfrac{\lambda}{T}\begin{pmatrix}
b_{11}^T\Delta U\\
b_{21}^T\Delta U\\
b_{31}^T\Delta U
 \end{pmatrix}
 -ie^{-i \omega \epsilon}\begin{pmatrix}
l_{1}^{T}\sigma\\
l_{2}^{T}\sigma\\
l_{3}^{T}\sigma
 \end{pmatrix}, \quad \quad \text{in which}\nonumber
\end{align}
\begin{align*}
b_{11}^T&=\Pi k[i\omega_n G_{11}q_1^\dagger-(G_{11}-G_{12})q_2^\dagger],\\
b_{21}^T&=\Pi k[i\omega_n G_{12}q_1^\dagger-(G_{12}-G_{11})q_2^\dagger],\\
b_{31}^T&=\Pi k[i\omega_n(G_{11}+\alpha G_{12})q_1^\dagger-(1-\alpha)(G_{11}-G_{12})q_2^\dagger]+q_2^\dagger,
\end{align*}
and the row vector containing thermal Gaussian noises and external Gaussian stochastic force in frequency domain is $\zeta^T_{n}=(\tilde{\eta}_A,\tilde{\eta}_B,\tilde{f}_A)$.

Therefore, we get
\begin{align}
\langle e^{E(\tau)}\rangle_{U_0}=&\left\langle \exp\left[-\frac{\lambda \tau}{2T} \zeta^{T}_{0} C_{0} \zeta_{0}+\zeta^{T}_{0}\alpha_{0}-\dfrac{\lambda \Pi k}{2T\tau}q_{0}^{2}  \right]\right\rangle\nonumber\\&\times\prod_{n=1}^{\infty}\left\langle \exp \left[-\frac{\lambda \tau}{T} \zeta^{T}_{n} C_{n} \zeta^{*}_{n}+\zeta^{T}_{n}\alpha_{n}+\alpha_{-n}^{T}\zeta^{*}_{n}-\dfrac{\lambda \Pi k}{T\tau}|q_{n}|^{2} \right]\right\rangle.
\end{align}
Here, the angular brackets show the average for each $n$$\geq$$1$ term according to the distribution given by 
\begin{equation}
P(\zeta_n)=
\begin{cases}
 \dfrac{\exp[-\zeta^{T}_{n} \Lambda^{-1}\zeta^{*}_{n}]}{\pi^{4} \det{\Lambda}},\quad\quad\text{for}\quad C=0,\\
 \vspace{0.005cm}\\
\dfrac{\exp[-\zeta^{T}_{n} \Lambda^{-1}\zeta^{*}_{n}]}{\pi^{3} \det{\Lambda}},\quad\quad \text{for}\quad C=1,
\end{cases}
\label{prob-n-1} 
\end{equation}
whereas for $n=0$ term, the average is taken with respect to the distribution
\begin{equation}
P(\zeta_0)=
\begin{cases}
\dfrac{ \exp[-\frac{1}{2}\zeta^{T}_{0} \Lambda^{-1}\zeta_{0}]}{\sqrt{(2 \pi)^{4}\det{\Lambda}}},\quad\quad \text{for}\quad C=0,\\
\vspace{0.005cm}\\
\dfrac{ \exp[-\frac{1}{2}\zeta^{T}_{0} \Lambda^{-1}\zeta_{0}]}{\sqrt{(2 \pi)^{3}\det{\Lambda}}},\quad\quad \text{for}\quad C=1.
\end{cases} 
\label{prob-n-2} 
\end{equation}
The diagonal matrix $\Lambda$ given above is 
\begin{equation*}
\Lambda=
\begin{cases}
(2D/\tau)\ \mathrm{diag}(1,1,\theta,\alpha^{2}\theta)\quad\quad\text{for}\quad C=0,\\
(2D/\tau)\ \mathrm{diag}(1,1,\theta)\hspace{0.8cm} \quad\quad\text{for}\quad C=1.
\end{cases} 
\end{equation*}
After computing the averages, we get
\begin{align}
\langle e^{E(\tau)}\rangle_{U_0}=e^{(\tau/\tau_{\gamma}) \mu(\lambda)}e^{\frac{1}{2}\sum_{n=-\infty}^{\infty}(\alpha_{-n}^{T}\Omega_{n}^{-1}\alpha_{n}-\frac{\lambda \Pi k}{T\tau}|q_{n}|^{2})},
\end{align}
where $\Omega_n=(\Lambda^{-1}+\lambda \tau\  C_{n}/T)$.

In the large time limit ($\tau/\tau_\gamma\to\infty$), we convert the summation into an integral, which gives
\begin{align}
\langle e^{E(\tau)}\rangle_{U_0}\approx e^{(\tau/\tau_{\gamma}) \mu(\lambda)}e^{-\frac{1}{2} \sigma^{T} H_{1}\sigma+i \Delta U^{T}H_{2} \sigma+\frac{1}{2}\Delta U^{T} H_{3}\Delta U},
\end{align}
where
\begin{align}
\mu(\lambda)&=-\dfrac{\tau_{\gamma}}{4 \pi}\int_{-\infty}^{\infty} d\omega\ \ln [\det{(\Lambda \Omega)}],\label{mu-int-1}\\
H_{1}&=\dfrac{\tau}{2 \pi}\int_{-\infty}^{\infty}d\omega\ \rho^{T}  \Omega^{-1} \phi,\label{H1}\\
H_{2}&=-\dfrac{\tau}{2 \pi}\int_{-\infty}^{\infty}d\omega\ e^{-i \omega\epsilon}a_1^{T}\Omega^{-1}\phi,\label{H2}\\
H_{3}&=\dfrac{\tau}{2 \pi}\int_{-\infty}^{\infty}d\omega\ \left[a_1^T\Omega^{-1}a_2-\dfrac{\lambda \Pi k}{T\tau}(q_{1}q_{2}^{\dagger}+q_{2}q_{1}^{\dagger})\right].\label{H3}
\end{align}
The vectors $\rho^T$, $a_1^T$, $\phi$, and $a_2$ are given in \tref{t1}, where
\begin{table}
\begin{tabular}{ |p{3cm}||p{5cm}|p{5cm}|}
 \hline
\hspace{1cm}Vectors&\hspace{1.cm}$C=0$&\hspace{1cm}$C=1$\\
 \hline 
\hspace{1.5cm}$\rho^T$&\hspace{0.5cm}$~~~~~~~(q_3^*,q_4^*,q_3^*,q_4^*)$&\hspace{0.7cm}$~~~~~~~(l_1^*,l_2^*,l_3^*)$\\
\hspace{1.5 cm}$a_1^T$&\hspace{0.001cm}$~~~~-\lambda/T(c_{11},c_{12},c_{13},c_{14})$&\hspace{0.2cm}$~~~-\lambda/T(d_{11},d_{12},d_{13})$\\
\hspace{1.5cm}$\phi$&\hspace{1.8cm}$\begin{pmatrix}
q_{3}^{T}\\
q_{4}^{T}\\
q_{3}^{T}\\
q_{4}^{T}
 \end{pmatrix}$&\hspace{1.5cm}$\begin{pmatrix}
l_{1}^{T}\\
l_{2}^{T}\\
l_{3}^{T}
 \end{pmatrix}$\\
  \hspace{1.5cm}$a_2$&\hspace{0.8cm} $-
 \dfrac{\lambda}{T}\begin{pmatrix}
a_{11}^T\\
a_{21}^T\\
a_{31}^T\\
a_{41}^T
 \end{pmatrix}$&\hspace{0.7cm}$-
 \dfrac{\lambda}{T}\begin{pmatrix}
b_{11}^T\\
b_{21}^T\\
b_{31}^T
 \end{pmatrix}$\\
   \hline
\end{tabular}
\caption{\label{t1} The vectors $\rho^T$, $a_1^T$, $\phi$, and $a_2$  for both choices of external forces.}
\end{table}
\begin{align*}
c_{11}&=-\Pi k[i\omega G^*_{11}q_1+(G^*_{11}-G^*_{12})q_2],\\
c_{12}&=-\Pi k[i\omega G^*_{12}q_1+(G^*_{12}-G^*_{11})q_2],\\
c_{13}&=-\Pi k[i\omega G^*_{11}q_1+(G^*_{11}-G^*_{12})q_2]+q_2,\\
c_{14}&=-\Pi k[i\omega G^*_{12}q_1+(G^*_{12}-G^*_{11})q_2],\\
d_{11}&=-\Pi k[i\omega G^*_{11}q_1+(G^*_{11}-G^*_{12})q_2],\\
d_{12}&=-\Pi k[i\omega G^*_{12}q_1+(G^*_{12}-G^*_{11})q_2],\\
d_{13}&=-\Pi k[i\omega (G^*_{11}+\alpha G_{12}^*)q_1+(G^*_{11}-G^*_{12})(1-\alpha)q_2]+q_2.
\end{align*}
Therefore, we write the restricted moment generating function for $W$ using \eref{ZW} as  
\begin{align}
Z_{W}(\lambda,U,\tau|U_{0})\approx e^{(\tau/\tau_{\gamma}) \mu(\lambda)} e^{\frac{1}{2}\Delta U^{T}H_{3}\Delta U}~\int \frac{d^{3}\sigma}{(2 \pi)^{3}}\ e^{i \sigma^{T} U}\  e^{-\frac{1}{2}\sigma^{T}H_{1}\sigma}\ e^{i \sigma^{T}H_{2}^{T}\Delta U}.
\end{align}
The integration over $\sigma$ yields
\begin{align}
Z_{W}(\lambda,U,\tau|U_{0})\approx \dfrac{e^{(\tau/\tau_{\gamma}) \mu(\lambda)}e^{\frac{1}{2}\Delta U^{T}H_{3}\Delta U}}{\sqrt{(2\pi)^{3}\det{H_{1}(\lambda)}}}~ e^{-\frac{1}{2}(U^{T}+\Delta U^{T}H_{2})H_{1}^{-1}(U+H_{2}^{T}\Delta U)}.
\end{align} 
Factorizing the above equation in terms of the initial $U_0$ and the final variable $U$ [see \eref{factor}] gives the condition $(H_{3}-H_{2}H_{1}^{-1}H_{2}^{T}-H_{1}^{-1}H_{2}^{T})+(H_{3}-H_{2}H_{1}^{-1}H_{2}^{T}-H_{2}H_{1}^{-1})^{T}=0$, therefore
\begin{align}
Z_{W}(\lambda,U,\tau|U_{0})\approx \dfrac{e^{(\tau/\tau_{\gamma}) \mu(\lambda)}e^{-\frac{1}{2}U^{T}L_{1}(\lambda)U}e^{-\frac{1}{2}U_{0}^{T}L_{2}(\lambda)U_{0}}}{\sqrt{(2 \pi)^{3}\det{H_{1}(\lambda)}}},
\label{factor-eqn}
\end{align}
where
\begin{align*}
&L_{1}(\lambda)=H_{1}^{-1}+H_{1}^{-1}H_{2}^{T},\\
&L_{2}(\lambda)=-H_{1}^{-1}H_{2}^{T}.
\end{align*}
From the above equation, one can find the steady state distribution for the coupled system  by substituting $\lambda=0$ and taking the limit $\tau\to\infty$:
\begin{equation}
Z_{W}(0,U,\tau \to\infty|U_{0})=P_{ss}(U)=\dfrac{\exp[-\frac{1}{2} U^{T}H_{1}^{-1}(0)U]}{\sqrt{(2 \pi)^{3}\det{H_{1}(0)}}}.
\end{equation}
Moreover, one can see from \eref{corr} that $\langle U(\tau) U^{T}(\tau)\rangle=H_{1}(0)$.

Using \eref{RCF}, the restricted moment generating function for $\Delta S_{tot}^A$ is given as 
\begin{align}
Z(\lambda,U,\tau|U_{0})\approx\dfrac{e^{(\tau/\tau_{\gamma}) \mu(\lambda)}e^{-\frac{1}{2}U^{T}\tilde{L}_{1}(\lambda)U}e^{-\frac{1}{2}U_{0}^{T}\tilde{L}_{2}(\lambda)U_{0}}}{\sqrt{(2 \pi)^{3}\det{H_{1}(\lambda)}}},
\label{stot-ch}
\end{align}
where the matrices $L_1(\lambda)$ and $L_2(\lambda)$ modify to
\begin{align*}
&\tilde{L}_{1}(\lambda)=L_{1}(\lambda)-\lambda\Sigma\bigg[\dfrac{m}{T}-\dfrac{1}{H}\bigg],\\
&\tilde{L}_{2}(\lambda)=L_{2}(\lambda)+\lambda\Sigma\bigg[\dfrac{m}{T}-\dfrac{1}{H}\bigg].
\end{align*}
The moment generating function $Z(\lambda)$ is obtained by integrating $Z(\lambda,U,\tau|U_0)$ over the initial variable $U_0$ with respect to the steady state distribution $P_{ss}(U_0)$ and the final variable $U$, 
\begin{align}
Z(\lambda)&=\int dU \int dU_{0}\ P_{ss}(U_{0})\ Z(\lambda,U,\tau|U_{0})\nonumber\\
&=g(\lambda)\ e^{(\tau/\tau_{\gamma}) \mu(\lambda)}+\dots,
\end{align}
where the prefactor $g(\lambda)$ is 
\begin{equation}
 g(\lambda)=\left[\det[{H_{1}(\lambda)H_{1}(0)\tilde{L}_{1}(\lambda)}]\det{[H_{1}^{-1}(0)+\tilde{L}_{2}(\lambda)]}\right]^{-1/2}.
 \label{g-lambda}
 \end{equation}


\section{Equations of contours and branch point singularities}
\label{BP-EQ-C}
 {\color{black} In this section, we discuss the equation of contours for \fref{phase-dig} which separates the regions of different branch point singularities in $\mu(\lambda)$.

For parameters $\Pi=1$, $C=0$, and $0<\delta<1$, we see that $\lambda_+^{\prime\prime}(0)>0$ and the sign of $\lambda_-^{\prime\prime}(0)$
is determined by a function $r_1(\theta,\alpha,\delta)$. Therefore, the equation of contour in this case is given by
\begin{equation}
r_{1}(\theta,\alpha,\delta)=0, \quad \quad \text{where}
\label{c1}
\end{equation}
 \begin{equation}
 r_{1}(\theta,\alpha,\delta)=-1-\theta(\alpha^{2}-3)(2+\theta+\theta\alpha^{2})+\delta[1+\theta(1+\alpha^{2})]^{2}\nonumber.
 \end{equation}
We plot the phase diagram in $(\alpha,\theta)$ plane using \eref{c1} in the weak coupling limit $(\delta\to0)$ as shown in \fref{phase-dig}(a).
In this case, the pair of branch point singularities in the limit $\delta\to0$ are $(\lambda_-,\tilde{\lambda}_+)$ and $(\tilde{\lambda}_-,\tilde{\lambda}_+)$ in regions I and II, respectively, of  \fref{phase-dig}(a) where
 \begin{align}
 \tilde{\lambda}_+&=1,\\
 \tilde{\lambda}_-&=-(1+\theta+\theta\alpha^2)^{-1}.
 \end{align}
For parameters $\Pi=0$, $C=0$, and $0<\delta<1$, the sign of $\lambda_+^{\prime\prime}(0)$ and $\lambda_-^{\prime\prime}(0)$ is determined
by functions $r_2(\theta,\alpha,\delta)$ and $r_3(\theta,\alpha,\delta)$, respectively.
Therefore, the equations of contours in this case are given by
\begin{align}
r_2(\theta,\alpha,\delta)=0,\label{c2}\\
r_3(\theta,\alpha,\delta)=0\label{c3},
\end{align}
where 
\begin{align}
r_2(\theta,\alpha,\delta)&=-4+\alpha^{4}\theta^{2}+4 \alpha^{2}\theta[\theta+\sqrt{\theta(2+\theta+\theta\alpha^{2})}]-\delta (2+\theta\alpha^{2})^{2},\nonumber\\
r_3(\theta,\alpha,\delta)&=4-\alpha^{4}\theta^{2}-4 \alpha^{2}\theta[\theta-\sqrt{\theta(2+\theta+\theta\alpha^{2})}]+\delta (2+\theta\alpha^{2})^{2}.\nonumber
\end{align}
Using \erefss{c2} and \erefs{c3}, we plot the phase diagram in $(\alpha,\theta)$ plane as shown in \fref{phase-dig}(b) in the limit $\delta\to0$.
In this case, the pair of branch point singularities in the limit $\delta \to 0$ are $(\lambda_-,\lambda_+)$, $(\lambda_-,\tilde{\lambda}_+)$, and $(\tilde{\lambda}_-,\tilde{\lambda}_+)$ in regions I, II and III, respectively, of \fref{phase-dig}(b) where
 \begin{align}
 \tilde{\lambda}_\pm=\dfrac{\theta\pm\sqrt{\theta(2+\theta+\theta\alpha^2)}}{2\theta+\theta^2\alpha^2}.
 \end{align}
For parameters $\Pi=1$, $C=1$, and $0<\delta<1$, we see that $\lambda_+^{\prime\prime}(0)>0$ and  the sign of $\lambda_-^{\prime\prime}(0)$
is determined by a function $r_4(\theta,\alpha,\delta)$. Therefore, the equation of contour in this case is given by
\begin{equation}
r_4(\theta,\alpha,\delta)=0, \quad \quad \text{where}
\label{c4}
\end{equation}
 \begin{align}
 r_4(\theta,\alpha,\delta)=-1-\theta(\alpha-1)(3+\alpha)[2+(1+\alpha)^2\theta]+2\delta[1-\theta(1-\alpha^2)\{2+(1+\alpha)^2\theta\}]\nonumber\\
 -\delta^2[1+\theta(1+\alpha)^2]^2.\nonumber
 \end{align}
We use \eref{c4} to plot the phase diagram in $(\alpha,\theta)$ plane in the limit $\delta\to0$ as shown in \fref{phase-dig}(c).
In this case, the pair of branch point singularities in the limit $\delta\to0$ are $(\lambda_-,\tilde{\lambda}_+)$ and $(\tilde{\lambda}_-,\tilde{\lambda}_+)$ in regions I and II, respectively, of \fref{phase-dig}(c) where
 \begin{align}
 \tilde{\lambda}_+&=1,\\
 \tilde{\lambda}_-&=-[1+\theta(1+\alpha)^2]^{-1}.
 \end{align}
Finally, for parameters $\Pi=0$, $C=1$, and $0<\delta<1$, we see that $\lambda_+^{\prime\prime}(0)<0$ and  the sign of $\lambda_-^{\prime\prime}(0)$
is determined by a function $r_5(\theta,\alpha,\delta)$. Therefore, the equation of contour in this case is given by
\begin{equation}
r_5(\theta,\alpha,\delta)=0, \quad \quad \text{where}
\label{c5}
\end{equation}
 \begin{align}
 r_5(\theta,\alpha,\delta)=1+2\alpha \theta(1+\alpha)-2\alpha\sqrt{\theta [2+\theta(1+\alpha)^2]}-\delta^2.\nonumber
 \end{align}
 In the limit $\delta\to0$, the phase diagram in $(\alpha,\theta)$ plane is shown using \eref{c5} in \fref{phase-dig}(d), and the pair of branch point singularities in the limit $\delta\to0$ are $(\lambda_-,\lambda_+)$ and $(\tilde{\lambda}_-,\lambda_+)$ in regions I and II, respectively, of \fref{phase-dig}(d) where
 \begin{align}
 \tilde{\lambda}_-=\dfrac{\theta(1+\alpha)-\sqrt{\theta[2+\theta(1+\alpha)^2]}}{2\theta}.
 \end{align}
In all above cases, $\lambda_\pm$ are given by \eref{lambda0-pm}.

}
\section{The large deviation function}
\label{section-LDF}
The large deviation function (LDF) is defined by 
\begin{equation}
I(s)=\lim_{(\tau/\tau_\gamma) \to\infty} \dfrac{1}{(\tau/\tau_\gamma)} \ln P(\Delta S^A_{tot}=s \tau/\tau_\gamma).
\end{equation}
Therefore, from \eref{saddle} we get
\begin{equation}
I(s):=h_s(\lambda^*)=\mu(\lambda^*) +\lambda^* s.
\label{LDF}
\end{equation}

First, consider the case when particle A is isolated from particle B ($\delta=0$) (see \sref{mu-delta-lambda}). 
In this case, $\mu_0(\lambda)$ is analytic only
within a finite region bounded by a pair of branch point singularities
at $\lambda_\pm$.  Here $\lambda_- <0$ and $\lambda_+ >1$ with
$\lambda_+ +\lambda_-=1$, where $\lambda_\pm$ are given in \eref{lambda0-pm}. In this case, $g_0(\lambda)$ is
also analytic within this region $\lambda \in (\lambda_-, \lambda_+)$.

The solution of the equation
\begin{equation}
\mu_0'(\lambda_0^*)=-s, 
\label{saddle-point condition0}
\end{equation}
gives the saddle point
\begin{equation} 
\lambda_0^*(s)=\dfrac{1}{2}\bigg[1-\dfrac{s}{\sqrt{s^2+\theta}}\sqrt{1+\dfrac{1}{\theta}}\bigg].
\label{sad-pt-0}
\end{equation}
It follows that
\begin{equation}
 \lambda_0^*(s) =
\begin{cases}
 \lambda_- +
 O(1/s^2) &\text{as $s\to +\infty$} ,\\
\lambda_+ -
 O(1/s^2) &\text{as $s\to -\infty$}.
\end{cases}
\end{equation}
Therefore, as $s$ decreases from $\infty$ to $-\infty$, the saddle
point $\lambda_0^*(s)$ moves from $\lambda_-$ to $\lambda_+$ on the
real $\lambda$ line. 

Thus, the LDF $I_0(s)$ is given by
\begin{align}
I_0(s)&=\mu_0(\lambda_0^*(s))+\lambda^*_0(s) s\nonumber\\&= 
\begin{cases}
\mu_0(\lambda_-) + \lambda_- s +
O(1/s) &\text{as
$s\to +\infty$}, \\
\mu_0(\lambda_+)+ \lambda_+ s -
O(1/s) &\text{as $s\to-\infty$}.
\end{cases}
\end{align}

In the presence of a nonzero coupling ($\delta > 0$), 
$\mu(\lambda)$ has branch points at $\lambda_\pm^{(\delta)}$. In this
case, the LDF $I(s)$ is related to $\mu(\lambda)$ by
\begin{equation*}
I(s)=\mu(\lambda^*) +\lambda^* s\quad\text{with}~ \mu'(\lambda^*)=-s,
\end{equation*}
where we have assumed that $g(\lambda)$ is analytic in the region
$\lambda\in\bigl(\lambda_-^{(\delta)}, \lambda_+^{(\delta)}\bigr)$.  In case
$g(\lambda)$ has a singularity within this range, it can change the
above LDF. However, we are interested in the $\delta\to0$
limit and in this limit we can write $g(\lambda)=g_0(\lambda)
+ \delta^{c} g_1(\lambda)$, with $c>0$, where the function
$g_1(\lambda)$ may have singularities. It is clear that the
singularities of $g_1(\lambda)$ are not going to contribute to the PDF
in the $\delta\to0$ limit \cite{Deepak}.

Now, we see from the \sref{mu-delta-lambda} in the limit $\delta \to 0$, $\lambda^{(\delta)}_+$$\to$$\tilde{\lambda}_+$ [see \fref{lamb-var}(a)] or $\lambda_+$ [see \fref{lamb-var}(b)] and $\lambda^{(\delta)}_-$$\to$$\tilde{\lambda}_-$ [see \fref{lamb-var}(c)] or $\lambda_-$ [see \fref{lamb-var}(d)]. In general, we can write the $\mu(\lambda)$ around these singularities. Consider a case when $\lambda^{(\delta)}_-\to\tilde{\lambda}_-$ and $\lambda^{(\delta)}_+\to\lambda_+$, near
$\tilde\lambda_-$, we can write
\begin{math}
\mu(\lambda)=\mu_\mathrm{a} (\lambda)
+\mu_\mathrm{s}(\lambda)
\end{math}
where $\mu_\mathrm{a}(\lambda)$ and $\mu_\mathrm{s}(\lambda)$ are,
respectively, the analytic and singular part of
$\mu(\lambda)$. Evidently, $\mu_\mathrm{a}(\lambda)\to\mu_0(\lambda)$
as $\delta\to0$. On the other hand, for the singular part near
$\tilde\lambda_-$, for small $\delta$, $\mu_\mathrm{s}(\lambda)
\propto - \delta\sqrt{\lambda-\tilde\lambda_-}$ [see \eref{asymp-mu}].  Note that, for $\delta\to 0$, if
$\lambda_-^{(\delta)}\to \tilde\lambda_-$ instead of $\lambda_-$, then
it is necessary that $\lambda_- < \tilde\lambda_- <0$. In the limit
$\delta\to 0$, when $s$ increases from $-\infty$ to $\infty$, the
saddle point $\lambda^*(s)$ moves from $\lambda_+$ to
$\tilde\lambda_-$ on the real $\lambda$ line. For
$[\lambda^*(s)-\tilde\lambda_- ]\gg \delta^{2}$, the
saddle point is dominated by the equation $\mu'_\mathrm{a}(\lambda^*)
= -s$, which in the limit $\delta\to 0$ reduces to \eref{saddle-point condition0}. Therefore, the LDF is the same $I_0(s)$ that has been
obtained for the uncoupled case. On the other hand, for
$[\lambda^*(s)-\tilde\lambda_-] \ll \delta^{2}$, the
saddle point is dominated by the singular part of the saddle point
equation, which results in
\begin{math}
\lambda^*(s)=\tilde\lambda_- +
O(\delta^2/s^2). 
\end{math}
This gives $I(s)=\mu_\mathrm{a}\bigl[\tilde\lambda_- +
O(\delta^2/s^2)\bigr]+
\tilde\lambda_- s +
O(\delta^2/s). $ Thus, in the
limit $\delta\to0$, we get
\begin{equation}
I(s)=
\begin{cases}
I_0(s)  &\text{for $s<s_1^*$},\\
\mu_0(\tilde\lambda_-) + \tilde\lambda_- s &\text{for $s>s_1^*$},
\end{cases}
\label{LDF2}
\end{equation}
where $s_1^*$ is given by 
\begin{equation}
\lambda_0^*(s_1^*)=\tilde\lambda_-.
\label{s* condition}
\end{equation}
with $\lambda_0^*(s)$ is the saddle point  given in \eref{sad-pt-0} for the uncoupled case ($\delta=0$).

A similar calculation can be done for the case when $\lambda^{(\delta)}_-$$\to$$\lambda_-$ and $\lambda^{(\delta)}_+$$\to$$\tilde{\lambda}_+$, in the limit $\delta\to0$
\begin{equation}
I(s)=
\begin{cases}
\mu_0(\tilde\lambda_+) + \tilde\lambda_+ s &\text{for $s<s_2^*$},\\[1mm]
I_0(s)  &\text{for $s>s_2^*$},
\end{cases}
\label{LDF3}
\end{equation}
where $s_2^*$ is given by 
\begin{equation}
\lambda_0^*(s_2^*)=\tilde\lambda_+,
\end{equation} 
with $\lambda_0^*(s)$ is the solution of \eref{saddle-point
condition0}.

Since $\lambda_0^*(s)$ is a monotonically decreasing
function of $s$ and $\tilde\lambda_+ > \tilde\lambda_-$, we have
$s_2^* < s_1^*$, and for the case when $\lambda^{(\delta)}_-$$\to$$\tilde{\lambda}_-$ and $\lambda^{(\delta)}_+$$\to$$\tilde{\lambda}_+$ , in the limit $\delta\to0$
\begin{equation}
I(s)=
\begin{cases}
\mu_0(\tilde\lambda_+) + \tilde\lambda_+ s &\text{for $s<s_2^*$},\\
I_0(s)  &\text{for $s_2^*<s<s_1^*$},\\
\mu_0(\tilde\lambda_-) + \tilde\lambda_- s &\text{for $s>s_1^*$}.
\end{cases}
\label{LDF4}
\end{equation}

Finally, when $\lambda^{(\delta)}_-$$\to$$\lambda_-$ and $\lambda^{(\delta)}_+$$\to$$\lambda_+$, in the limit $\delta\to0$, we get \begin{equation}
I(s)=I_0(s) \hspace{0.5cm}\text{for all $s$}
\end{equation}
\section{Second order discontinuity of the large deviation function}
\label{cont-LDF}
Consider a case where the LDF given by \eref{LDF2} has the following form
\begin{equation*}
I(s)=
\begin{cases}
\mu_0(\lambda_0^*) + \lambda_0^* s &\text{for $s<s_1^*$}, \\
\mu_0(\tilde\lambda_-) + \tilde\lambda_- s &\text{for $s>s_1^*$}, 
\end{cases}
\end{equation*}
where $\lambda_0^*(s)$ is the solution of \eref{saddle-point
 condition0} and $s_1^*$ is given by \eqref{s* condition}.
Evidently, $I(s^*_{1-})= I(s^*_{1+})$, where $s^*_{1\pm}=\lim_{\epsilon\to 0}
(s_1^*\pm \epsilon)$.

Taking a derivative with respective to $s$, for $s>s_1^*$ we have
$I'(s)=\tilde{\lambda}_-$. On the other hand, for $s< s_1^*$ we get
\begin{equation*}
I'(s)=\lambda_0^*(s) + \frac{d \lambda_0^*}{d s}
\left[\mu_0'(\lambda_0^*)+s\right].
\end{equation*}
Note that the prime $'$ represents the derivative with respect
to $s$ ($\lambda_0^*$) on the left- (right-) hand side of above equation.
Now using Eqs.~\erefs{saddle-point condition0} and \erefs{s*
condition}, we have $I'(s_{1-}^*)= I'(s_{1+}^*)$.

For the second derivatives, for $s>s_1^*$ we have $I''(s)=0$. On the
other hand, for $s<s_1^*$, we get 
\begin{equation*}
I''(s)= \frac{d \lambda_0^*}{d s}=-\frac{1}{\mu_0''(\lambda_0^*)}. 
\end{equation*}
Therefore, $I''(s_{1-})=-1/\mu''_0(\tilde{\lambda}_-)$ whereas
$I''(s_{1+})=0$ --- the second derivative is discontinuous across $s=s_1^*$.

Similarly, one can show that the LDF has second order discontinuities
across $s_1^*$ and $s_2^*$ for the other cases also.
\section{The asymmetry function and its discontinuity}
\label{cont-asymm-function}
In this section, we analyze the asymmetry function which is defined as follows
\begin{equation}
f(s)=I(s)-I(-s),
\label{asymmetry function}
\end{equation}
in the limit $\delta\to 0$.  We analyze below $f(s)$ only for $s>0$, as
for $s<0$ it can be obtained from the relation $f(-s)=-f(s)$.

When $\lambda^{(\delta)}_-$$\to$$\lambda_-$ and $\lambda^{(\delta)}_+$$\to$$\lambda_+$, in the limit $\delta\to0$, evidently $f(s)=s$ for all $s$. Now, for the
situation when $\lambda^{(\delta)}_-$$\to$$\tilde{\lambda}_-$ and $\lambda^{(\delta)}_+$$\to$$\lambda_+$ in the limit $\delta\to0$, we analyze the asymmetry function using the expression of the LDF given
by \eref{LDF2}. Here, for brevity, we denote
$$I_1(s)= \mu_0(\tilde\lambda_-) + \tilde\lambda_- s.$$ From \eref{sad-pt-0}, $\lambda_0^*(s) + \lambda_0^*(-s)=1$, and 
$\lambda_0^*(0)=1/2$. Since $\lambda_0^*(s)$ is a monotonically
decreasing function of $s$, we have $\lambda^*_0 > 1/2 $ for $s<0$ and
$\lambda^*_0 < 1/2 $ for $s>0$. Therefore, $s_1^*$ is always positive as
$\tilde{\lambda}_- <0$. Now, for
$s_1^* >0$, we get
\begin{equation*}
f(s)=\begin{cases} 
I_0(s) - I_0(-s)=s  &\text{for $0 \le s < s_1^*$},\\
I_1(s) -
I_0(-s) &\text{for $s> s_1^*$}.
\end{cases}
\end{equation*}
Since the LDF has a second order discontinuity at $s_1^*$, it is
evident that $f(s)$ also has a second order discontinuity at
$s_1^*$. The asymptotic expression of $f(s)$, as $s\to\infty$, is
given by
\begin{equation*}
f(s)= \bigl[\mu_0(\tilde\lambda_-) -\mu_0(\lambda_+)\bigr] +
[\tilde\lambda_- +\lambda_+]s +\dotsb
\end{equation*}

When $\lambda^{(\delta)}_-$$\to$$\lambda_-$ and $\lambda^{(\delta)}_+$$\to$$\tilde{\lambda}_+$ in the limit $\delta\to0$, we again for brevity, denote $$I_2(s)=
[\mu(\tilde\lambda_+) +\tilde\lambda_+ s]$$ in \eref{LDF3}.  Since
$\tilde\lambda_+>0$, we get $s_2^* < 0$ for $ \tilde\lambda_+>1/2$ and
$s_2^* > 0$ for $ 0< \tilde\lambda_+< 1/2$. Now when $s_2^* < 0$ we
get
\begin{equation*}
f(s)=\begin{cases}
 I_0(s)-I_0(-s)=s &
\text{for $0\le s < -s_2^*$},\\
I_0(s) -I_2(-s) &\text{for $s> -s_2^*$}.
\end{cases}
\end{equation*}
 On the other hand,
when $s_2^* > 0$, we get
\begin{equation*}
f(s)=\begin{cases}
 I_2(s)-I_2(-s)=2\tilde\lambda_+ s &
\text{for $0\le s < s_2^*$},\\
I_0(s) -I_2(-s) &\text{for $s> s_2^*$}.
\end{cases}
\end{equation*}
Again, from the second order discontinuity of the LDF, it is evident
that $f(s)$ also exhibits a second order discontinuity at
$|s_2^*|$. The asymptotic expression of $f(s)$, as $s\to\infty$, is
given by
\begin{equation*}
f(s)= \bigl[\mu_0(\lambda_-) -\mu_0(\tilde\lambda_+)\bigr] 
+ [\tilde\lambda_+ +\lambda_-]s +\dotsb
\end{equation*}

When $\lambda^{(\delta)}_-$$\to$$\tilde{\lambda}_-$ and $\lambda^{(\delta)}_+$$\to$$\tilde{\lambda}_+$ in the limit $\delta\to0$, for the case $s_2^* >0$, we get
\begin{equation*}
f(s)=\begin{cases}
I_2(s) - I_2(-s)=2\tilde\lambda_+ s & \text{for $0 \le s < s_2^*$},\\
I_0(s) - I_2(-s) & \text{for $s_2^* < s < s_1^*$},\\
I_1(s) - I_2(-s) & \text{for $ s > s_1^*$}.
\end{cases}
\end{equation*}

On the other hand, for $s_2^* <0$, there may be two cases: (1) $-s_2^*
< s_1^*$ and (2) $-s_2^* > s_1^*$. In the first case, we have
\begin{equation*}
f(s)=\begin{cases}
I_0(s) - I_0(-s)=s & \text{for $0 \le s < -s_2^*$},\\
I_0(s) - I_2(-s) & \text{for $-s_2^* < s < s_1^*$},\\
I_1(s) - I_2(-s) & \text{for $ s > s_1^*$},
\end{cases}
\end{equation*}
whereas for the second case, we get
\begin{equation*}
f(s)=\begin{cases}
I_0(s) - I_0(-s)=s & \text{for $0 \le s < s_1^*$},\\
I_1(s) - I_0(-s) & \text{for $s_1^* < s < -s_2^*$},\\
I_1(s) - I_2(-s) & \text{for $ s > -s_2^*$}.
\end{cases}
\end{equation*}
From, the second order discontinuities of the LDF at the points $s_1^*$
and $s_2^*$, it is evident that $f(s)$ also exhibits second order
discontinuity at the points $s_1^*$ and $|s_2^*|$.  Moreover, in all
cases, for $s > \max(s_1^*, |s_2^*|)$, we have
\begin{equation*}
f(s)=\bigl[\mu_0(\tilde\lambda_-) -\mu_0(\tilde\lambda_+)\bigr] +
[\tilde\lambda_+ +\tilde\lambda_-]s. 
\end{equation*}

\section{Calculation of moment generating function: Model 2}
\label{appendix}
In this section, we use the method developed in \cite{Kundu} and recently used in \cite{SS-1,SS-2,Apal-1,Apal-2} to compute the moment generating function for partial and apparent entropy production for model 2. The Langevin \erefss{full-dynamics-1}--\erefs{full-dynamics-4} for the coupled system shown in \fref{system-pic}, can be written in the matrix form as
\begin{align}
&\dot{X}=V(t),\label{mat-dynamics-1}\\
&m\dot{V}=-\gamma V(t)-\bar{\Phi} X(t)+\xi(t)+F(t),
\label{mat-dynamics-2}
\end{align}
where $X(t)=(x_A(t),x_B(t))^T$, $V(t)=(v_A(t),v_B(t))^T$, $\xi(t)=(\eta_A(t),\eta_B(t))^T$, $F(t)=(f_A(t),f_B(t))^T$, and the matrix $\bar{\Phi}$ is given by
\[
\bar{\Phi}=\begin{pmatrix}
k_0+k&&-k\\
-k&&k_0+k
\end{pmatrix}.
\]
Using the finite time Fourier transform defined in \aref{appendix-1}, one can write \erefss{mat-dynamics-1} and \erefs{mat-dynamics-2} in frequency domain as
\begin{align}
\tilde{X}(\omega_n)&=\bar{G}[\tilde{F}(\omega_n)+\tilde{\xi}(\omega_n)]-\dfrac{\bar{G}}{\tau}[(im\omega_n+\gamma)\Delta X+m\Delta V],\label{X-fourier}\\
\tilde{V}(\omega_n)&=i\omega_n \bar{G}[\tilde{F}(\omega_n)+\tilde{\xi}(\omega_n)]+\dfrac{\bar{G}}{\tau}[\bar{\Phi} \Delta X-im\omega_n \Delta V],\label{V-fourier}
\end{align}
where $\Delta X=X(\tau)-X(0)$, $\Delta V=V(\tau)-V(0)$, and  $\bar{G}(\omega_n)=[(-m\omega_n^2+i\gamma \omega_n){ \mathrm{I}}+\bar{\Phi}]^{-1}$ is the Green's function symmetric matrix in which I is the identity matrix.

We calculate $\mathcal{U}^{T}(\tau)=[X^T(\tau),V^T(\tau)]$ as
\begin{equation}
\mathcal{U}^T(\tau)=\lim_{\epsilon \to 0}\sum_{n=-\infty}^\infty e^{-i\omega_n \epsilon} \big[\tilde{X}^T(\omega_n),\tilde{V}^T(\omega_n)\big].
\end{equation}
Using \erefss{X-fourier} and \erefs{V-fourier} in the above equation, we see that the following terms
\begin{align}
&\lim_{\epsilon \to 0}\frac{1}{2\pi}\int_{-\infty}^\infty d\omega\ e^{-i\omega \epsilon}[(im\omega+\gamma)\Delta X^T+m\Delta V^T] \bar{G}^{T}\rightarrow 0 \nonumber\\
&\lim_{\epsilon \to 0}\frac{1}{2\pi}\int_{-\infty}^\infty d\omega\ e^{-i\omega \epsilon}[\Delta X^T\bar{\Phi}^T-im\omega \Delta V^T] \bar{G}^T\rightarrow 0. \nonumber
 \end{align}
This is because, in the large time limit ($\tau\to\infty$), we can convert the summation into integration over $\omega$, and the contour of integration (clockwise) is a semicircle in the lower half of the complex $\omega$-plane  with center at the origin and all the poles in the upper half of complex $\omega$-plane. Therefore, $\mathcal{U}^T(\tau)$ becomes 
\begin{align}
\mathcal{U}^T(\tau)=\lim_{\epsilon \to 0}\sum_{n=-\infty}^{\infty}e^{-i\epsilon\omega_n}\big[(1-C)\{(\tilde{\eta}_A+\tilde{f}_A)\mathcal{Q}^T_{1}+(\tilde{\eta}_B+\tilde{f}_B)\mathcal{Q}^T_{2}\}\nonumber\\+C(\tilde{\eta}_A\mathcal{R}_1^T+\tilde{\eta}_B\mathcal{R}_2^T+\tilde{f}_A\mathcal{R}_3^T)\big],
\label{U-tau}
\end{align}
where
\begin{align}
\mathcal{Q}^T_1&=\mathcal{R}_1^T=(\bar{G}_{11},\bar{G}_{12},i\omega_n \bar{G}_{11},i\omega_n \bar{G}_{12}),\nonumber\\
\mathcal{Q}^T_2&=\mathcal{R}_2^T=(\bar{G}_{12},\bar{G}_{11},i\omega_n \bar{G}_{12},i\omega_n \bar{G}_{11}),\nonumber\\
\mathcal{R}_3^T&=[\bar{G}_{11}+\alpha \bar{G}_{12},\bar{G}_{12}+\alpha \bar{G}_{11},i\omega_n(\bar{G}_{11}+\alpha \bar{G}_{12}),\nonumber\\&\hspace{3cm}i\omega_n (\bar{G}_{12}+\alpha \bar{G}_{11})],\nonumber
\end{align}
and $\bar{G}_{ij}=[\bar{G}(\omega_n)]_{ij}$. From \eref{U-tau}, one can easily find the mean and correlation of $\mathcal{U}(\tau)$, and these are
\begin{align}
\langle \mathcal{U}(\tau)\rangle&=0,\label{mean-variance-1}\\
\langle \mathcal{U}(\tau)\mathcal{U}^T(\tau)\rangle=&\dfrac{T\gamma}{\pi}\int_{-\infty}^{\infty}d\omega\big[(1-C)\{(1+\theta)\mathcal{Q}_1\mathcal{Q}_1^\dagger+(1+\theta \alpha^2)\mathcal{Q}_2\mathcal{Q}_2^\dagger\}+C\{\mathcal{R}_1\mathcal{R}_1^\dagger\nonumber\\&+\mathcal{R}_2\mathcal{R}_2^\dagger+\theta \mathcal{R}_3\mathcal{R}_3^\dagger\}\big].
\label{mean-variance-2}
\end{align}
Since $\mathcal{U}$ is linear in thermal Gaussian noises and external stochastic Gaussian forces, the steady state distribution can be written using the mean and correlation given above: 
\begin{equation}
P(\mathcal{U},\tau\to\infty|\mathcal{U}_0)=P^{full}_{ss}(\mathcal{U})=\dfrac{e^{-\frac{1}{2}\mathcal{U}^T\mathcal{M}^{-1}\mathcal{U}}}{\sqrt{(2\pi)^4\det\mathcal{M}}},
\label{full-ss-dist}
\end{equation}
where $\mathcal{M}_{ij}=\langle \mathcal{U}(\tau)\mathcal{U}^T(\tau)\rangle_{ij}.$

Using \erefss{X-fourier} and \erefs{V-fourier}, we write $\tilde{x}_B(\omega_n)$ and $\tilde{v}_{A}(\omega_n)$ as
\begin{align}
\tilde{x}_{B}(\omega_n)=&\bar{G}_{12}(\tilde{\eta}_{A}+\tilde{f}_{A})+\bar{G}_{11}(\tilde{\eta}_{B}+\tilde{f}_{B})-\dfrac{1}{\tau}\mathcal{Q}_3^T\Delta U,\label{x-small}\\
\tilde{v}_{A}(\omega_n)=&i\omega[\bar{G}_{11}(\tilde{\eta}_{A}+\tilde{f}_{A})+\bar{G}_{12}(\tilde{\eta}_{B}+\tilde{f}_{B})]+\dfrac{1}{\tau}\mathcal{Q}_4^T\Delta U\label{v-small},
\end{align}
where
\begin{align}
\mathcal{Q}_3^T=&[(\gamma+i\omega_n m)\bar{G}_{12},(\gamma+i\omega_n m)\bar{G}_{11},m \bar{G}_{12},m\bar{G}_{11}],\nonumber\\
\mathcal{Q}_{4}^T=&\big([\bar{G}\bar{\Phi}]_{11},[\bar{G}\bar{\Phi}]_{12},-i\omega_n m \bar{G}_{11},-i\omega_n m \bar{G}_{12}\big).\nonumber
\end{align}
From \eref{W}, $\mathcal{W}$ can be written as a sum of $\mathcal{W}_1$ and $\mathcal{W}_2$:
\begin{align}
\mathcal{W}=\mathcal{W}_1+\mathcal{W}_2,\label{sum-W} \quad \quad \text{where}\\
\mathcal{W}_1=\dfrac{1}{T}\int_0^\tau dt\ f_A(t)v_A(t),\\
\mathcal{W}_2=\dfrac{\Pi k}{T}\int_0^\tau dt\ x_B(t)v_A(t).
\end{align}
Using the finite time Fourier transform, we write $\mathcal{W}_1$ as
\begin{equation}
\mathcal{W}_1=\dfrac{\tau}{2T}\sum_{n=-\infty}^\infty [\tilde{f}_A(\omega_n)\tilde{v}_A(-\omega_n)+\tilde{f}_A(-\omega_n)\tilde{v}_A(\omega_n)].
\end{equation}
Substituting $\tilde{v}_A(\omega_n)$ from \eref{v-small} in the above equation, we get
\begin{align}
\mathcal{W}_1=\dfrac{\tau}{2T}\sum_{n=-\infty}^\infty \bigg[i\omega_n\{\bar{G}_{11}(\tilde{\eta}_A+\tilde{f}_A)\tilde{f}^*_A+\bar{G}_{12}(\tilde{\eta}_B+\tilde{f}_B)\tilde{f}^*_A
\nonumber\\-\bar{G}^*_{11}(\tilde{\eta}^*_A+\tilde{f}^*_A)\tilde{f}_A-\bar{G}^*_{12}(\tilde{\eta}^*_B+\tilde{f}^*_B)\tilde{f}_A\}+\dfrac{\tilde{f}_A\mathcal{Q}_4^\dagger \Delta \mathcal{U}}{\tau}
\nonumber\\+\dfrac{\tilde{f}^*_A\Delta \mathcal{U}^T \mathcal{Q}_4}{\tau}\bigg],
\end{align}
where $\bar{G}_{ij}^*=[\bar{G}(-\omega_n)]_{ij}$.

Similarly, we can write $\mathcal{W}_2$ as
\begin{align}
\mathcal{W}_2=\dfrac{\Pi k \tau}{2T}\sum_{n=-\infty}^\infty [\tilde{x}_B(\omega_n)\tilde{v}_A(-\omega_n)+\tilde{x}_B(-\omega_n)\tilde{v}_A(\omega_n)].
\end{align}
Substituting $\tilde{x}_B(\omega_n)$ and $\tilde{v}_A(\omega_n)$ from \erefss{x-small} and \erefs{v-small}, respectively, in the above equation, we get
\begin{align}
\mathcal{W}_2=&\dfrac{\Pi k \tau}{2T}\sum_{n=-\infty}^\infty\bigg[i\omega_n\big\{[\bar{G}_{11}(\tilde{\eta}_A+\tilde{f}_A)+\bar{G}_{12}(\tilde{\eta}_B+\tilde{f}_B)][\bar{G}^*_{12}(\tilde{\eta}_A^*+\tilde{f}^*_A)+\bar{G}_{11}^*(\tilde{\eta}^*_B+\tilde{f}^*_B)]-[\bar{G}_{12}(\tilde{\eta}_A+\tilde{f}_A)\nonumber\\&+\bar{G}_{11}(\tilde{\eta}_B+\tilde{f}_B)]
[\bar{G}^*_{11}(\tilde{\eta}_A^*+\tilde{f}^*_A)+\bar{G}_{12}^*(\tilde{\eta}^*_B+\tilde{f}^*_B)]\big\}+\dfrac{\mathcal{Q}_4^\dagger \Delta \mathcal{U}}{\tau}[\bar{G}_{12}(\tilde{\eta}_A+\tilde{f}_A)+\bar{G}_{11}(\tilde{\eta}_B+\tilde{f}_B)]
\nonumber\\&+\dfrac{\Delta \mathcal{U}^T\mathcal{Q}_4}{\tau}[\bar{G}^*_{12}(\tilde{\eta}^*_A+\tilde{f}^*_A)+\bar{G}^*_{11}(\tilde{\eta}^*_B+\tilde{f}^*_B)]+\dfrac{i\omega_n \Delta \mathcal{U}^T \mathcal{Q}_3}{\tau}[\bar{G}_{11}^*(\tilde{\eta}^*_A+\tilde{f}^*_A)+\bar{G}^*_{12}(\tilde{\eta}^*_B+\tilde{f}^*_B)]
\nonumber\\&-\dfrac{i\omega_n \mathcal{Q}_3^\dagger \Delta \mathcal{U}}{\tau}[\bar{G}_{11}(\tilde{\eta}_A+\tilde{f}_A)+\bar{G}_{12}(\tilde{\eta}_B+\tilde{f}_B)]-\dfrac{\Delta \mathcal{U}^T(\mathcal{Q}_3\mathcal{Q}_4^\dagger+\mathcal{Q}_4\mathcal{Q}_3^\dagger)\Delta \mathcal{U}}{\tau^2}\bigg].
\end{align}
The restricted moment generating function for $\mathcal{W}$ is given as
\begin{align}
Z^{\kappa}_\mathcal{W}(\lambda,\mathcal{U},\tau|\mathcal{U}_0)&=\langle e^{-\lambda \mathcal{W}}\delta[\mathcal{U}-\mathcal{U}(\tau)]\rangle_{\mathcal{U}_0}\nonumber\\&=\int \frac{d^4\bar\sigma}{(2\pi)^4}e^{i\bar\sigma^T \mathcal{U}}\langle e^{\mathcal{E}(\tau)} \rangle_{\mathcal{U}_0},
\end{align}
where we have use the integral representation of the Dirac delta function. In the above equation, $\mathcal{E}(\tau)=-\lambda \mathcal{W}-i\bar\sigma^T \mathcal{U}(\tau)$. Using \erefss{U-tau} and \erefs{sum-W}, we write $\mathcal{E}(\tau)$ as
\begin{align}
\mathcal{E}(\tau)=\sum_{n=1}^{\infty}\left[-\dfrac{\lambda \tau}{T}\zeta^{T}_{n}\mathcal{C}_{n}\zeta^{*}_{n}+\zeta^{T}_{n}\mathbb{\beta}_{n}+\beta_{-n}^{T}\zeta^{*}_{n}+\dfrac{\lambda \Pi k}{T\tau}|\mathcal{Q}_{n}|^{2} \right]-\dfrac{\lambda \tau}{2T}\zeta_{0}^{T}\mathcal{C}_{0}\zeta_{0}+\zeta^{T}_{0}\beta_{0}+\dfrac{\lambda \Pi k}{2T\tau}\mathcal{Q}_{0}^{2},
\end{align}
where $\mathcal{C}_n=\mathcal{C}_n^\mathrm{I}+\Pi k \mathcal{C}_n^{\mathrm{II}}$ and $|\mathcal{Q}_n|^2=\Delta \mathcal{U}^T(\mathcal{Q}_3\mathcal{Q}_4^\dagger+\mathcal{Q}_4\mathcal{Q}_3^\dagger)\Delta \mathcal{U}$.

For \emph{uncorrelated forces} ($\langle f_A(t)f_{B}(t^\prime)\rangle=0$ for all $t$, $t^\prime$), 
the row vector $\zeta_n^T=(\tilde{\eta}_A,\tilde{\eta}_B,\tilde{f}_A,\tilde{f}_B)$, the matrix $\mathcal{C}_n^\mathrm{I}$ is 
\[
\mathcal{C}^{\mathrm{I}}_{n}=\begin{pmatrix}
   0&0&i\omega_n \bar{G}_{11}&0\\
   0&0&i\omega_n \bar{G}_{12}&0\\
    -i\omega_n \bar{G}^*_{11}&-i\omega_n \bar{G}^*_{12}&i\omega_n [\bar{G}_{11}- \bar{G}^*_{11}]&-i\omega_n \bar{G}^*_{12}\\
0&0&i\omega_n \bar{G}_{12}&0\\
  \end{pmatrix},
\]and the matrix $\mathcal{C}_n^\mathrm{II}$ is 
\begin{equation*}
\mathcal{C}^{\mathrm{II}}_n=
\begin{pmatrix}
\mathcal{C}_{11}&&\mathcal{C}_{12}&&\mathcal{C}_{13}&&\mathcal{C}_{14}\\
\mathcal{C}^*_{12}&&\mathcal{C}_{22}&&\mathcal{C}_{23}&&\mathcal{C}_{24}\\
\mathcal{C}_{13}^*&&\mathcal{C}_{23}^*&&\mathcal{C}_{33}&&\mathcal{C}_{34}\\
\mathcal{C}_{14}^*&&\mathcal{C}_{24}^*&&\mathcal{C}_{34}^*&&\mathcal{C}_{44}
\end{pmatrix}
\end{equation*}
whose matrix elements are
\begin{align*}
\mathcal{C}_{11}=&-\mathcal{C}_{22}=\mathcal{C}_{33}=-\mathcal{C}_{44}=\mathcal{C}_{13}=i\omega_n[\bar{G}_{11}\bar{G}_{12}^*-\bar{G}_{12}\bar{G}_{11}^*],\\
\mathcal{C}_{12}=&\mathcal{C}_{14}=\mathcal{C}_{34}=-\mathcal{C}_{23}=i\omega_n[|\bar{G}_{11}|^2-|\bar{G}_{12}|^2],\\
\mathcal{C}_{24}=&-\mathcal{C}_{11},\\
\mathcal{C}_{ij}^*=&\mathcal{C}_{ij}(-\omega_n).
\end{align*}
The column vector $\beta_n$ is given by
\begin{equation*}
\beta_n=-\frac{\lambda}{T}
\begin{pmatrix}
\bar{a}_{11}^T\Delta \mathcal{U}\\
\bar{a}_{21}^T\Delta \mathcal{U}\\
\bar{a}_{31}^T\Delta \mathcal{U}\\
\bar{a}_{41}^T\Delta \mathcal{U}
\end{pmatrix}
-ie^{-i\epsilon \omega_n}\begin{pmatrix}
\mathcal{Q}_{1}^T\bar{\sigma}\\
\mathcal{Q}_{2}^T\bar{\sigma}\\
\mathcal{Q}_{1}^T\bar{\sigma}\\
\mathcal{Q}_{2}^T\bar{\sigma}
\end{pmatrix},  \quad\text{in which}
\end{equation*}
\begin{align*}
\bar{a}_{11}^T&=-\Pi k(i\omega_n \mathcal{Q}_3^\dagger \bar{G}_{11}-\mathcal{Q}_4^\dagger \bar{G}_{12}),\\
\bar{a}_{21}^T&=\bar{a}_{41}^T=-\Pi k(i\omega_n \mathcal{Q}_3^\dagger \bar{G}_{12}-\mathcal{Q}_4^\dagger \bar{G}_{11}),\\
\bar{a}_{31}^T&=-\Pi k(i\omega_n \mathcal{Q}_3^\dagger \bar{G}_{11}-\mathcal{Q}_4^\dagger \bar{G}_{12})+\mathcal{Q}_4^\dagger.\\
\end{align*}
For the second choice of external forces [i.e., $f_B(t)=\alpha f_A(t)$], the row vector
$\zeta_n^T=(\tilde{\eta}_A,\tilde{\eta}_B,\tilde{f}_A)$, the matrix $\mathcal{C}_n^\mathrm I$ is
\[
\mathcal{C}^{\mathrm{I}}_{n}=\begin{pmatrix}
   0&0&i\omega_n \bar{G}_{11}\\
   0&0&i\omega_n \bar{G}_{12}\\
    -i\omega_n \bar{G}^*_{11}&-i\omega_n \bar{G}^*_{12}&i\omega_n[(\bar{G}_{11}-\bar{G}^*_{11})+\\
    &&\alpha(\bar{G}_{12}-\bar{G}^*_{12})]\\
  \end{pmatrix},
\]
and the matrix $\mathcal{C}_n^{\mathrm{II}}$ is
\[
\mathcal{C}^{\mathrm{II}}_{n}=\begin{pmatrix}
   \mathcal{C}_{11}&&\mathcal{C}_{12}&&\mathcal{C}_{13}\\
   \mathcal{C}_{12}^*&&\mathcal{C}_{22}&&\mathcal{C}_{23}\\
    \mathcal{C}^*_{13}&&\mathcal{C}_{23}^*&&\mathcal{C}_{33}
  \end{pmatrix}
\]
whose matrix elements are
\begin{align*}
\mathcal{C}_{11}=&-\mathcal{C}_{22}=i\omega_n[ \bar{G}_{11}\bar{G}_{12}^*-\bar{G}^*_{11}\bar{G}_{12}],\\
\mathcal{C}_{12}=&i\omega_n [|\bar{G}_{11}|^2-|\bar{G}_{12}|^2],\\
\mathcal{C}_{13}=&-\mathcal{C}_{23}=\mathcal{C}_{11}+\alpha \mathcal{C}_{12},\\
\mathcal{C}_{33}=&(1-\alpha^2)\mathcal{C}_{11},\\
\mathcal{C}_{ij}^*=&\mathcal{C}_{ij}(-\omega_n).\end{align*}
The column vector $\beta_n$ in this case is given by
\begin{equation*}
\beta_n=-\frac{\lambda}{T}
\begin{pmatrix}
\bar{c}_{11}^T\Delta \mathcal{U}\\
\bar{c}_{21}^T\Delta \mathcal{U}\\
\bar{c}_{31}^T\Delta \mathcal{U}
\end{pmatrix}
-ie^{-i\epsilon \omega_n}\begin{pmatrix}
\mathcal{R}_{1}^T\bar\sigma\\
\mathcal{R}_{2}^T\bar\sigma\\
\mathcal{R}_{3}^T\bar\sigma
\end{pmatrix}, \quad\text{in which}
\end{equation*}
\begin{align*}
\bar{c}_{11}^T=&-\Pi k(i\omega_n \mathcal{Q}_3^\dagger \bar{G}_{11}-\mathcal{Q}_4^\dagger \bar{G}_{12}),\\
\bar{c}_{21}^T=&-\Pi k(i\omega_n \mathcal{Q}_3^\dagger \bar{G}_{12}-\mathcal{Q}_4^\dagger \bar{G}_{11}),\\
\bar{c}_{31}^T=&-\Pi k[i\omega_n \mathcal{Q}_3^\dagger(\bar{G}_{11}+\alpha \bar{G}_{12})-\mathcal{Q}_4^\dagger(\bar{G}_{12}+\alpha \bar{G}_{11})]+\mathcal{Q}_4^\dagger.
\end{align*}

Therefore, we get
\begin{align}
&\langle e^{\mathcal{E}(\tau)}\rangle_{\mathcal{U}_{0}}=\bigg\langle \exp\bigg[-\frac{\lambda \tau}{2T} \zeta^{T}_{0} \mathcal{C}_{0} \zeta_{0}+\zeta^{T}_{0}\beta_{0}+\dfrac{\lambda \Pi k}{2T\tau}\mathcal{Q}_{0}^{2} \bigg]\bigg\rangle\nonumber\\
&\times\prod_{n=1}^{\infty}\bigg\langle \exp \bigg[-\frac{\lambda \tau}{T} \zeta^{T}_{n} \mathcal{C}_{n} \zeta^{*}_{n}+\zeta^{T}_{n}\beta_{n}+\beta_{-n}^{T}\zeta^{*}_{n}+\dfrac{\lambda \Pi k}{T\tau}|\mathcal{Q}_{n}|^{2} \bigg]\bigg\rangle.
\end{align}
In the above equation, the angular brackets represent the average over the joint Gaussian distribution of thermal and external noises $\zeta_n$.  For terms $n\geq 1$,  the average is done independently on each term using the distribution \eref{prob-n-1}. Similarly, for $n=0$, the average is computed with respect to the distribution \eref{prob-n-2}.

Computation of averages yields 
\begin{align}
\langle e^{\mathcal{E}(\tau)}\rangle_{\mathcal{U}_{0}}=e^{(\tau/\tau_{\gamma})\mu_\kappa(\lambda)}e^{\frac{1}{2}\sum_{n=-\infty}^{\infty}(\beta_{-n}^{T}\bar\Omega_{n}^{-1}\beta_{n}+\frac{\lambda \Pi k}{T\tau}|\mathcal{Q}_{n}|^{2})},
\end{align}
where $\bar\Omega_n=(\Lambda^{-1}+\lambda \tau \mathcal{C}_{n}/T)$.
In the large time limit ($\tau/\tau_\gamma\to\infty$), we convert the summation into integration. Therefore, we get
\begin{align}
\langle e^{\mathcal{E}(\tau)}\rangle_{\mathcal{U}_{0}}\approx e^{(\tau/\tau_{\gamma}) \mu_\kappa(\lambda)}e^{-\frac{1}{2} \bar\sigma^{T} \bar{H}_{1}\bar\sigma+i \Delta \mathcal{U}^{T}\bar{H}_{2} \bar\sigma+\frac{1}{2}\Delta \mathcal{U}^{T} \bar{H}_{3}\Delta \mathcal{U}},
\end{align}
where
\begin{align}
\mu_\kappa(\lambda)=&-\dfrac{\tau_{\gamma}}{4 \pi}\int_{-\infty}^{\infty} d\omega\ \ln \big[\det{(\Lambda \bar\Omega)}\big],\label{mu-k}\\
\bar{H}_{1}=&\dfrac{\tau}{2 \pi}\int_{-\infty}^{\infty}d\omega\ \bar\rho^{T}  \bar\Omega^{-1} \bar\phi,\label{h1-k}\\
\bar{H}_{2}=&-\dfrac{\tau}{2 \pi}\int_{-\infty}^{\infty}d\omega\ e^{-i \omega\epsilon}\bar{a}_1^{T}\bar\Omega^{-1}\bar\phi,\label{h2-k}\\
\bar{H}_{3}=&\dfrac{\tau}{2 \pi}\int_{-\infty}^{\infty}d\omega\ \bigg[\bar{a}_1^T\bar\Omega^{-1}\bar{a}_2+\dfrac{\lambda \Pi k}{T\tau}(\mathcal{Q}_{3}\mathcal{Q}_{4}^{\dagger}+\mathcal{Q}_{4}\mathcal{Q}_{3}^{\dagger})\bigg].\label{h3-k}
\end{align}
The vectors $\bar\rho^T$, $\bar{a}_1^T$, $\bar\phi$, and $\bar{a}_2$ are given in \tref{table-1}, where
\begin{table}[!h]
\begin{tabular}{|p{3cm}||p{5cm}|p{5cm}|}
 \hline
\hspace{0.5cm}Vectors&\hspace{2cm}$C=0$&\hspace{2.0cm}$C=1$\\
 \hline 
\hspace{1cm}$\bar\rho^T$&\hspace{0.5cm}$~~~~(\mathcal{Q}_1^*,\mathcal{Q}_2^*,\mathcal{Q}_1^*,\mathcal{Q}_2^*)$&\hspace{0.7cm}$~~~~(\mathcal{R}_1^*,\mathcal{R}_2^*,\mathcal{R}_3^*)$\\
\hspace{1cm}$\bar{a}_1^T$&\hspace{0.001cm}$~~~~-\lambda/T(\bar{b}_{11},\bar{b}_{12},\bar{b}_{13},\bar{b}_{14})$&\hspace{0.05cm}$~~~~-\lambda/T(\bar{d}_{11},\bar{d}_{12},\bar{d}_{13})$\\
\hspace{1cm}$\bar\phi$&\hspace{1cm}$~~~~~~\begin{pmatrix}
\mathcal{Q}_{1}^{T}\\
\mathcal{Q}_{2}^{T}\\
\mathcal{Q}_{1}^{T}\\
\mathcal{Q}_{2}^{T}
 \end{pmatrix}$&\hspace{1.cm}$~~~~~\begin{pmatrix}
\mathcal{R}_{1}^{T}\\
\mathcal{R}_{2}^{T}\\
\mathcal{R}_{3}^{T}
 \end{pmatrix}$\\
  \hspace{1cm}$\bar{a}_2$&\hspace{0.4cm} $~~~-
 \dfrac{\lambda}{T}\begin{pmatrix}
\bar{a}_{11}^T\\
\bar{a}_{21}^T\\
\bar{a}_{31}^T\\
\bar{a}_{41}^T
 \end{pmatrix}$&\hspace{0.4cm}$~~~-
 \dfrac{\lambda}{T}\begin{pmatrix}
\bar{c}_{11}^T\\
\bar{c}_{21}^T\\
\bar{c}_{31}^T
 \end{pmatrix}$\\
   \hline
\end{tabular}
\caption{\label{table-1} The vectors $\bar\rho^T$, $\bar{a}_1^T$, $\bar\phi$, and $\bar{a}_2$ are shown.}
\end{table}
\begin{align*}
\bar{b}_{11}&=\Pi k[i\omega \bar{G}^*_{11}\mathcal{Q}_3+\bar{G}^*_{12}\mathcal{Q}_4],\\
\bar{b}_{12}&=\Pi k[i\omega \bar{G}^*_{12}\mathcal{Q}_3+\bar{G}^*_{11}\mathcal{Q}_4]=\bar{b}_{14},\\
\bar{b}_{13}&=\Pi k[i\omega \bar{G}^*_{11}\mathcal{Q}_3+\bar{G}_{12}^*\mathcal{Q}_4]+\mathcal{Q}_4,\\
\bar{d}_{11}&=\Pi k[i\omega \bar{G}^*_{11}\mathcal{Q}_3+\bar{G}^*_{12}\mathcal{Q}_4],\\
\bar{d}_{12}&=\Pi k[i\omega \bar{G}^*_{12}\mathcal{Q}_3+\bar{G}^*_{11}\mathcal{Q}_4],\\
\bar{d}_{13}&=\Pi k[i\omega (\bar{G}^*_{11}+\alpha \bar{G}_{12}^*)\mathcal{Q}_3+(\bar{G}_{12}^*+\alpha \bar{G}_{11}^*)\mathcal{Q}_4]+\mathcal{Q}_4.
\end{align*}
The restricted moment generating function for $\mathcal{W}$  can be written as  
\begin{align}
Z^\kappa_\mathcal{W}(\lambda,\mathcal{U},\tau|\mathcal{U}_{0})&\approx e^{(\tau/\tau_{\gamma}) \mu_\kappa(\lambda)} e^{\frac{1}{2}\Delta \mathcal{U}^{T}\bar{H}_{3}\Delta \mathcal{U}}\nonumber\\&\int \dfrac{d^{4}\bar\sigma}{(2 \pi)^{4}}\ e^{i \bar\sigma^{T} \mathcal{U}}\  e^{-\frac{1}{2}\bar\sigma^{T}\bar{H}_{1}\bar\sigma}\ e^{i \bar\sigma^{T}\bar{H}_{2}^{T}\Delta \mathcal{U}}.
\end{align}
Computing the integration over $\bar\sigma$, we get
\begin{align}
Z^\kappa_{\mathcal{W}}(\lambda,\mathcal{U},\tau|\mathcal{U}_{0})\approx \dfrac{e^{\frac{1}{2}\Delta \mathcal{U}^{T}\bar{H}_{3}\Delta \mathcal{U}}}{\sqrt{(2\pi)^{4}\det{\bar{H}_{1}(\lambda)}}}~e^{(\tau/\tau_{\gamma}) \mu_\kappa(\lambda)}e^{-\frac{1}{2}(\mathcal{U}^{T}+\Delta \mathcal{U}^{T}\bar{H}_{2})\bar{H}_{1}^{-1}(\mathcal{U}+\bar{H}_{2}^{T}\Delta \mathcal{U})}.
\end{align}
We factorize the above equation in terms of the initial and the final variable (see \aref{appendix-1}) which implies $(\bar{H}_{3}-\bar{H}_{2}\bar{H}_{1}^{-1}\bar{H}_{2}^{T}-\bar{H}_{1}^{-1}\bar{H}_{2}^{T})+(\bar{H}_{3}-\bar{H}_{2}\bar{H}_{1}^{-1}\bar{H}_{2}^{T}-\bar{H}_{2}\bar{H}_{1}^{-1})^{T}=0$. Therefore, we get
\begin{align}
Z^\kappa_{\mathcal{W}}(\lambda,\mathcal{U},\tau|\mathcal{U}_{0})&\approx
\dfrac{\ e^{(\tau/\tau_{\gamma}) \mu_\kappa(\lambda)} e^{-\frac{1}{2}\mathcal{U}^{T}\bar{L}_{1}(\lambda)\mathcal{U}}e^{-\frac{1}{2}\mathcal{U}_{0}^{T}\bar{L}_{2}(\lambda)\mathcal{U}_{0}} }{\sqrt{(2 \pi)^{4}\det{\bar{H}_{1}(\lambda)}}},
\label{factor-eqn}
\end{align} 
where
\begin{align*}
&\bar{L}_{1}(\lambda)=\bar{H}_{1}^{-1}+\bar{H}_{1}^{-1}\bar{H}_{2}^{T},\\ &\bar{L}_{2}(\lambda)=-\bar{H}_{1}^{-1}\bar{H}_{2}^{T}.
\end{align*}
In the case of partial entropy production and for both choices of external forces, the steady state distribution $P_{ss}(\tilde{\mathcal{U}})$ can be obtained by integrating $P^{full}_{ss}(\mathcal{U})$ given in \eref{full-ss-dist}, over $x_B$ and\ $v_B$. Therefore, we get
\begin{equation}
P_{ss}(\tilde{\mathcal{U}})=\dfrac{\exp[-\frac{1}{2}\tilde{\mathcal{U}}^T\tilde{\mathbb{H}}^{-1}_P\tilde{U}]}{\sqrt{(2\pi)^2 \det\tilde{\mathbb{H}}_P}}.
\label{ss-1}
\end{equation}
Similarly, in the case of apparent entropy production, the steady state distribution $\tilde P_{ss}(\tilde{\mathcal{U}})$ can be obtained from \erefss{full-dynamics-1} and \erefs{full-dynamics-3} at $k=0$ and given by
\begin{equation}
\tilde{P}_{ss}(\tilde{\mathcal{U}})=\dfrac{\exp[-\frac{1}{2}\tilde{\mathcal{U}}^T\tilde{\mathbb{H}}^{-1}_A\tilde{\mathcal{U}}]}{\sqrt{(2\pi)^2 \det\tilde{\mathbb{H}}_A}}.
\label{ss-2}
\end{equation}
In \erefss{ss-1} and \erefs{ss-2}, $\tilde{\mathcal{U}}=(x_A,v_A)^T$, and the matrices $\tilde{\mathbb{H}}_P$ and $\tilde{\mathbb{H}}_A$ are given in \eref{HP-HA}.

The system entropy production of particle A in the coupled system is given by 
\begin{align}
\Delta \mathcal{S}^A_{sys}&=\dfrac{\Pi}{2}[\mathcal{U}^T\mathbb{H}_P^{-1} \mathcal{U}-\mathcal{U}_0^T\mathbb{H}_P^{-1} \mathcal{U}_0]\nonumber\\&
+\dfrac{1-\Pi}{2}[\mathcal{U}^T\mathbb{H}_A^{-1} \mathcal{U}-\mathcal{U}_0^T\mathbb{H}_A^{-1} \mathcal{U}_0],
\end{align}
where
\begin{align}
\mathbb{H}_P^{-1}&=\mathrm{diag}(1/\mathbb{H}^{11}_P,0,1/\mathbb{H}^{33}_P,0),\nonumber\\
\mathbb{H}_A^{-1}&=\mathrm{diag}(1/\mathbb{H}^{11}_A,0,1/\mathbb{H}^{33}_A,0),\nonumber\\ 
\mathcal{U}^T&=(x_A,x_B,v_A,v_B).\nonumber
\end{align}
Total entropy production of particle A in the coupled system given in \eref{general-EP} can be written as
\begin{equation}
\Delta \mathcal{S}^A_{tot}=\mathcal{W}-\dfrac{1}{2}\mathcal{U}^T\mathbb{H}^{-1}\mathcal{U}+\dfrac{1}{2}\mathcal{U}_0^T\mathbb{H}^{-1}\mathcal{U}_0, 
\label{Stot}
\end{equation}
where
\begin{equation*}
\mathbb{H}^{-1}=\Pi(\Xi_P-\mathbb{H}_P^{-1})+(1-\Pi)(\Xi_A-\mathbb{H}_A^{-1}).
\end{equation*}
The diagonal matrices $\Xi_P$ and $\Xi_A$ are given by
\begin{align*}
&\Xi_P=\mathrm{diag}((k+k_0)/T,0,m/T,0),\\&\Xi_A=\mathrm{diag}(k_0/T,0,m/T,0).
\end{align*}
Therefore, the restricted moment generating function for $\Delta \mathcal{S}_{tot}^A$ [see \eref{Stot}] is given as
\begin{align}
Z_\kappa(\lambda,\mathcal{U},\tau|\mathcal{U}_{0})&\approx\dfrac{\ e^{(\tau/\tau_{\gamma}) \mu_\kappa(\lambda)} e^{-\frac{1}{2}\mathcal{U}^{T}\tilde{\bar{L}}_{1}(\lambda)\mathcal{U}}e^{-\frac{1}{2}\mathcal{U}_{0}^{T}\tilde{\bar{L}}_{2}(\lambda)\mathcal{U}_{0}} }{\sqrt{(2 \pi)^{4}\det{\bar{H}_{1}(\lambda)}}},\end{align} with
\begin{align*}
&\tilde{\bar{L}}_{1}(\lambda)=\bar{L}_{1}(\lambda)-\lambda \mathbb{H}^{-1},\\&\tilde{\bar{L}}_{2}(\lambda)=\bar{L}_{2}(\lambda)+\lambda \mathbb{H}^{-1}.
\end{align*}
The moment generating function is obtained by integrating over the initial steady state distribution and the final variable
\begin{align}
Z_\kappa(\lambda)&=\int d\mathcal{U} \int d\mathcal{U}_{0}P^{full}_{ss}(\mathcal{U}_{0})Z_\kappa(\lambda,\mathcal{U},\tau|\mathcal{U}_{0})\nonumber\\&\approx g_\kappa(\lambda) e^{(\tau/\tau_{\gamma}) \mu_\kappa(\lambda)}, 
\end{align} 
where
\begin{equation}
 g_\kappa(\lambda)=\left[\det[{\bar{H}_{1}(\lambda)\bar{H}_{1}(0)\tilde{\bar{L}}_{1}(\lambda)}]\det{[\bar{H}_{1}^{-1}(0)+\tilde{\bar{L}}_{2}(\lambda)]}\right]^{-1/2}.
 \label{glambda-eqn}
 \end{equation}

\vskip 2cm

\bibliographystyle{acm}
\bibliography{ref}
\end{document}